\documentclass[aps,prx,twocolumn,showpacs,superscriptaddress,preprintnumbers,longbibliography]{revtex4-2} 
\usepackage{graphicx}  
\usepackage{dcolumn}   
\usepackage{bm}        
\usepackage{amssymb}   
\usepackage{amsmath}
\usepackage{comment}
\usepackage{graphicx}
\usepackage{color}
\usepackage{slashed}
\usepackage{braket}
\usepackage{amssymb,amsmath,graphicx,color}
\usepackage[tight]{subfigure}
\usepackage[export]{adjustbox}
\usepackage[colorlinks=true,citecolor=blue,linkcolor=red]{hyperref}
\usepackage{multirow}
\usepackage{rotating,booktabs}
\usepackage[verbose]{placeins}
\usepackage{mathrsfs}
\usepackage{tikz}
\usetikzlibrary{quantikz}

\newcommand{\Eq}[1]{Eq.~(\ref{#1})}

\newcommand{\Eqs}[2]{Eqs.~(\ref{#1}-\ref{#2})}
\newcommand{\Fig}[1]{Fig.~\ref{#1}}

\newcommand{\Figs}[2]{Fig.~\ref{#1}-\ref{#2}}
\newcommand{\ua}{\mathord{\uparrow}}
\newcommand{\da}{\mathord{\downarrow}}

\makeatletter
\def\blfootnote{\xdef\@thefnmark{}\@footnotetext}
\makeatother
\begin{document}
\preprint{UMD-PP-022-09, IQuS@UW-21-032}

\title{
Quantum computation of dynamical quantum phase transitions \\ and entanglement tomography in a lattice gauge theory
}

\author{Niklas~Mueller}
\email{niklasmu@uw.edu}
\thanks{corresponding author. All other authors contributed equally and are listed in alphabetical order.}
\affiliation{Maryland Center for Fundamental Physics and Department of Physics, University of Maryland, College Park, MD 20742, USA}
\affiliation{Joint Quantum Institute, NIST/University of Maryland, College Park, MD 20742, USA.}
\affiliation{InQubator for Quantum Simulation (IQuS), Department of Physics, University of Washington, Seattle, WA 98195, USA.}

\author{Joseph~A.~Carolan}
\affiliation{Department of Computer Science, University of Illinois at Urbana-Champaign, Urbana, Illinois 61801, USA}

\author{Andrew~Connelly}
\affiliation{Department of Physics, North Carolina State University, Raleigh, North Carolina 27695, USA}

\author{Zohreh~Davoudi}
\email{davoudi@umd.edu}
\affiliation{Maryland Center for Fundamental Physics and Department of Physics, University of Maryland, College Park, MD 20742, USA}
\affiliation{Joint Center for Quantum Information and Computer Science, NIST/University of Maryland, College Park, MD 20742 USA}

\author{Eugene~F.~Dumitrescu}
\email{dumitrescuef@ornl.gov}
\affiliation{Computational Sciences and Engineering Division, Oak Ridge National Laboratory, Oak Ridge, Tennessee 37831, USA}

\author{K\"ubra~Yeter-Aydeniz}
\affiliation{The MITRE Corporation, Emerging Engineering and Physical Sciences Department, 7515 Colshire Drive, McLean, VA 22102-7539, USA}
\date{\today}

\begin{abstract}
Strongly-coupled gauge theories far from equilibrium may exhibit unique features that could illuminate the physics of the early universe and of hadron and ion colliders. Studying real-time phenomena has proven challenging with classical-simulation methods, but is a natural application of quantum simulation. To demonstrate this prospect, we quantum compute non-equal time correlation functions and perform entanglement tomography of non-equilibrium states of a simple lattice gauge theory, the Schwinger model, using a trapped-ion quantum computer by IonQ Inc. As an ideal target for near-term devices, a recently-predicted [Zache \textit{et al.}, Phys. Rev. Lett. 122, 050403 (2019)] dynamical quantum phase transition in this model is studied by preparing, quenching, and tracking the subsequent non-equilibrium dynamics in three ways: i) overlap echos signaling dynamical transitions, ii) non-equal time correlation functions with an underlying topological nature, and iii) the entanglement structure of non-equilibrium states, including entanglement Hamiltonians. These results constitute the first observation of a dynamical quantum phase transition in a lattice gauge theory on a quantum computer, and are a first step toward investigating topological phenomena in nuclear and high-energy physics using quantum technologies. 

\end{abstract}

\blfootnote{- This manuscript has been authored by UT-Battelle, LLC, under Contract No.~DE-AC0500OR22725 with the U.S.~Department of Energy (DOE). The United States Government retains and the publisher, by accepting the article for publication, acknowledges that the United States Government retains a non-exclusive, paid-up, irrevocable, world-wide license to publish or reproduce the published form of this manuscript, or allow others to do so, for the United States Government purposes. The DOE will provide public access to these results of federally sponsored research in accordance with the DOE's Public Access Plan.}
\blfootnote{- \copyright 2022 The MITRE Corporation. All rights reserved. Approved for public release. Distribution unlimited PR$\_$21-03848-4.}

\maketitle

\section{Introduction
\label{sec:intro}}
\noindent
Understanding how strongly-coupled quantum many-body systems behave far from equilibrium is a common goal across many physics disciplines. In nuclear- and high-energy physics, evolution of matter and the mechanisms for its thermalization and hydrodynamization, prevalent in early-universe~\cite{micha2004turbulent} and probed in ultra-relativistic heavy-ion collisions~\cite{baier2001bottom,berges2002controlled,arrizabalaga2005equilibration,balasubramanian2011holographic,grozdanov2016strong}, are not yet fully understood~\cite{berges2020thermalization}. Unfortunately, first-principles studies of such dynamical non-equilibrium processes grounded in the fundamental quantum field theories of nature have generally evaded classical-computing methods~\cite{NSAC-QIS-2019-QuantumInformationScience,davoudi2022quantum}. Addressing classically intractable problems on analog quantum simulators and digital quantum computers has been a driver of innovation and development for emergent hardware platforms, based on e.g., atomic~\cite{jaksch1998cold,lewenstein2012ultracold,bloch2012quantum}, molecular~\cite{carr2009cold}, optical~\cite{o2009photonic, noh2016quantum,atature2018material}, and solid-state or hybrid systems~\cite{kjaergaard2020superconducting,blais2021circuit,carusotto2020photonic,ozguler2022dynamics,ozguler2022numerical,cho2023direct}, see Ref.~\cite{altman2021quantum} for a review. It has further propelled progress in theory, algorithm, and co-design~\cite{cirac2012goals,hauke2012can,georgescu2014quantum,montanaro2016quantum,preskill2018quantum,cerezo2021variational,bharti2022noisy,bauer2020quantum,alexeev2021quantum,altman2021quantum}. In particular, aspects of quantum simulation of quantum field theories, including scalar field theory~\cite{jordan2012quantum,klco2019digitization,yeter2019scalar,barata2021single,tong2022provably,kurkcuoglu2021quantum,yeter2022quantum,li2023simulating} and Abelian and non-Abelian gauge theories~\cite{banerjee2012atomic,zohar2013simulating, zohar2015quantum,mildenberger2022probing,yang2016analog,zache2018quantum, davoudi2020towards,surace2020lattice, luo2020framework, andrade2022engineering,klco2018quantum, lu2019simulations, barbiero2019coupling, Chakraborty:2020uhf, shaw2020quantum, stryker2021shearing, davoudi2021toward, homeier2021z, Pederiva:2021tcd, Rajput:2021khs,martinez2016real,nguyen2022digital, mil2020scalable,zhou2021thermalization,byrnes2006simulating, zohar2013quantum,zohar2013cold,tagliacozzo2013simulation,de2021quantum,klco20202,atas20212,rahman20212,haase2021resource,kan2021lattice,davoudi2021search,davoudi2022general,Ciavarella:2021nmj,alam2022primitive,gustafson2022primitive,Paulson:2020zjd,halimeh2022gauge,Ciavarella:2021lel,lamm2019general,cohen2021quantum, gonzalez2022hardware, atas2022real,farrell2022preparations,farrell2022preparations,Murairi:2022zdg,clemente2022strategies} have been advanced considerably in recent years, see Refs.~\cite{banuls2020simulating,klco2022standard,davoudi2022quantum,bauer2023quantum} for reviews.

Simulating complex theories, such as the theory of the strong force, quantum chromodynamics (QCD), remains beyond the capabilities of the Noisy-Intermediate-Scale-Quantum (NISQ)~\cite{Preskill:2018} hardware. However, present-day access to these systems is essential in developing and tailoring optimized algorithms, obtaining a deeper understanding of hardware capabilities and limitations, and informing the upcoming hardware design of the requirements of nuclear- and high-energy physics simulations~\cite{davoudi2022quantum,catterall2022report, beck2023quantum,carlson2018quantum,cloet2019opportunities,NSAC-QIS-2019-QuantumInformationScience,davoudi2022quantum,humble2022snowmass,alam2022quantum}. With this aim, we formulate, explore, and analyze non-equilibrium phenomena in a lattice gauge theory (LGT) on IonQ's 11-qubit trapped-ion quantum computer~\cite{wright2019benchmarking, kawashima2021optimizing}, accessed through Google Cloud~\cite{GCP} and Microsoft Azure~\cite{Azure} Services. Using a workhorse of quantum exploration, the simple and well-studied Schwinger model~\cite{Banerjee:2012pg, Zohar:2015hwa, Yang:2016hjn, davoudi2020towards, surace2020lattice, luo2020framework, andrade2022engineering,yang2020observation, Chakraborty:2020uhf, shaw2020quantum, stryker2021shearing, kan2021lattice, davoudi2021toward, Pederiva:2021tcd,martinez2016real,kokail2018self,joshi2021probing,mil2020scalable,zhou2021thermalization,klco2018quantum,lu2019simulations,ferguson2021measurement,de2021quantum,nguyen2022digital,thompson2022quantum,honda2022classically,belyansky2023high}, we focus on a novel direction: exploring the non-equilibrium dynamics and topology of LGTs from their entanglement structure and from non-equal time correlation functions. Entanglement and entanglement structure~\cite{kharzeev2022quantum,cervera2017maximal,beane2019entanglement,beane2021geometry,beane2021entanglement,klco2021geometric,klco2021entanglement,klco2021entanglementspheres,klco2023entanglement,casini2014remarks,aoki2015definition,ghosh2015entanglement,lin2020comments,chen2020strong,mueller2021thermalization} are important probes of thermalization or the characterization of quantum phases, while non-equal time correlation functions are crucial tools in studying not only transport phenomena in QCD~\cite{florkowski2010phenomenology,gale2013hydrodynamic,bzdak2020mapping,meyer2009transport,Kogut:2004su,cohen2021quantum}, but also the structure of nuclei and nucleons in electron scattering experiments such as at the Electron Ion Collider (EIC)~\cite{accardi2016electron}, in neutrino-nucleus scattering~\cite{roggero2020quantum,roggero2018linear,baroni2022nuclear} at Deep Underground Neutrino Experiment (DUNE)~\cite{Dune}, and in nuclear reaction processes~\cite{farrell2023preparations}.

In particular, focusing on far-from-equilibrium dynamics after a quantum quench, we investigate dynamical quantum phase transitions (DQPTs), first proposed and observed in a range of quantum many-body systems in Refs.~\cite{heyl2013dynamical,flaschner2016observation,jurcevic2017direct,heyl2018dynamical,budich2016dynamical,zhang2017observation,tian2018direct,xu2018measuring,dumitrescu2021realizing}, and later predicted to occur in gauge theories as well~\cite{Zache:2018cqq}, see also Refs.~\cite{huang2019dynamical,halimeh2021achieving,van2022dynamical,van2022anatomy}. Importantly, this phenomenon is shown to be an ideal candidate for exploration with existing analog and digital devices, since it persists in small systems and on short time scales~\cite{Zache:2018cqq,jensen2022dynamical,pedersen2021lattice}.
DQPTs are manifest in non-equal time correlation functions and non-equal time wavefunction overlaps (Loschmidt echos), and in the Schwinger model they are related to an underlying topological transition~\cite{Zache:2018cqq}. In fact, from a phenomenological standpoint, there may be interesting connections between potential non-equilibrium dynamics that could change the value of the topological $\theta$ angle in the early universe and the strong CP problem, a possibility that quantum simulations of real-time phenomena in simple models, and eventually in QCD, may shed light on.

To achieve a first realization of a DQPT in a lattice gauge theory on quantum hardware, we perform real-time evolution of the Schwinger model in a digital scheme in a manner that is partly distinct from previous Schwinger-model simulations. The algorithm developed and implemented proceeds with minimal Trotter error (vanishing in the non-interacting limit) by transforming between the eigenbases of free and interacting parts of the  Hamiltonian, using a composite Bogoliubov-staggered  and Fourier transformation. Dynamical non-equal time correlation functions, Loschmidt echos, and a topological order parameter are obtained using a Ramsey interferometry scheme~\cite{somma2002simulating,knap2013probing,yao2016interferometric,tonielli2020ramsey}.
Utilizing random-measurement-based entanglement- (and `shadow'-) tomography methods~\cite{van2012measuring,ohliger2013efficient,pichler2016measurement,dalmonte2018quantum,elben2018renyi,vermersch2018unitary,brydges2019probing,elben2019statistical,elben2020mixed,kokail2021entanglement,huang2020predicting,zhou2020single,anshu2021sample,kokail2021entanglement,huang2021efficient,rath2021importance,huang2021demonstrating,zhao2021fermionic,neven2021symmetry,kunjummen2021shadow,elben2022randomized} R\'enyi entropies and fidelities are  quantum computed with the protocol developed in Refs.~\cite{elben2018renyi,vermersch2018unitary,brydges2019probing}. We apply a generalization of the Bisognano-Wichmann theorem~\cite{bisognano1975duality,bisognano1976duality} to extract Entanglement Hamiltonians (EHs) and entanglement spectra (ES) of non-equilibrium states~\cite{kokail2021entanglement}, see Ref. \cite{dalmonte2022entanglement} for a recent review on entanglement Hamiltonians and random-measurement protocols. 

This manuscript is organized as follows: Sec.~\ref{sec:DQPTbasics} contains i) a brief overview of the topological DQPT exhibited by the Schwinger model,  our strategy for ii) simulating the system on a quantum computer, and iii) identifying the topological phase transition by analyzing dynamical non-equal time correlation-function holonomy. In Sec.~\ref{sec:EHT}, entanglement tomography is employed to build a classical representation of the systems' entanglement structure for non-equilibrium states. Our results are summarized in Sec.~\ref{sec:conclusions}. The manuscript is supplemented by several Appendices. Appendices~\ref{app:analytics} and \ref{app:quantumalgo} include analytical details and quantum algorithms for real-time evolution of the Schwinger model, including the scheme to obtain non-equal time observables. Appendix~\ref{app:hardware} contains experimental details of the IonQ device used in this study. Appendices~\ref{app:symmerr} and \ref{app:error} provide a description of a symmetry-based error-mitigation scheme employed in this work, and a discussion of device imperfections. Appendix~\ref{app:EHTdetails} contains more details on the entanglement-tomography results. All circuits, raw data and the data shown in figures can be found at \url{https://gitlab.com/Niklas-Mueller1988/dqpt_2210.03089}.
\begin{figure*}[t]
\centering{
\includegraphics[scale=0.66]{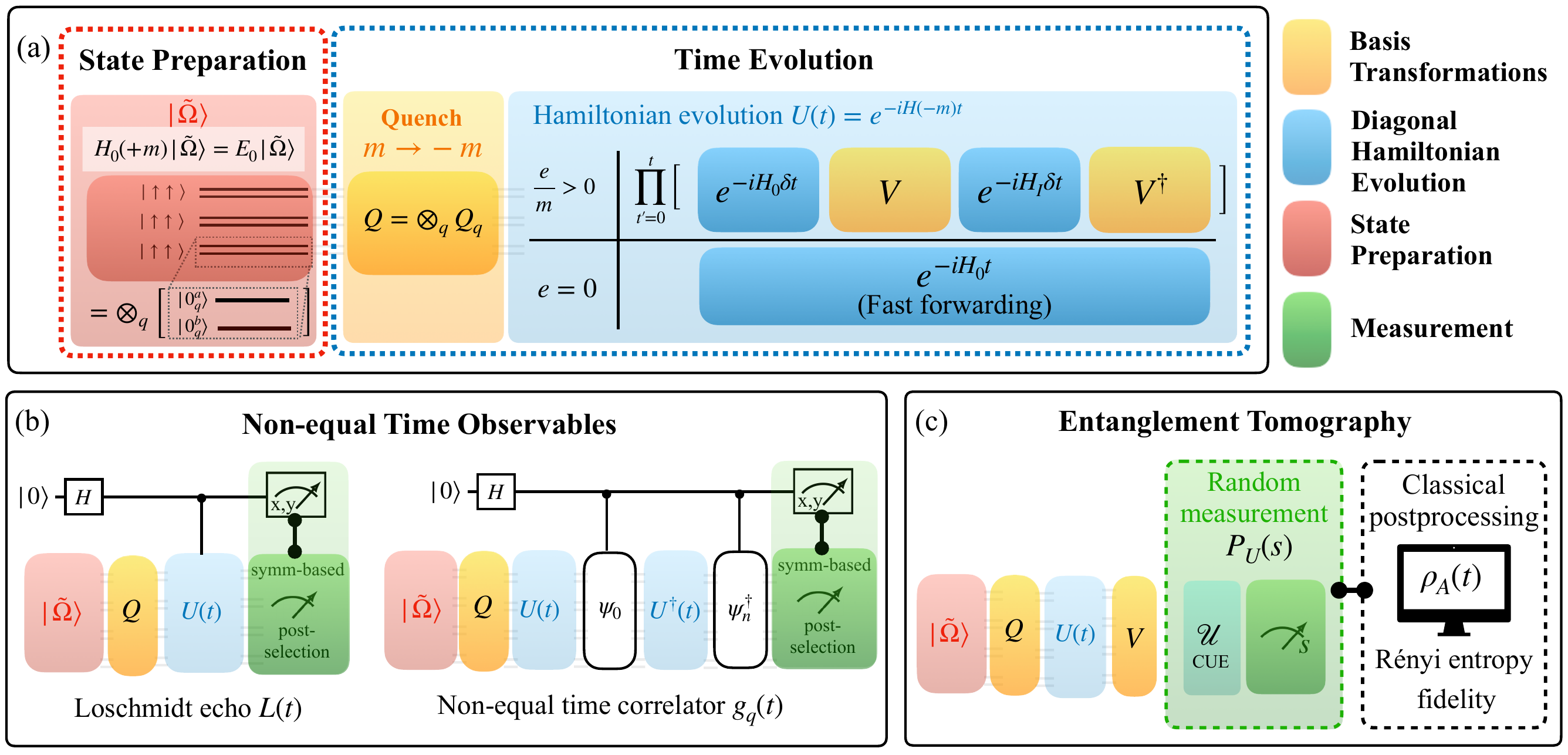}}
\caption{(a) Implementation strategy to prepare ground state of the non-interacting Schwinger model and simulate non-equal time evolution after a quantum quench, involving  basis changes $V$ from position to momentum space. Free ($H_0$) and interacting ($H_I$) parts of the time evolution are performed in a diagonal basis. The quench from $m$ to $-m$ is achieved via a basis transformation from the vacuum of the free theory in momentum-space computational basis with mass $m$ to that with mass $-m$. (b)  Interferometry schemes, employed to compute Loschmidt echo $L(t)$ (\Eq{Eq:loschmidt}) and NECFs $g_q(t)$ (\Eq{eq:correlator}), include a symmetry-based error-mitigation scheme. (c) entanglement tomography scheme to extract R\'enyi entropies, fidelities, as well as the reduced density matrix $\rho_A(t)$ from an entanglement Hamiltonian ansatz that is constrained by a classical optimization based on a number of random measurements.}\label{fig:overview}
\end{figure*}

\section{Quantum Computing Dynamical Quantum Phase Transitions\label{sec:DQPTbasics}}
\noindent
Consider the massive lattice Schwinger model with the Hamiltonian
\begin{align}\label{eq:Hamiltonian}
H(m) = &\frac{1}{2a}\sum_{n=0}^{N-1} (\psi^\dagger_n \mathcal{U}_n \psi_{n+1} +\text{H.c.})
\nonumber\\+&m\sum_{n=0}^{N-1} (-1)^n\psi^\dagger_n \psi_n
+ \frac{ae^2}{2} \sum_{n=0}^{N-1} E_n^2
\end{align}
on a one-dimensional (spatial) lattice with $N$ sites and periodic boundary conditions (PBCs), with mass $m$, electric coupling $e$, and lattice spacing $a$. $\psi_n^\dagger$ and $\psi_n$ denote creation and annihilation operators for the (staggered) fermions, respectively. $\mathcal{U}_n$ is the link  and $E_n$ the electric-field operator, satisfying the commutation relation $[E_n, \mathcal{U}_m]= \delta_{nm} \mathcal{U}_n$. 
The Hamiltonian commutes with Gauss's law operator at each site, $G_n \equiv E_n -E_{n-1} -  \mathcal{Q}_n$, where 
\begin{align}
\mathcal{Q}_n\equiv \psi_n^\dagger \psi_n -\frac{1}{2} [1-(-1)^n]
\end{align}
is the staggered fermion charge. The gauge-invariant physical Hilbert space contains states that satisfy $G_n | \psi\rangle^{\rm phys}=0$.

Adding a topological $\theta$ term $\frac{e\theta}{2\pi}\sum_n E_n$ to \Eq{eq:Hamiltonian}, and upon performing a chiral transformation to absorb the $\theta$ parameter into a (complexified) fermion mass term, yields the Hamiltonian~\cite{coleman1975charge,coleman1976more,manton1985schwinger,kharzeev2020real,chakraborty2020digital,ikeda2021real} 
\begin{align}\label{eq:Hamiltoniantheta}
\tilde H(m,\theta) = &\sum_{n=0}^{N-1}\frac{1-ma \, \sin (\theta )\,(-1)^n}{2a} (\psi^\dagger_n \mathcal{U}_n \psi_{n+1} +\text{H.c.})
\nonumber\\+&\sum_{n=0}^{N-1} m \, \cos (\theta) \, (-1)^n \psi^\dagger_n \psi_n
+ \frac{ae^2}{2} \sum_{n=0}^{N-1} E_n^2.
\end{align}
This model is considered a prototype model for CP violation in QCD~\cite{tHooft:1976snw,Jackiw:1976pf,Callan:1979bg}. With this motivation, a quench of the $\theta$ parameter,  mimicking a rapidly changing background axion field~\cite{Weinberg:1977ma,Wilczek:1977pj,Peccei:1977hh}, was considered in Ref.~\cite{Zache:2018cqq}, and a novel non-equilibrium topological transition was discovered in the far-from-equilibrium response of the system. The transition was also identified as a manifestation of the more general phenomenon of DQPTs ~\cite{heyl2013dynamical,flaschner2016observation,jurcevic2017direct,heyl2018dynamical,budich2016dynamical,tian2018direct,xu2018measuring,huang2019dynamical}, and recognized as an ideal target for quantum simulators and computers because of short time scales involved and potential for its realization in small systems. This phenomenon will be explored in this work using a digital trapped-ion quantum computer.

In particular, we employ the maximum possible quench of the $\theta$ parameter of the Schwinger model, $\Delta \theta = \pi$, corresponding to changing the sign of the mass term. This is achieved by preparing the ground state $ | \Psi(t=0) \rangle \equiv | \rm{GS}({m})\rangle$ of $H(m) \equiv  \tilde H(m,\theta = 0)$ and then time evolving
according to $H(-m)\equiv \tilde H(m,\theta=\pi)$,
\begin{align}
    | \Psi(t) \rangle =e^{-i H(-m) \, t} | \rm{GS}({m}) \rangle\,.
\end{align}

Two different non-equal time observables will be computed. The first is the Loschmidt echo, the overlap of the time-evolved state with the initial state,
\begin{align}
\label{Eq:loschmidt}
L(t)\equiv \langle\Psi(0)   | \Psi(t) \rangle = \langle {\rm GS}(m) |e^{-i H(-m) \, t} |{\rm GS}(m)\rangle \,,
\end{align}
from which one can define an intensive rate function,
\begin{align}
\label{Eq:RateFunction}
    \Gamma(t) \equiv \lim_{N \to \infty}\left\{- \frac{1}{N} \log ( |L(t)| )\right\}\,.
\end{align}
The non-analyticities of \Eq{Eq:RateFunction} correspond to DQPTs and can be extracted on lattices as small as four to eight sites with small finite-volume effects in the $e/|m| \lesssim 1$ regime~\cite{Zache:2018cqq}. The second quantity is a set of non-equal time correlation functions (NECFs),  defined in the staggered lattice formulation as
\begin{align}
\label{eq:correlator}
   g_q(t) = \sum_{j=0}^{N/2-1} e^{-i \frac{ 2\pi q j }{N/2}} \Big[ g^{\rm even}_{j}(t)+g^{\rm odd}_{j}(t)\Big]\,,
   \end{align}
where $q \in[-\frac{N}{4}, \frac{N}{4}-1]$ and
\begin{align}
\label{eq:defcorrstaggered}
g^{\rm even\,(odd)}_j(t) \equiv \langle  \psi^\dagger_{2j(2j+1)}(t)\,  \mathcal{U}_{2j(2j+1),0(1)}(t) \, \psi_{0(1)}(0) \rangle\,,
\end{align}
with $\langle \dots \rangle \equiv \langle {\rm GS}(m) | \dots |{\rm GS}(m)  \rangle$ and $\mathcal{U}_{n,m}(t)\equiv \prod_{k=m}^{n-1}\mathcal{U}_k(t)$ ($\mathcal{U}_{n,n}=1$). From $g_q(t)$, an integer-valued topological order parameter can be extracted~\footnote{Note here a relative minus compared to the convention in Ref.~\cite{Zache:2018cqq} so that $\nu(t)\ge 0$. Explicitly, Ref.~\cite{Zache:2018cqq} adopts the definition $\nu (t) \equiv n_+(t) - n_-(t)$ so that $\nu(t)\le 0$.}
\begin{align}
\nu (t) \equiv n_-(t) - n_+(t)\,, \label{eq:deftopologicalorderparamnu}
\end{align} 
where
\begin{align}
\label{eq:deftopologicalorderparam}
n_\pm(t) \equiv 
\frac{1}{2 \pi} \oint_{\mathcal{C}_\pm(t)} d\mathbf{z} \,\tilde{g}^\dagger_{\mathbf{z}} \partial_\mathbf{z} \tilde{g}_{\mathbf{z}}\,.
\end{align}
Here, $\mathbf{z}\equiv(q,t')$, ${g}_\mathbf{z} \equiv g_q(t')$, and $\tilde{g}_\mathbf{z} \equiv {g}_\mathbf{z} / |{g}_\mathbf{z}|$. Concretely, $\mathcal{C}_\pm(t)$ runs clockwise (counter-clockwise) in the positive (negative) half of wavenumbers $q$ in the $(q,t')$-plane, i.e., $\mathcal{C}_+(t): (0,0)\rightarrow ( N/4-1 ,0) \rightarrow( N/4-1,t)\rightarrow(0,t)\rightarrow (0,0)$ and similarly for $\mathcal{C}_-(t)$. Note that $t'$ is continuous and $q$ is discrete, hence integral and derivative become a sum and finite difference  along the $q$-sections of $\mathcal{C}_\pm(t)$. Equation~(\ref{eq:deftopologicalorderparamnu}), which is valid at arbitrary coupling $e$, changes by an integer whenever the system undergoes a dynamical quantum phase transition. We refer the reader to Ref.~\cite{Zache:2018cqq} for more details, and here we focus on a quantum-computational representation of the topological phenomenon.

The gauge degrees of freedom can be integrated out using Gauss's law in a way that maintains translation invariance with PBCs, see Ref.~\cite{Zache:2020qny} for details. A gauge-field zero mode (average electric field)  remains untreated in this case, but since our goal is to demonstrate the quantum computation of topological NECFs and utilizing entanglement tomography on an existing device, instead of engineering gauge-field digitizations, we leave the treatment of this term to future studies~\footnote{DQPTs occur in the infinite-volume (and continuum) limit of this model~\cite{Zache:2018cqq} where boundary conditions do not matter. For example, in the case of open boundary conditions (OBCs), the gauge-field degrees of freedom can be eliminated completely. One can show that the DQPT persists in this case. In fact, in the non-interacting limit, the DQPT is due to a critical mode $q_c$ in momentum space. Compared with the OBCs, PBCs lead to additional boundary excitations which do not effect the $q_c$ mode responsible for the DQPT. From an algorithmic perspective, the OBCs case is slightly more involved (but straightforward) because of the more complicated circuit implementation of the momentum-to-position-space basis transformation. Therefore, the PBCs are adopted in this study to simplify state preparation, time evolution, and measurement based on the simpler basis-transformation properties.}.

Explicitly, in the purely fermionic form, $\mathcal{U}_n$ is set to one in Eqs.~(\ref{eq:Hamiltonian}) and (\ref{eq:defcorrstaggered}), and the electric-field term in \Eq{eq:Hamiltonian} is replaced by a (translation-invariant) long-range fermionic density-density interaction~\cite{Zache:2018cqq,Zache:2020qny}
\begin{align}
 \frac{ae^2}{2} \sum_{n=0}^{N-1} E_n^2 \rightarrow    ae^2\sum_{n,m=0}^{N-1} \nu(d_{nm}) \mathcal{Q}_{n} \mathcal{Q}_m \equiv H_I\,,
 \label{eq:positioninteractingH}
\end{align}
where $d_{nm} = \min(|n-m|,N-|n-m|)$, and
\begin{align}\label{eq:defnu}
    \nu(d) \equiv \frac{3-N}{4(N-2)} \times \begin{cases} d & \text{if }d=0,1\\
    d+\frac{d^2 - 3d +2}{3-N}& \text{if }2 \le d\le \frac{N}{2}-1\,.\\
    \frac{N^2-8}{4(N-3)} & \text{if }d=\frac{N}{2}
    \end{cases}
\end{align}
The resulting Hamiltonian reads
\begin{align}
\label{eq:HamiltonianFermion}
H(m) = H_0(m)+H_I\,,
\end{align}
with $H_I$ defined above and
\begin{align}
\label{eq:H0-position}
H_0=\frac{1}{2a}\sum_{n=0}^{N-1} (\psi^\dagger_n  \psi_{n+1} +\text{H.c.})+m\sum_{n=0}^{N-1} (-1)^n\psi^\dagger_n \psi_n
\end{align}
being the non-interacting fermionic part of the Hamiltonian.

The fermionic degrees of freedom can be represented in two different bases. Firstly, they can be represented by fermionic Fock states in position space which are given by spin states, identifying an occupied (unoccupied) state with $|\eta_n=1 \rangle \equiv | 1_n\rangle =   |\ua \rangle$ ($|\eta_n=0 \rangle \equiv | 0_n \rangle =  |\da \rangle$), where $n\in [0,N-1]$ labels a lattice site. The operator-ordering convention is adopted from left to right in ascending order, i.e., $\psi^\dagger_{n_0}\psi^\dagger_{n_1} \dots \psi^\dagger_{n_{M-1}}  \prod_{n=0}^{N-1} |0_n \rangle$ with $n_0 < n_1 <\dots n_{M-1}$ for an $M$-fermion state, where $\prod_{n=0}^{N-1} |0_n \rangle$ is the zero-fermion state. Secondly, fermions can be represented by fermionic Fock states in momentum space, which form an eigenbasis of $H_0$,
\begin{align}
\label{eq:free}
    H_0 = \sum_{q=-N/4}^{N/4-1} \omega_q \big( a_q^\dagger a_q - b_{\scriptsize -q} b_{\scriptsize -q}^\dagger \big)\,.
\end{align}
Here, $\omega_q \equiv \sqrt{ m^2 +
|\tilde{p}(q)|^2 }$ with $\tilde{p}(q)\equiv a^{-1}e^{i2 \pi q /N} \cos({2 \pi q}/{N} ) $, with modes $a_q$ and $b_{\scriptsize -q}$ for each $q\in [-N/4,  N/4-1]$. Fock states are defined with the same convention as in position space, with $\ket{\ua}$ ($\ket{\da}$) identified as the occupied (unoccupied) state, and with modes ascending in $q$, i.e., $a_{q_0}^\dagger b_{-q_0}^\dagger $$\dots a_{q_{M-1}}^\dagger b_{-q_{M-1}}^\dagger \prod_{q=-N/4}^{N/4-1} | 0^a_q \rangle | 0^b_q \rangle $ for an $M$-fermion state. Note that the local momentum ($q$-) Hilbert space has four states, $\{ a_q^\dagger b_{-q}^\dagger \ket{0_q^a }\ket{0_q^b} $, $ a_q^\dagger \ket{0_q^a }\ket{0_q^b}$, $ b_{-q}^\dagger \ket{0_q^a} \ket{ 0_q^b} $, $ |  0_q^a  \rangle |0_q^b \rangle \}$.
In this manuscript, fermionic parity is explicitly realized within the unitary operations; our mode ordering is exactly equivalent to a Jordan-Wigner transformation starting at site $n=0$ (momentum $q=-\frac{N}{4}$), i.e., $\psi_n^{(\dagger)}\rightarrow \big[ \prod_{n'=0}^{n-1}(-\sigma^z_{n'})\big] \sigma_n^{-(+)}$.

Position- and momentum-space computational bases are related by composite fermionic Bogoliubov and {staggered}  Fourier transforms~\cite{ferris2014fourier,epple2017implementing,babbush2018low,cervera2018exact,kivlichan2020improved} ($\mathfrak{n}\in [0,N/2-1]$ labeling a supercell with $n=2\mathfrak{n}$ for even sites, $n=2\mathfrak{n}+1$ for odd sites),
\begin{align}\label{eq:Fourierdef}
\Psi_{\mathfrak{n}} = \frac{1}{\sqrt{N/2}}
    \sum_{\scriptsize q=-{N}/{4}}^{{N}/{4}-1}
    \Psi_q\, 
    e^{-i\frac{2\pi q \mathfrak{n} }{N/2}}\,,
\end{align}
where $\Psi_{\mathfrak{n}} \equiv (\psi^{\rm even}_{\mathfrak{n}},\psi^{\rm odd}_{\mathfrak{n}})^T$  and $\Psi_q\equiv(\psi_q^{\rm even},\psi_q^{\rm odd} )^T$. Here, the even (odd) superscript on the position-space fields denote that they are acting on even(odd)-indexed sites $n$, while even (odd) superscripts on momentum-space fields indicate that the fields are obtained from their even (odd) position-space counterparts. The Bogoliubov transformation transforms creation and annihilation operators as follows
\begin{align}
\label{eq:Bog1}
\Psi_q = U_q \begin{pmatrix}
    a_q \\b_{-q}^\dagger
    \end{pmatrix}\,,
\end{align}
with 
\begin{align}
\label{eq:Bog2}
    U_q=\begin{pmatrix}
    \cos(\beta) & -e^{i\alpha} \sin(\beta)\\
    e^{-i\alpha} \sin(\beta) & \cos(\beta)
    \end{pmatrix}\,,
\end{align}
with $\alpha \equiv \frac{2\pi q}{N}$ and $\beta\equiv \arctan \big( \big[\frac{\omega_q -m}{\omega_q+m} \big]^{{1}/{2}}\big)$. Details and the corresponding circuits are discussed in Appendices~\ref{app:analytics} and ~\ref{app:quantumalgo}. A detailed gate and qubit count, as well as a summary of all data taken,  is presented in Sec.~\ref{app:data_circuitcounts} of the Appendix.

\subsection{Non-equal time correlation functions}\label{sec:nonequaltime}
\begin{figure}[t]
\begin{center}
\includegraphics[scale=0.525]{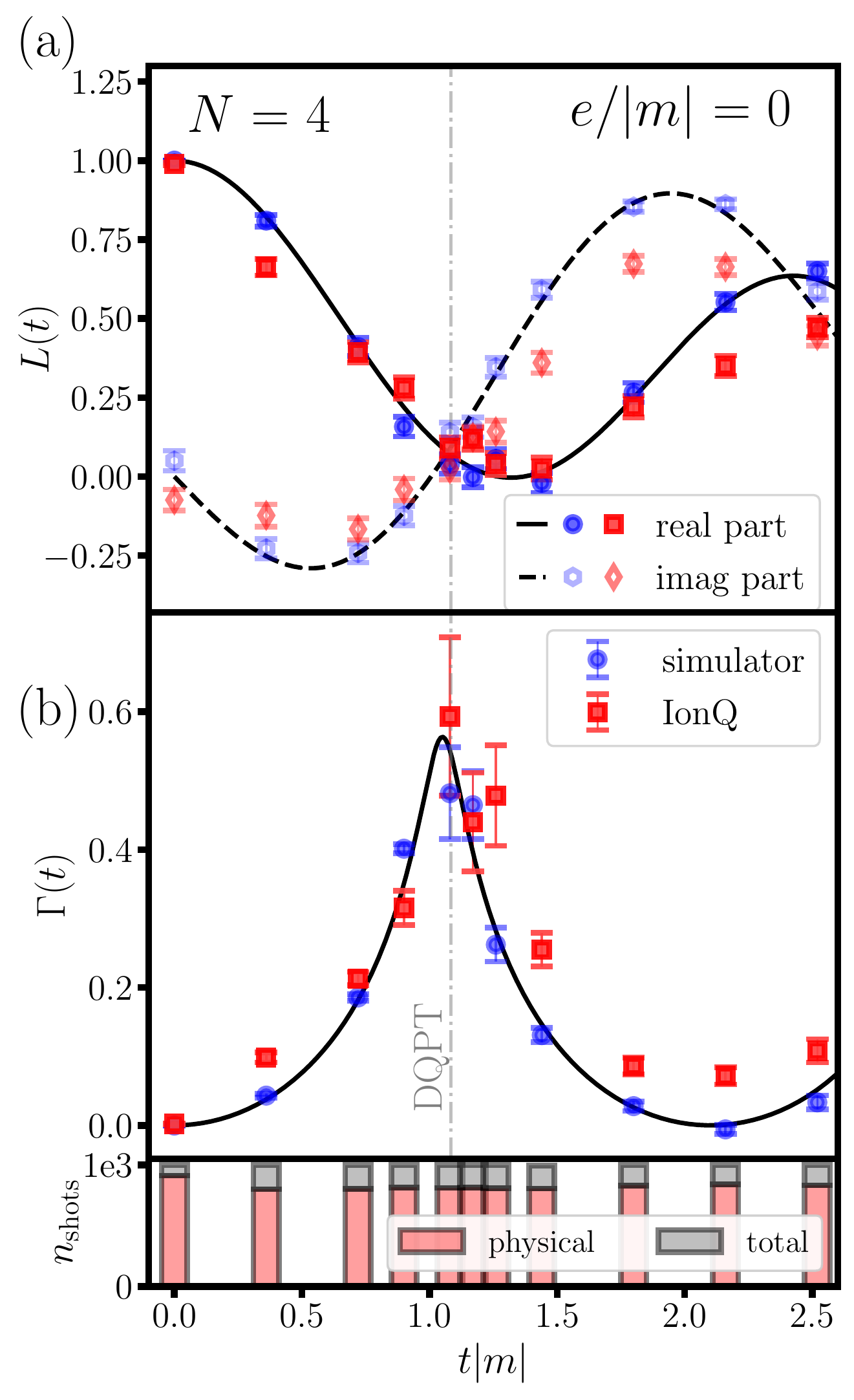}
\caption{(a) Real (solid line, dark-colored symbols) and imaginary (dashed line, light-colored symbols) parts of the Loschmidt echo $L(t)$ from an ideal-simulator (blue circles) versus error-mitigated results from IonQ Harmony (red squares), for $N=4$ sites, $e=0$ and $|m|  a =0.9$. (b) Rate function $\Gamma(t)$ reconstructed from the same data. The bottom panels shows the number of shots resulting in a physical, i.e., occupation-number symmetry-preserving result (red bars) versus all results (gray bars).}
\label{fig:LoschmidtN4free}
\end{center}
\end{figure}
To uncover the DQPT, we study the non-equilibrium dynamics  after a quench of the topological angle by $\Delta \theta=\pi$, i.e., from $+m$ to $-m$, by quantum computing non-equal time observables, focusing first on the Loschmidt echo $L(t)$ and the rate function $\Gamma(t)$, defined in \Eqs{Eq:loschmidt}{Eq:RateFunction}. Our general approach is summarized in \Fig{fig:overview} (a) and the schemes to measure non-equal time observables are shown in (b).

To do so, first the (non-interacting) ground state of $H_0(m)$, $| {\rm GS}(m) \rangle =\prod_{q=-N/4}^{N/4-1} | \bar{0}_q^a \rangle | \bar{0}_q^b \rangle $, where $| \bar{0}^{a/b}_q\rangle$ are zero-fermion number eigenstates of $H_0(+m)$, is prepared in a computational-basis state in the (prequenched) momentum-space representation (a bar distinguishes them from eigenstates of the postquench $H_0(-m)$). Then a Ramsey interferometry scheme is applied, depicted in \Fig{fig:overview}(b) and discussed in greater detail in App.~\ref{sec:RS}, with an ancilla in Hadamard superposition representing the two interferometric paths of the qubits encoding the fermions. We begin with the non-interacting case, $e=0$, where the initial state is time evolved in momentum space
\begin{align}
U(t) = e^{-i H_0(-m) t},
\end{align}
with no Trotter error.  Since the system is prepared in the ground state of $H_0(+m)$ with positive mass, $+m$ (\Eq{eq:free}), but is evolved with $H_0(-m)$ with negative mass, $-m$, and the dispersion $\omega_q$ is independent of the sign of the mass, $H_0(+m)$ and $H_0(-m)$ are formally identical in momentum space and the quench is not realized by simply changing a parameter in $H_0$. Instead, a change of basis, from $H_0(+m)$ to $H_0(-m)$, is performed, see Appendix~\ref{app:analytics} for details. At $e/|m|>0$, a Trotter scheme is applied, where the time-evolution operator is split into a free part ($H_0$ in \Eq{eq:free}) realized in momentum space, and an interacting part ($H_I$ in \Eq{eq:positioninteractingH}) realized in position space,
\begin{align}
\label{eq:trotterscheme}
    {U}(t) \equiv 
    \big(V^\dagger e^{-iH_I \delta t }   V   e^{-iH_0(-m) \delta t } \big)^{N_T}\,,
\end{align}
where $\delta t \equiv t/N_T$, with $N_T$ being the number of Trotter steps. $V \equiv F B$ contains the fermionic Fourier, $F$, and Bogoliubov, $B$, transforms defined in \Eqs{eq:Fourierdef}{eq:Bog2}. In their respective bases, the time-evolution operators $e^{-iH_I \delta t }$ and $e^{-iH_0 (-m) \delta t }$ are diagonal and  their control by the ancilla is  easy to implement, see Appendix \ref{app:quantumalgo} for details. The basis transformations $V$ and $V^\dagger$ need not be controlled. The quench in the mass parameter is still performed via a basis transformation~\footnote{A different Trotter scheme, evolving with a position-space Hamiltonian was explored in Ref.~\cite{martinez2016real}. To study the DPQT with such a scheme, one would have to approximately prepare the initial state, e.g., adiabatically, using a much deeper circuit.  The initial state is simply a computational basis state in our (momentum-space) approach and hence more straightforward to prepare. Additionally, when evaluating non-equal-time correlation functions using the Ramsey protocol, our Trotter scheme avoids controlled operation of non-diagonal unitaries, which would have been necessary in the position-space scheme.}

Figure~\ref{fig:LoschmidtN4free} shows the quantum-computational results for $N=4 $ sites and $e=0$, where panel (a) contains  real and imaginary parts of $L(t)$ defined in \Eq{Eq:loschmidt} simulated on ideal quantum hardware (blue circles) versus IonQ's 11-qubit machine (red squares), all with $n_{\rm shots}$=1000 shots per data point. Black dotted and dashed line are  results from exact numerical diagonalization.   Panel (b) displays $\Gamma(t)$ obtained from $L(t)$ according to \Eq{Eq:RateFunction}. The peak (non-analyticity) of the rate function at $t |m| \approx 1.1$ (gray dashed-dotted vertical line) marking the DQPT is well reproduced by the quantum computation. Errorbars in (a) are from statistical uncertainty assuming a binomial distribution for the outcome of single-qubit measurements, see Appendix~\ref{app:staterror}. The same uncertainty is propagated to (b) where, because of the logarithm in \Eq{Eq:RateFunction}, deviations are largest in the peak region.
\begin{figure}[t]
\begin{center}
\includegraphics[scale=0.525]{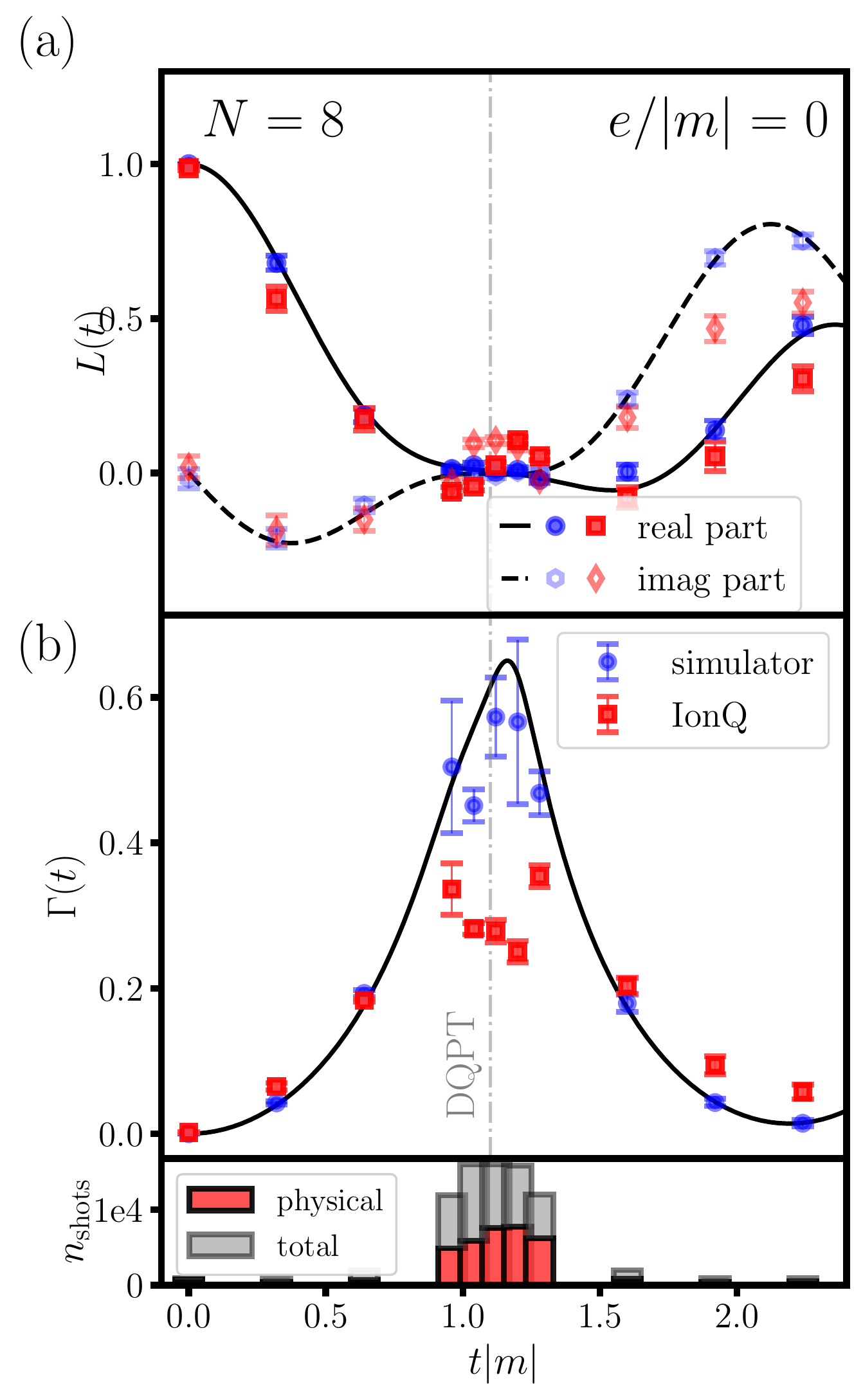}
\caption{(a) Real (solid line, dark-colored symbols) and imaginary (dashed line, light-colored symbols) parts of the Loschmidt echo $L(t)$ from an ideal-simulator (blue circles) versus error-mitigated results from IonQ Harmony (red squares), for $N=8$ sites, $e=0$ and $|m|  a =0.8$. (b) Rate function $\Gamma(t)$ reconstructed from the same data. The bottom panels shows the number of shots resulting in a physical, i.e., occupation-number symmetry-preserving result (red bars) versus all results (gray bars).
}
\label{fig:LoschmidtN8free}
\end{center}
\end{figure}

An error-mitigation scheme is used in this work, relying on the measurement of the system qubits in addition to the ancilla qubit. Because of the particle-number conservation in the model, events where the measured system bit sequence violates this constraint can be identified and eliminated at the cost of reduced statistics for a fixed number of shots. The bottom panel of \Fig{fig:LoschmidtN4free} shows the fraction of physical (symmetry-preserving) results (red) versus all results (gray), with a symmetry-violating device error occurring in about 25\% of the events. Machine errors are not fully eliminated by this procedure, as e.g.,  errors that conserve particle number are not mitigated. Details, including a comparison of the unmitigated and error-mitigated data can be found in Appendix~\ref{app:symmerr}.
\begin{figure}[t]
   \begin{center}
        \includegraphics[scale=0.525]{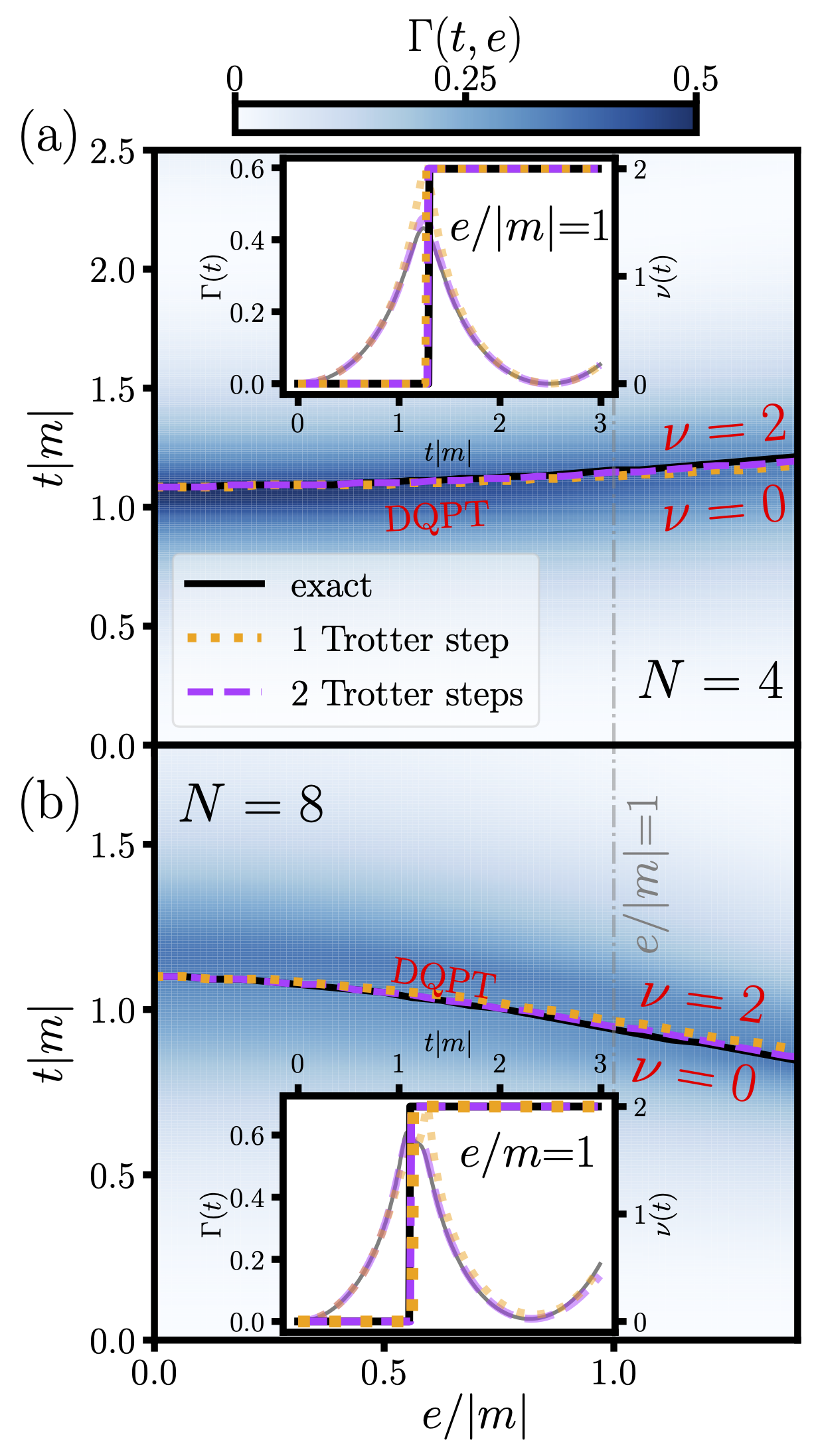}
        \caption{Negligible Trotterization effects in the DQPT computation, for the quench $(m,e=0)\rightarrow(-m,e>0) $. Shown is the rate function $\Gamma(t)$ as a function of time $t  m$ and coupling $e/|m|$. The position of the DQPT is marked for exact time evolution (black lines), 1 (orange dotted dashed) and 2 (purple dashed) Trotter steps. The insets show the topological parameter $\nu(t)$ (thick lines), see \Eq{eq:deftopologicalorderparamnu}, as well as the  rate function $\Gamma(t)$ for $e/|m|=1$. (a) $N=4$, $|m|  a =0.9$ (b) $N=8$, $|m|  a=0.8$. \label{fig:trotterover}  }
   \end{center}
\end{figure}

Both the Loschmidt echo $L(t)$ and the rate function $\Gamma(t)$ for $N=8$ lattice sites are shown in~\Fig{fig:LoschmidtN8free}. To resolve the DQPT, significantly more shots, up to $n_{\rm shots}=16,000$, are required in the peak region at $t |m| \approx 1.1$ (gray dashed line). The quantum hardware performs to the extent that a feature is discernible, reproducing, with some deviation, the simulator results after error mitigation. Errorbars are from the statistical shot-noise uncertainty effecting $\Gamma(t)$ mostly in the peak region.
\begin{figure}[t]
\begin{center}
\includegraphics[scale=0.525]{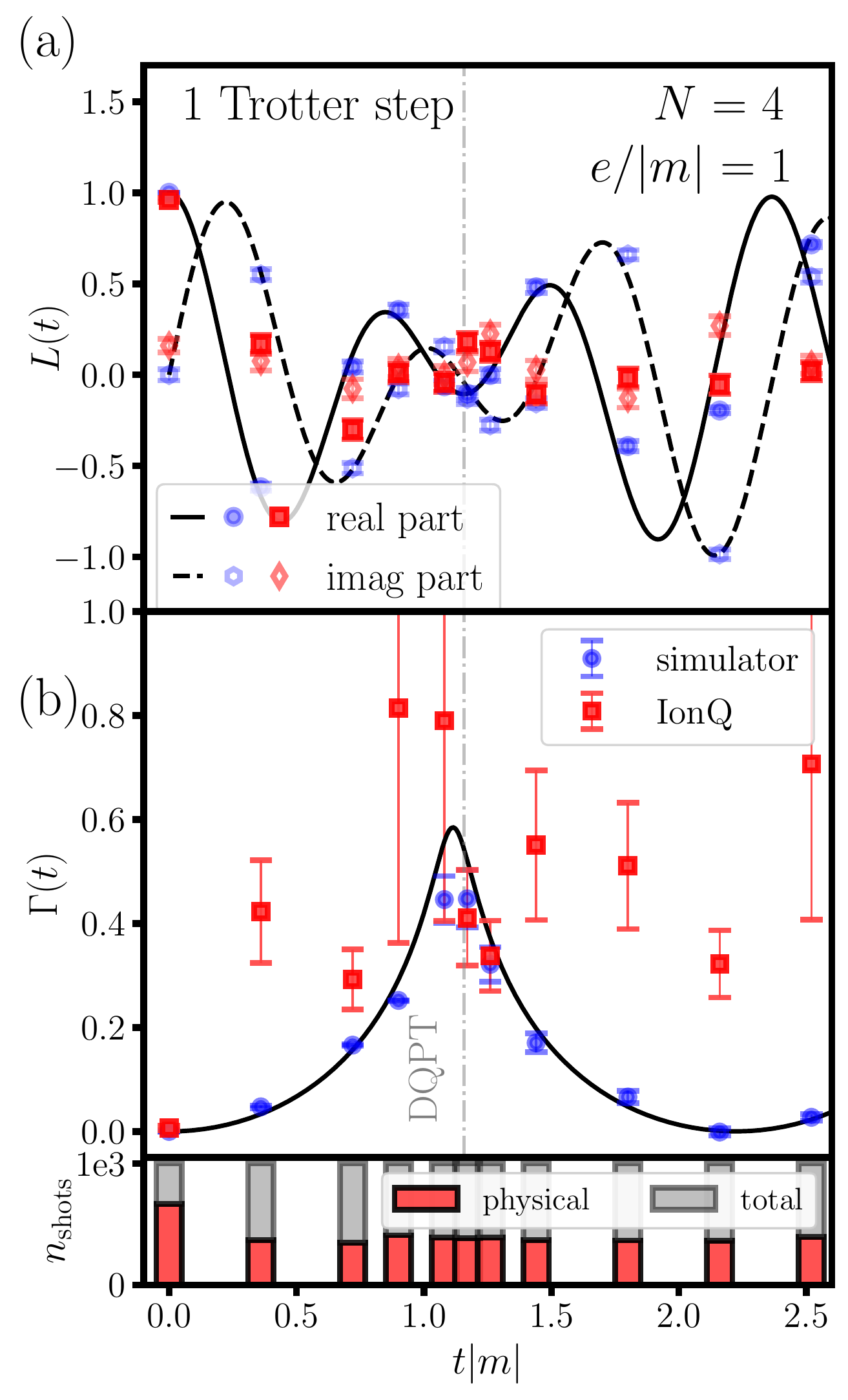}
\caption{(a)  Loschmidt echo $L(t)$ from an ideal-simulator (blue) versus error-mitigated results from IonQ Harmony (red), for $N=4$ sites, $|m|  a =0.9$ at finite coupling $e/|m|=1$, using a one step Trotter scheme. (b) Rate function $\Gamma(t)$, with bottom panels showing physical (red), i.e., occupation number symmetry respecting results, versus all results (gray). }
\label{fig:LoschmidtInteracting}
\end{center}
\end{figure}

Because it is more difficult to prepare the ground state of the interacting theory, $e/|m|>0$, a double quench is considered: First, the ground state of the non-interacting theory with $H(m,e=0)$ is transformed to that with $H(-m,e=0)$, then time evolution is performed with $H(-m,e)$, at finite $e/|m|$. Using the Trotter scheme, \Eq{eq:trotterscheme} and given basis transformations results in a significantly greater circuit complexity compared to the non-interacting case.

Figure~\ref{fig:trotterover} investigates the effect of Trotterization. The rate function $\Gamma(t)$, computed exactly, as a function of time $t |m|$ and coupling $e/|m|$ is shown for (a) $N=4$, $|m|  a =0.9$ and (b) $N=8$, $|m|  a=0.8$. The position of the DQPT at $t=t_c$ is further shown, indicated by the change of the topological order parameter $\nu(t)$ in an integer step from $\nu=0$ to $\nu=2$, for the exact time evolution (black lines), with 1 (orange dotted lines) or 2 (purple dashed lines) Trotter steps. While agreement is expected at $e=0$, where the Trotter error vanishes, the transition can be quantitatively reproduced even at large couplings $e/|m|\gtrsim 1$ with only one Trotter step.

Figure~\ref{fig:LoschmidtInteracting} summarizes the results of quantum computing (a)  $L(t)$ and (b) $\Gamma(t)$ at  $e/|m|=1$, $N=4$, and with $N_{\rm T}=1$ Trotter step. IonQ results (red symbols) are noisier when compared to ideal-simulator results (blue symbols), due to an approximately four-fold larger gate count relative to $e=0$~\footnote{The constant terms of $H_I$, resulting in the absence of a global (time-dependent) phase in the circuits, will not be implemented. For comparison, we simply multiply $L(t)=\langle \Omega(m) |e^{-i H(-m) \, t} | \Omega(m)\rangle$ obtained on the device by this phase, which is included in the solid and dashed lines in \Fig{fig:LoschmidtInteracting}(a) obtained from exact diagonalization. The rate function $\Gamma(t)$ is independent of this phase.}. 

Symmetry-based error-mitigated results are shown in~\Fig{fig:LoschmidtInteracting}, and the bottom panel of figure shows the fraction of physical (red bars) compared to all measurements (gray bars), with about 70\% of events containing an identifiable symmetry-violating error. Unmitigated errors are significant, and the position of the DQPT is not statistically confirmed with the presented data set. Coherence and error models, which could potentially robustly resolve the DQPT and optimize this calculation, are discussed in Appendix \ref{app:error}. 
We show statistical shot-noise uncertainty, noting a severe degradation of the machine results (red) because of the reduced statistics due to symmetry-based postselection.

\subsection{Quantum computing topological transitions}\label{sec:TO}
\begin{figure}[t]
\begin{center}
\includegraphics[scale=0.485]{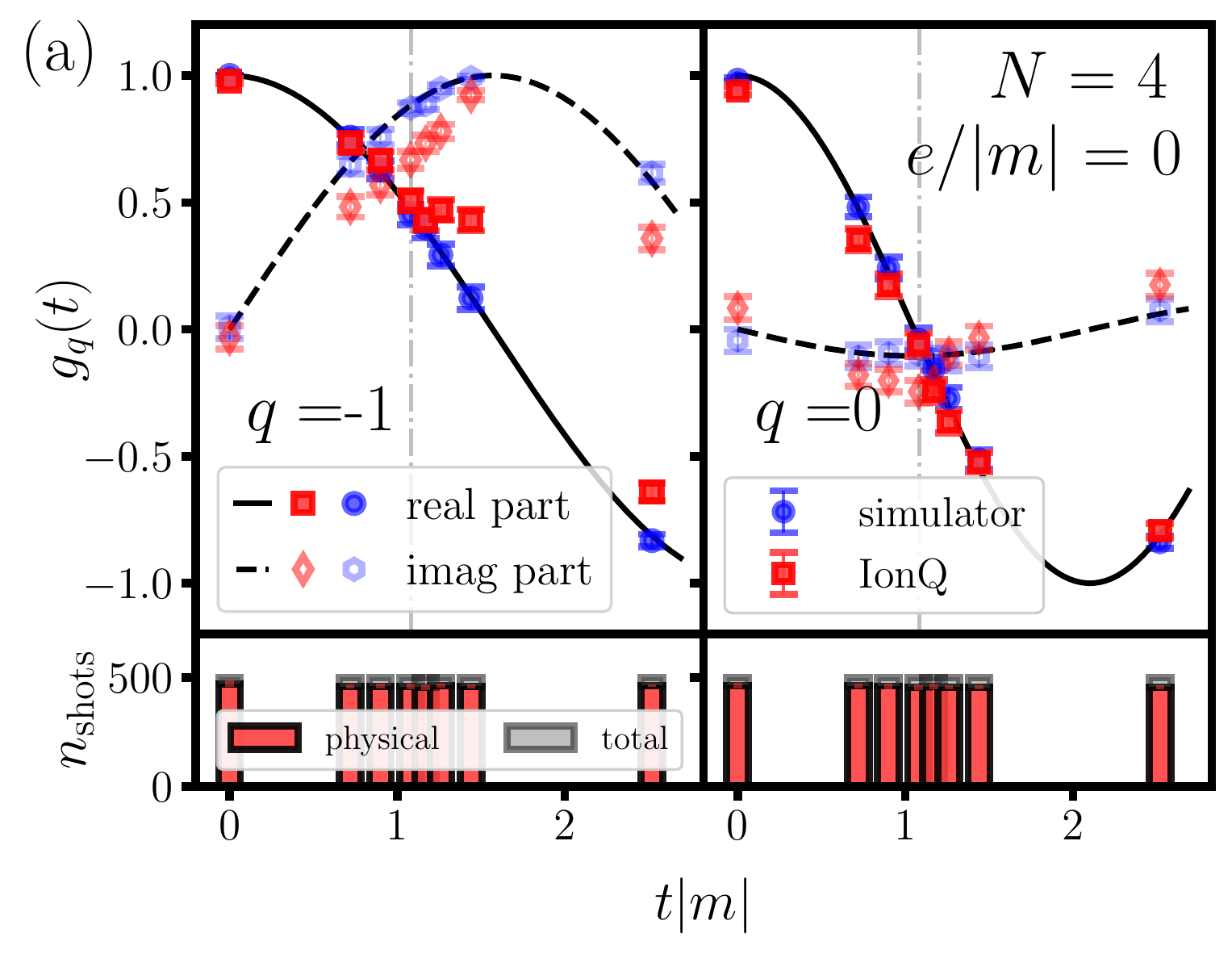}
\includegraphics[scale=0.485]{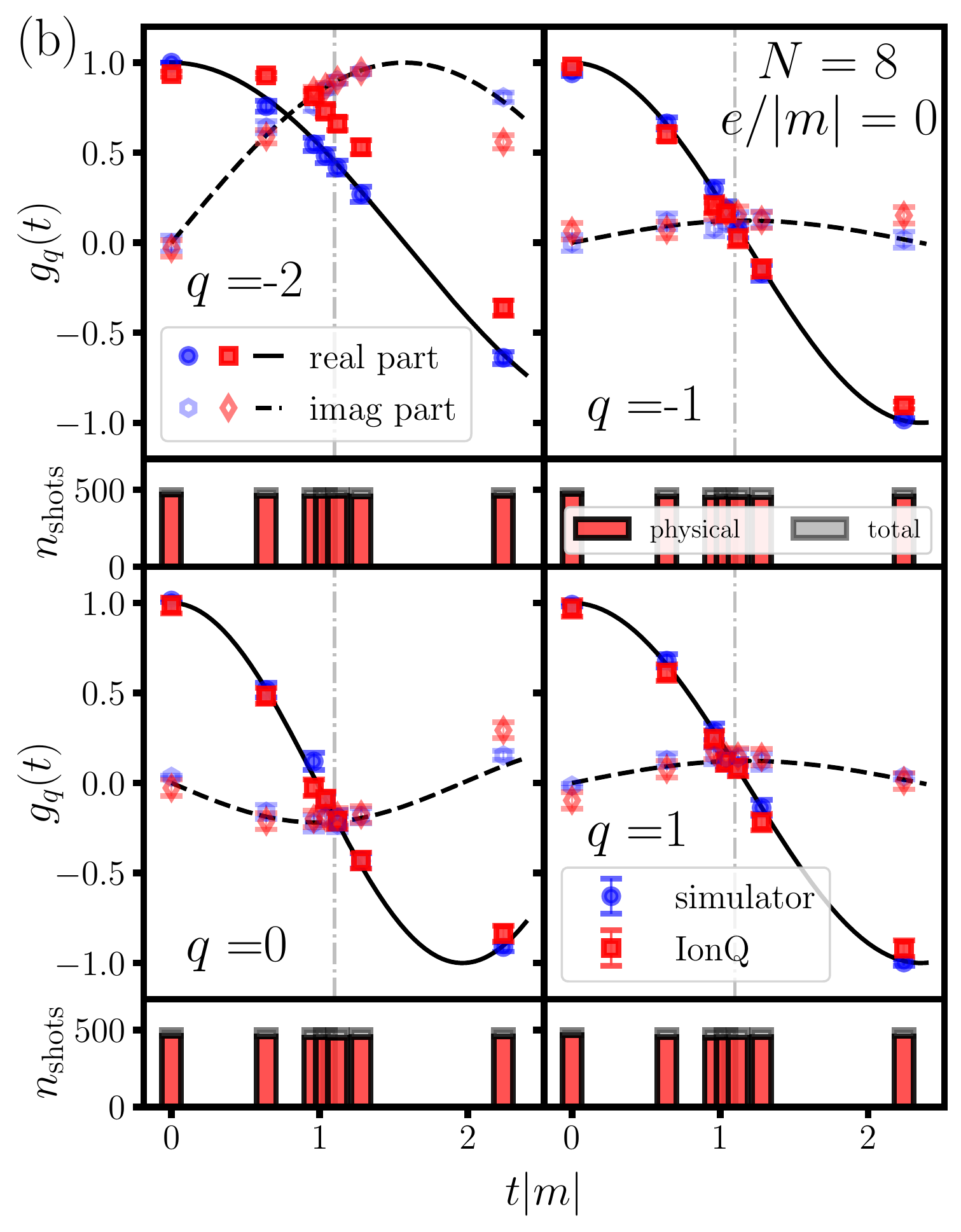}
\caption{(a) Real (solid lines) and imaginary (dashed lines) parts of the NECF $g_q(t)$ in \Eq{eq:correlator} for $N=4$, $|m|  a =0.9$, and zero coupling $e=0$, and for each Fourier mode $q\in[-N/4,N/4-1]$. Simulator results (blue symbols) are compared with symmetry-based, error-mitigated results from IonQ (red symbols), both obtained with $n_{\rm shots}=500$ (average events over real and imaginary parts of the correlators of unitary operators making up $g_q(t)$, see text for details). Bottom panels show the physical results, i.e., occupation-number conserving (red bars) versus all results (gray bars). (b) NECF $g_q(t)$ for $N=8$, $|m|  a=0.8$, and $e=0$.}
\label{fig:corr}
\end{center}
\end{figure}

A time-dependent topological index $\nu(t)$, as defined in \Eq{eq:deftopologicalorderparamnu}, is associated with the DQPT in this theory, as discussed in Sec.~\ref{sec:DQPTbasics} and in Ref.~\cite{Zache:2018cqq}. Errorbars are from the statistical shot-noise uncertainty which is nearly negligible. The topological index can be obtained by quantum computing the non-equal time correlation functions $g_q(t)$ in \Eq{eq:correlator}  using \Eqs{eq:deftopologicalorderparamnu}{eq:deftopologicalorderparam}. This is done by applying a Ramsey interferometry scheme using an ancilla in Hadamard superposition, \Fig{fig:overview}(b), involving (uncontrolled) forward time evolution, followed by the ancilla controlling an annihilation operator in position space, $\psi_n$, followed by (uncontrolled) backward time evolution, and the control of a creation operator in position space,  $\psi_n^\dagger$, see Appendix~\ref{app:quantumalgo} for details. 
\begin{figure}[t]
\begin{center}
\includegraphics[scale=0.525]{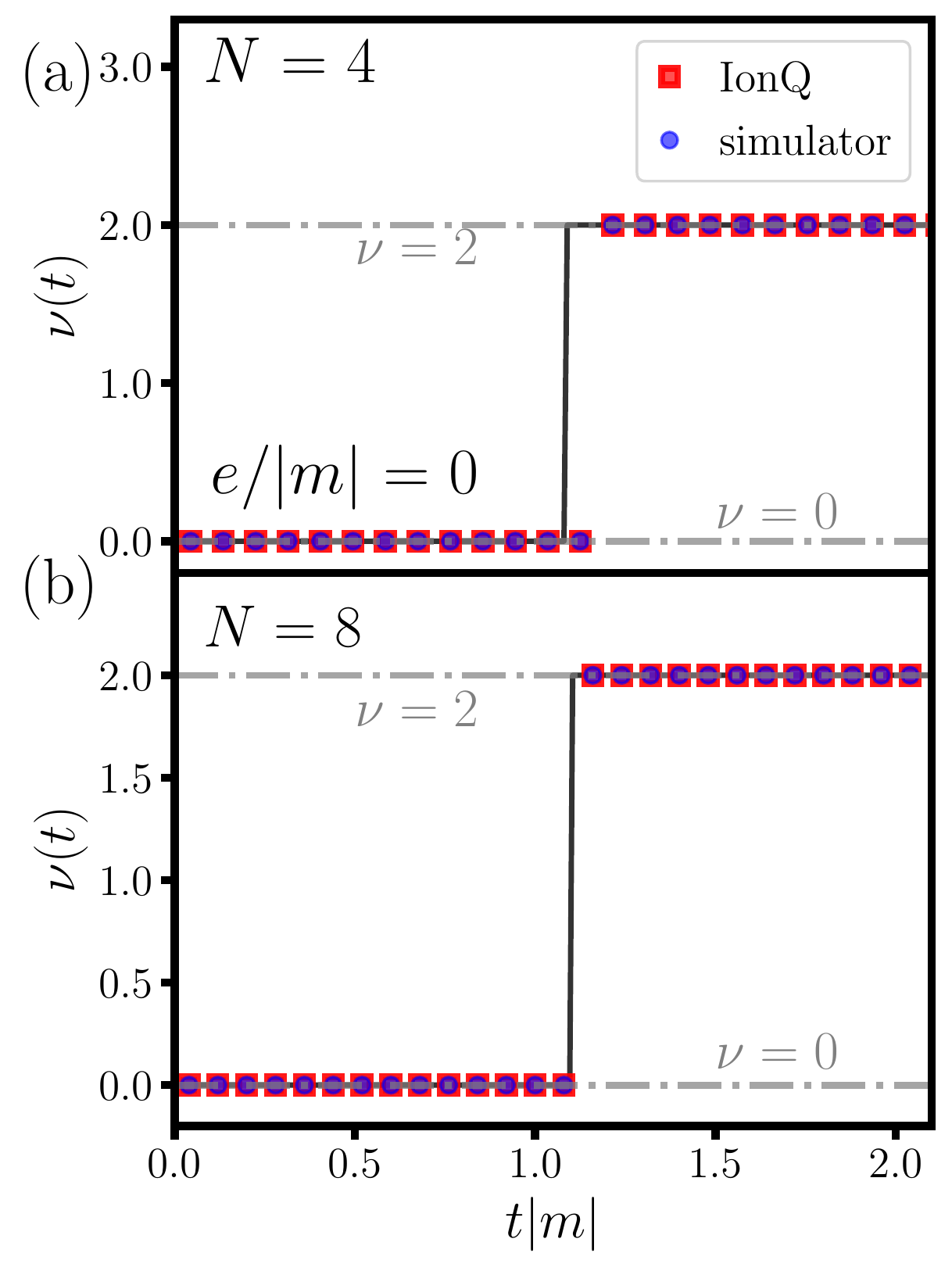}
\caption{Topological index $\nu(t)$, computed via \Eqs{eq:deftopologicalorderparamnu}{eq:deftopologicalorderparam} from $g_q(t)$ (\Fig{fig:corr}), for (a) $N=4$, $|m|  a =0.9$, and (b) $N=8$, $|m|  a=0.8$. Exact results  are black lines,  simulator results  are blue symbols, and IonQ results  are red symbols.  Horizontal dotted-dashed gray lines indicate the possible integer values $\nu(t)$ can take.
}
\label{fig:TO}
\end{center}
\end{figure}

Because of the significant gate cost at $e\neq 0$ resulting in reduced performance, we focus on $e=0$ where $g_q(t)$ can be computed separately for each $q$, hence a lower gate cost. Explicitly,
\begin{align}
\label{eq:corr_mom}
    g_q(t)
    &=
    \langle  e^{iH_0(-m)t}\Psi^\dagger_q(0)e^{-iH_0(-m)t}\, \Psi_q(0) \rangle
\nonumber\\
    &=
    \langle e^{iH_0(-m)t}\,  a_q^\dagger \, e^{-iH_0(-m)t}  \, a_q \rangle 
  \nonumber\\  &+\langle e^{iH_0(-m)t}\,  b_{-q} \, e^{-iH_0(-m)t}  \, b^\dagger_{-q} \rangle \,.
\end{align}
where the unitarity of $U_q$ in \Eq{eq:Bog2} is used. Here, $\langle \cdot \rangle \equiv  \langle {\rm GS}(m) | \cdot | {\rm GS}(m)\rangle$ refers to the expectation values in ground state of $H_0$ with mass $m$, while $a_q$ and $b_{-q}$ are Fock operators with respect to the eigenstates of $H_0(-m)$ in momentum space. 

We make use of a $q$-local Jordan-Wigner transformation to map fermion creation and annihilation operators onto spin raising and lowering operators, 
$a_q^\dagger = \sigma^+_{q,a} 
$, and $b_{-q}^\dagger = -\sigma^z_{q,a}\sigma^+_{q,b} $. Because spin raising and lowering operators are not unitary, they cannot directly be realized in the circuit and are decomposed into unitaries $\sigma^x=\sigma^++\sigma^-$ and $\sigma^y=i(-\sigma^++\sigma^-)$, with  eight combinations $\langle \sigma^{x/y}_{q,a}(t) \sigma_{q,a}^{x/y}(0) \rangle$ and $\langle \sigma^{x/y}_{q,b}(t) \sigma_{q,b}^{x/y}(0)\rangle$ which are computed separately~\footnote{The sum of all four components could be directly computed using two ancillas in a Hadamard superposition and corresponding controls. Unfortunately, the entangling-gate count with such a scheme goes beyond the coherence time of the hardware.}. The circuits use three qubits to compute real and imaginary parts of \Eq{eq:corr_mom} (for both $N=4$ and $N=8$ since each momentum mode is computed individually). We combine the circuits
for real and imaginary part into one six-qubit circuit that computes real and imaginary parts independently and in parallel, i.e., with no gates connecting the two circuits. In principle, one may need to worry about cross talks between target and neighboring ``spectator'' qubits (which occur typically due to spillover of control signals) as well as a relative increase in ion heating, motional noise, mode frequency drift, and axial motion due to parallel operations. Nonetheless, since the circuits here are modest six-qubit circuits with relatively shallow gate depth, we expect no significant differences compared to separate executions of three-qubit circuits, hence adopting this computational-cost-saving option.

The results of quantum computing the NECF, 
$g_q(t)$, for (a) $N=4$ and (b) $N=8$ lattice sites and zero coupling $e=0$ are displayed in \Fig{fig:corr}.  Exact results for both real (solid lines) and imaginary (dashed lines) parts of $g_q(t)$ are shown in the plots, along with the simulator (blue symbols) and IonQ results (red symbols), the latter computed with $n_{\rm shots}=500$ shots.  Exact results are reproduced well at $N=4$ and $N=8$,  for all $q$-modes. A symmetry-based error-mitigation scheme is applied, where results that violate a conservation law are dropped, see the bottom panels of \Fig{fig:corr}(a) and (b)~\footnote{We plot the average number of physical results for real and imaginary parts for the correlators $\langle \sigma^{x/y}_{q,a}(t) \sigma_{q,a}^{x/y}(0) \rangle$ and $\langle \sigma^{x/y}_{q,b}(t) \sigma_{q,b}^{x/y}(0) \rangle$. Fluctuations in the number of events between different components are negligible.}. Because of the small system size and the relatively small unphysical part of the 2-qubit Hilbert space, the noise-mitigation scheme has little effect. Errorbars are from the statistical shot-noise uncertainty.

Furthermore, the time-dependent topological index $\nu(t)$, extracted from $g_q(t)$ via \Eqs{eq:deftopologicalorderparamnu}{eq:deftopologicalorderparam}, is plotted in \Fig{fig:TO} for (a) $N=4$ and (b) $N=8$ lattice sites. To compute \Eq{eq:deftopologicalorderparam}, the data in \Fig{fig:corr} is linearly interpolated in time. The topological index $\nu(t)$ is successfully reconstructed from the IonQ data (red symbols) for (a) $N=4$ and (b) $N=8$~\footnote{Note that $N=4$ and $N=8$ results are computed using circuits with identical gate complexity.} and reliably indicates the position of the DQPT. Because $\nu(t)$ is topological, essentially measuring a branch cut in the phase of the complex non-equal time correlation function $g_q(t)$, the device noise seen in \Fig{fig:corr} does not effect the identification of the step-wise change in the topological index at all. If the signal is good enough to contain the holonomy of $g_q(t)$, perfect recovery of the DQPT is guaranteed (or not at all if the quality threshold is not met).

\section{Entanglement Tomography}\label{sec:EHT}
\begin{figure}[t]
\begin{centering}
\includegraphics[scale=0.525]{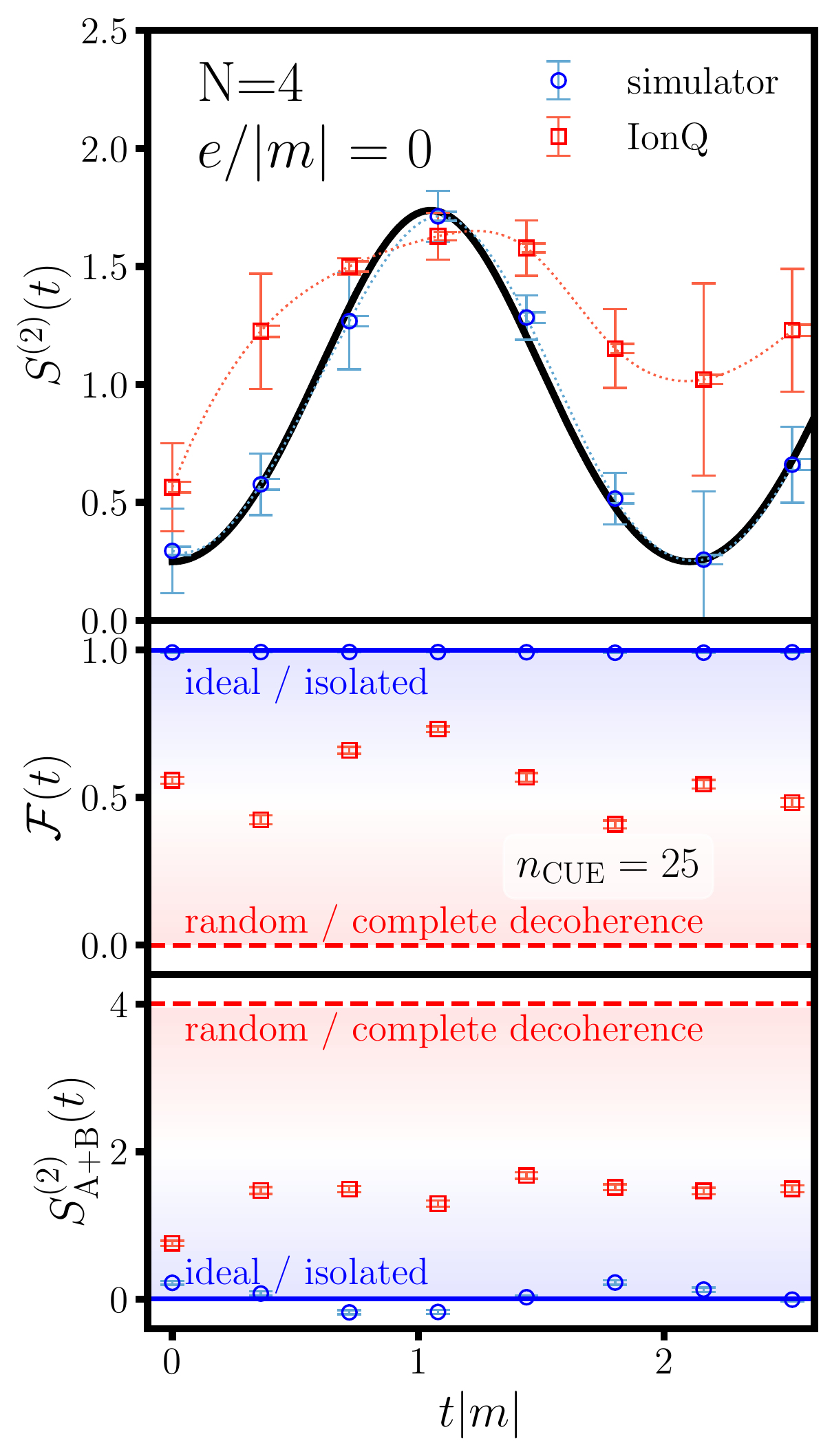}
\caption{The top panel shows R\'enyi entropy averaged over subsystems $A$ and $B$, $S^{(2)}(t)\equiv\frac{1}{2}(S_A^{(2)}(t)+S_B^{(2)}(t))$ (with $S_A^{(2)}$ and similarly $S_B^{(2)}$ defined in \Eq{eq:defrenyientropy}), for $N=4$, $|m|  a =0.9$, $e=0$, $n_{\rm CUE}=25$, and $n_{\rm shots}=1000$, including simulator (blue symbols) and  IonQ (red symbols) results. The middle panel depicts  fidelity $\mathcal{F}(t)$ (\Eq{eq:fidelity}). The bottom panel shows the  R\'enyi entropy of the full system, $S^{(2)}_{\rm A+B}$ (relative to the environment). Blue horizontal lines in the middle and bottom panels indicate ideal results, and a horizontal red line indicates zero fidelity or maximal entropy ($\rho(t) = \mathbb{I}/2^N$), respectively.}
\label{fig:RenyiFidelityN4}
\end{centering}
\end{figure}
\noindent
Entanglement structure and state fidelity of non-equilibrium states can be obtained following Refs.~\cite{van2012measuring,elben2018renyi,vermersch2018unitary,brydges2019probing,kokail2021entanglement}, based on random measurement. Our circuit-based approach is summarized in \Fig{fig:overview}(c). One can compute R\'enyi entropies and fidelities, and reconstruct the reduced density matrix using a generalization of the Bisognano-Wichmann (BW) theorem~\cite{bisognano1975duality,bisognano1976duality}. The  second order bipartite R\'enyi entropy is
\begin{align}
\label{eq:defrenyientropy}
    S^{(2)}_A (t) \equiv - \log_2\{ \text{Tr}_A(\rho_A^2(t) ) \}\,,
\end{align}
where the reduced density matrix of the first $N/2$ sites (`system $A$') is $\rho_A(t) = \text{Tr}_B [\rho(t)]$ with  $\rho(t) = | \psi(t)\rangle \langle \psi(t)|$. For the quantum experiments, one may define a fidelity (of the whole system),
\begin{align}
\label{eq:fidelity}
    \mathcal{F}(t)\equiv \frac{1}{\mathcal{N}(t)} \text{Tr}(\rho(t)\,   {\rho}_{\rm exact}(t))\,,
\end{align}
to assess the performance of the implementation. Here, $\mathcal{N} \equiv ([ \text{Tr}[\rho^2(t)]   \text{Tr}[{\rho}^2_{\rm exact}(t)]  )^{1/2} $, with $\rho(t)$ being the result extracted from experiment and $\rho_{\rm exact}$ being the exact density matrix, classically computed using exact diagonalization. 

To compute \Eq{eq:defrenyientropy} and \Eq{eq:fidelity}, random single-qubit basis rotations $u_i$ on each qubit $i$ are performed using a shallow circuit,  $| \psi(t) \rangle\rightarrow \{\mathscr{U}| \psi(t) \rangle\}_{n_{\rm CUE} }$, where $\mathscr{U} = \otimes_{i=0}^{N-1} u_i$ with $u_i$ drawn from a circular unitary ensemble (CUE)~\cite{mezzadri2006generate}, followed by measurement of the probabilities $P_{\mathscr{U}}(s)$ of  a bit sequence $s$ in the resulting basis. Both  bipartite R\'enyi entropy, \Eq{eq:defrenyientropy}, and fidelity, \Eq{eq:fidelity}, can be obtained using the relation~\cite{brydges2019probing}
\begin{align}
\label{eq:EHTprotocol}
    \text{Tr}(\rho_1 \rho_2) =\langle 2^N\sum_{s_1,s_2}(-2)^{D({s_1,s_2})}  P_{\mathscr{U}}^{1}(s_1)P_{\mathscr{U}}^{2}(s_2)\rangle_{\mathscr{U}}\,,
\end{align}
where $D(s_1,s_2)$ is the Hamming distance between $s_1$ and $s_2$, and $\langle \dots \rangle_{\mathscr{U}} \equiv \frac{1}{n_{\rm CUE}}\sum_{\{\mathscr{U}\}} \dots $ is the average over $n_{\rm CUE}$ random circuits.

Entropy and fidelity measurements are determined in position space, so the corresponding circuit involves transforming the time-evolved state from momentum to position space, as shown in \Fig{fig:overview}(c) and detailed in Appendix~\ref{app:quantumalgo}. The top panel of \Fig{fig:RenyiFidelityN4} shows the R\'enyi entropy $S_{A}^{(2)}(t)$ of system $A$ as a function of time, with $N=4$ sites and $e=0$. A solid black line denotes exact results, blue circles are simulator 
and red squares IonQ data (no error mitigation applied), with $n_{\rm shots}=1000$ and $n_U=25$ random CUE rotations for each data point. The R\'enyi entropies of both partitions, $A$ and $B$, are determined and the average is shown. Two sets of errorbars are shown for all machine data points (red). Smaller errorbars, offset slightly to the right, are the statistical uncertainties due to the shot noise. This was determined from a bootstrap procedure outlined in Appendix \ref{app:staterror}.  The second, larger errorbars are the difference of the entropy measured in system A versus B.
Both should be equal in the limit $n_{\rm CUE}\rightarrow \infty$, but are not because of finite $n_{\rm CUE}$. This is an accurate estimate for the simulator data, but note that neither this difference nor the statistical shot-noise uncertainty account fully for the deviations on IonQ, which are mostly systematic.

The middle panel of \Fig{fig:RenyiFidelityN4} shows the fidelity $\mathcal{F}$, extracted from the same data. Here, $n_{\rm shots}=1000$ and $n_{\rm CUE}=25$ result in near-perfect fidelity with the simulator. The device returns a fidelity of  $\sim 50 \%$ (as the transpiled circuit consists of 26 CNOT and 84 single-qubit operations, see Table \ref{table:complexity}, and hence decoherence is expected).  The bottom panel of \Fig{fig:RenyiFidelityN4}  shows the R\'enyi entropy $S^{(2)}_{A+B}(t)$ of the whole system, expected to vanish if the system is an ideal/pure state. The IonQ result deviates from zero, and shows an entropy equivalent to 1 to 2 bits of entanglement (with the environment). Statistical uncertainties are shown in panels (b) and (c) respectively, following the procedure outlined in Appendix~\ref{app:staterror}. 
\begin{figure}[t!]
\begin{centering}
\includegraphics[scale=0.525]{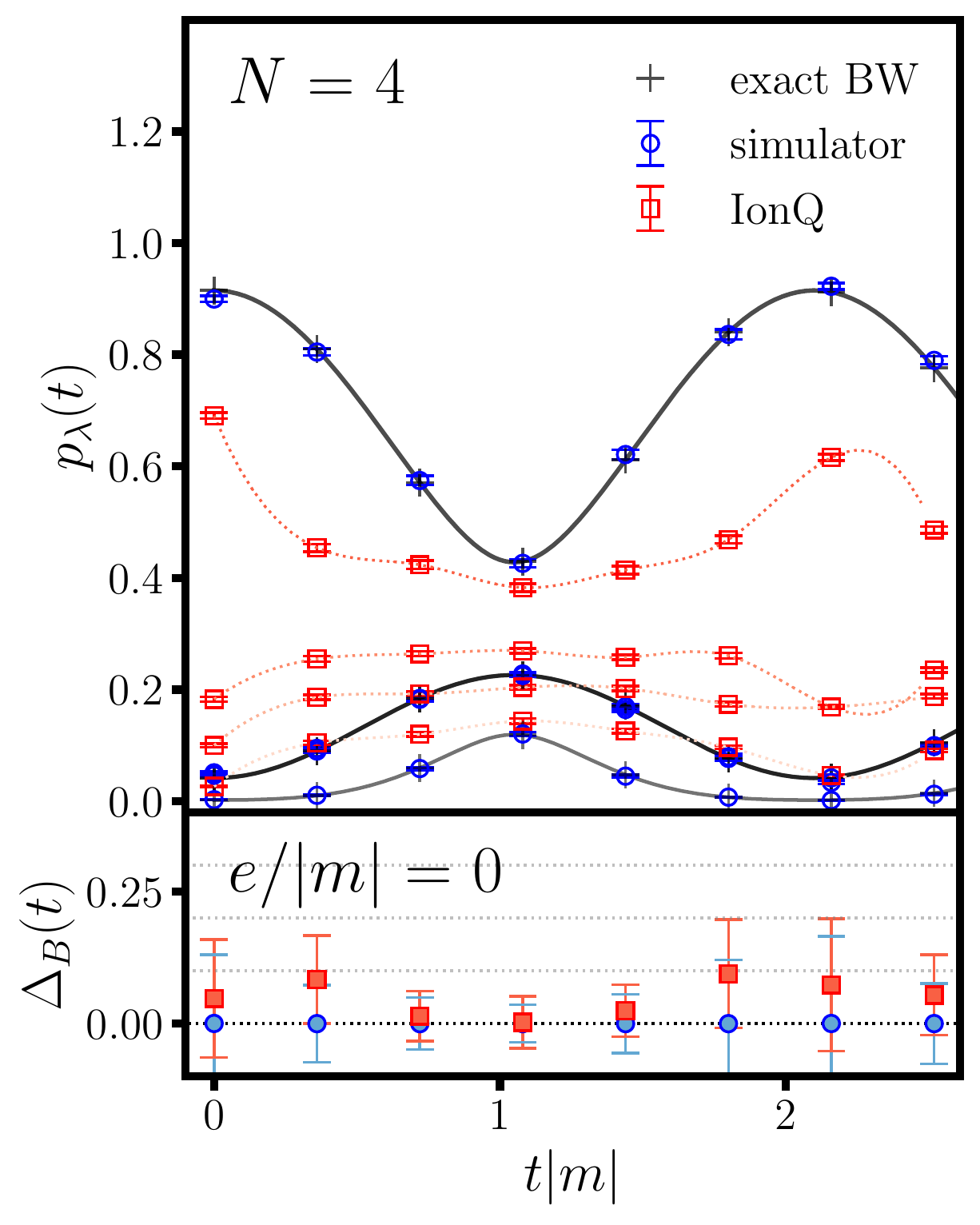}
\caption{ The top panel shows the eigenvalue spectrum $P_\lambda(t)$ of the reduced density matrix $\rho_{A}(t)$ of a bipartition, with $N=4$ and $e=0$, as a function of time, compared with the exact results (black lines), exact results reconstructed from the BW ansatz (black plus symbols),
simulator (blue  circles), and IonQ (red squares) data, with $n_{\rm shots}=1000$  and $n_U=25$. Note that the degeneracy between the two middle states in the spectrum is broken due to imperfections of the hardware implementation.
The bottom panel shows the Bhattacharyya distance $\Delta_B(t)$ between exact BW results (crosses), and simulator (blue circles) versus IonQ data (red squares).}
\label{fig:EHTN4}
\end{centering}
\end{figure}

The reduced density matrix $\rho_A(t)$ of system A is reconstructed using the entanglement hamiltonian tomography (EHT) protocol of Ref.~\cite{kokail2021entanglement}, based on the BW theorem~\cite{bisognano1975duality,bisognano1976duality} generalized to non-equilibrium states. To do so,
$\rho_A(t)$ is parametrized as~\cite{li2008entanglement}
\begin{align}
\label{eq:HA-def}
 \rho_A(t) = e^{-H_A(t)}\,,
\end{align}
 in terms of an entanglement hamiltonian (EH),
\begin{align}\label{eq:BWansatz}
   H_A(t)\equiv H_A(t;\{\beta_i,\mu_i\}) =  \sum_{j \in A} \beta_j(t) H_j + \sum_{j \in A} \mu_j(t) T_j\,.
\end{align}
Inspired by the BW theorem, $H_j$ are the local operators in \Eq{eq:HamiltonianFermion} (`energy densities') and $T_i$  are commutators of the latter. In practice, at $e=0$, these are
\begin{align}
    H_j &= \{ (-1)^n \psi^\dagger_n \psi_n, \, \psi_n^\dagger \psi_{n+1} + \text{H.c.} \}\nonumber\,,
    \\
    T_{j} &= \{ i ( \psi_n^\dagger \psi_{n+1} - \text{H.c.}),\, \psi_n^\dagger \psi_{n+2} + \text{H.c.} \} \,,\label{eq:ansatzT}
\end{align}
with sites $n,n+1,n+2$, etc., in system $A$~\footnote{Note that any staggering factor is absorbed in the coefficients of the terms in $T_j$.}. Extracting $\rho_A(t)$ from  $P_{\mathscr{U}}(s)$ at every $t$ requires minimizing a functional~\cite{kokail2021entanglement}
\begin{align}
\label{eq:variation}
    \chi^2 =  \sum_{s} \langle  \big[P_{\mathscr{U}}(s) - \text{Tr}[\rho_A(t,\{\beta_i,\mu_i\}) {\mathscr{U}} | s \rangle \langle s| {\mathscr{U}}^\dagger ] \big]^2  \rangle_{\mathscr{U}}
\end{align}
to obtain optimal parameters $\{\beta_i,\mu_i\}$. All the parameters are kept independent in the fit,  not assuming a specific functional form as done in Ref.~\cite{kokail2021entanglement}. The optimization is done classically and will become costly for much larger (sub-)systems. Nonetheless, scalable, efficient shadow and entanglement tomography schemes are currently under investigation, see e.g., Refs.~\cite{bairey2019learning,huang2020predicting,kokail2021quantum}.

The result of the optimization is shown in \Fig{fig:EHTN4} for $N=4$ ($N_A=N/2=2$), $|m|  a =0.9$, and $e=0$, displaying the reconstructed 
Schmidt spectrum $p_\lambda(t)$ of $\rho_A(t) \equiv \sum_\lambda  p_\lambda(t)  | \lambda_A(t) \rangle \langle \lambda_A(t) | $, where 
$ |\lambda_A(t)\rangle$ are the Schmidt vectors.
Shown are simulator results (blue symbols) versus quantum-computed results (red symbols). Black crosses denote perfect BW ansatz results, i.e., with infinite measurement and shots, while solid lines are (BW-ansatz independent) exact results. The bottom panel shows the Bhattacharyya distance,
\begin{align}
    \Delta B (t)\equiv-\log\left[\sum_\lambda \sqrt{p_\lambda(t) p_{\lambda}^{
\rm BW}(t)}\right]\,,
\end{align}
between the exact and BW simulator results, and between BW simulator and the BW IonQ results. The hierarchy and time dependence of $p_\lambda(t)$ is qualitatively reproduced, but the largest eigenvalues are systematically too small, consistent with  $\rho_A(t)$ appearing more mixed. In the presence of noise, time evolution is described by an effective but unknown Hamiltonian that is not reflected in the BW ansatz, leading to a systematic error in the reconstruction of density matrices. Errorbars are from the shot-noise uncertainty. The `exact BW' results (black crosses) are what would be measured in the $n_{\rm CUE}\rightarrow \infty$ and infinite shot limits; more discussion can be found in Appendix~\ref{app:staterror}.  In Appendix~\ref{app:error}, we discuss device errors and use the entanglement protocol to investigate sources of decoherence. Using simple ans\"atze for local depolarization, bit-flip, and phase error channels, it is found that such a simple error model does not account fully for the deviations observed, and the noise mechanism in the hardware needs a more complex modeling to accurately describe the features observed in experiment~\footnote{Reconstructing Eqs.~(\ref{eq:defrenyientropy}) and (\ref{eq:fidelity}) depends strongly on the accuracy of $\mathscr{U}$, involving single-qubit rotations at arbitrary angle, as examined in Refs.~\cite{brydges2019probing, kokail2021entanglement}. The 2-qubit errors, nonetheless, are the dominant source of error in the experiments performed, and such sensitivity to single-qubit imperfections is not considered here.}.

In Appendix~\ref{app:EHTdetails}, an EHT computation for $N=8$ is presented, but the observed fidelity is essentially zero given the large number of gates used. Therefore, the entanglement structure cannot be reliably reconstructed in this case. Finally, connecting the two main aspects of this manuscript, non-equal time observables and entanglement tomography, it is shown in  Appendix \ref{app:EHTdetails} how the ancilla-based interferometry scheme in Sec.~\ref{sec:nonequaltime} can be avoided by reconstructing $| L(t) |^2 $ from random measurements.

\section{Conclusions}\label{sec:conclusions}
\noindent
With the advancement of quantum-computing hardware and growing access to such systems, a primary goal among domain scientists is to identify those applications of quantum computing that can still benefit from near-term noisy and small- to intermediate-scale hardware, but are beyond capabilities of the most advanced classical computers. In the domain of lattice gauge theory, i.e., the numerical suite for simulating fundamental forces in nature such as quantum chromodynamics, computations that suffer from severe sign and signal-to-noise degradation problems appear to not be classically tractable, but they may be advanced using quantum simulators and quantum computers. Nonetheless, the size of required simulations is yet far beyond capabilities of near-term hardware. As such, it is critical to establish near-term goals of the quantum-simulation program for QCD. One such area is computing qualitative features of simpler gauge theories with similar features as QCD to discover hints of unexplored phenomena, e.g., exhibited in out of equilibrium states, that could impact our interpretation of experimental and observational data in the field of hot and dense QCD~\cite{berges2020thermalization}. Arguably, topological phenomena with robust signatures in presence of noise and other systematics, as well as static and dynamical (quantum) phases and phase transitions, and the mechanism for equilibration and thermalization are among those near-term goals. We emphasize that, because of a typical non-logarithmic volume-law entanglement growth in these cases, most non-equilibrium phenomena are out of reach for tensor-network techniques~\cite{banuls2015thermal,banuls2016chiral,banuls2017density,buyens2016hamiltonian}, even for simple 1+1-dimensional models. In this manuscript, within a simple prototype model of QCD, we target quantities with such characteristics, and concretely lay out the path to obtain non-equilibrium and topological features using quantum hardware. The first results on these features are obtained by performing the simulation on commercial trapped-ion quantum hardware accessible through Cloud services. This allowed for testing the robustness of the expected features to realistic hardware noise.

The first part of this manuscript focused on  non-equal time observables and correlation functions to unravel a far-from-equilibrium dynamical quantum phase transition in the lattice Schwinger model with topological features. An optimal time-evolution scheme was employed, based on a combined Fourier and Bogoliubov basis transformations, with zero Trotter error in the non-interacting limit. The scheme also allowed preparing (ground and excited) eigenstates of free-fermion Hamiltonian exactly. The algorithm can be easily generalized to a wide class of fermion models in any dimension~\footnote{Albeit with growing complexity of the fermionic swap operations of the fermion Fourier transform in higher than 1+1 dimensions.}, e.g., in condensed matter and atomic, molecular and optical systems. Further, a Ramsey interferometry scheme was used to quantum compute real-time observables such as Loschmidt echos and non-equal time correlation functions, at both zero and strong couplings. The underlying topological origin of the DQPT was revealed by the noise-robust holonomy of the obtained NECFs. The second part of this manuscript utilizes a random-measurement-based entanglement-tomography protocol~\cite{van2012measuring,ohliger2013efficient,pichler2016measurement,dalmonte2018quantum,elben2018renyi,vermersch2018unitary,brydges2019probing,elben2019statistical,elben2020mixed,kokail2021entanglement,huang2020predicting,zhou2020single,anshu2021sample,kokail2021entanglement,huang2021efficient,rath2021importance,huang2021demonstrating,zhao2021fermionic,neven2021symmetry,kunjummen2021shadow,elben2022randomized} to extract, for the first time for a lattice gauge theory, R\'enyi entropies and fidelities, and, based on a generalization of the Bisognano-Wichmann theorem, reduced density matrices and entanglement Hamiltonians. These techniques  are  applicable to other digitized models and in higher dimensions.

The quality of results obtained using IonQ's device is promising given the indirect, and thus not fully-optimized, access to the device. The all-to-all connectivity of the trapped-ion systems were beneficial to the scheme of this work, especially for non-equal time interferometry where a single ancillary qubit is used to control evolution on all qubits. The classical entanglement tomography studied in this work allowed the quantification, based on a few test cases, of the performance of the machine for a given gate count and entanglement of the state. Because of the indirect access, hence limited knowledge of the calibration time frame and the exact operations compiled at the hardware level, as well as trap, laser, ions, and environment characteristics, a detailed modeling of noise was not possible, and a simple error model based on single-qubit depolarization, bit-flip, and phase errors appear not to be sufficient to explain deviations from an ideal simulation. The observed deviations are likely mainly from imprecise 1- and 2-qubit gates, noting that especially the latter involves long pulse operations and are thus susceptible to device parameter drifts, changes in the position and heating of the ions, and other decoherence effects. Furthermore, the employed entanglement protocol is very sensitive to the precision of 1-qubit rotations~\cite{brydges2019probing}, and no optimization was applied in our tomography protocol against this effect. More direct access to machine parameters and control may allow for the mitigation of these effects significantly. It may be valuable to repeat and extend the presented computations to other available quantum hardware in future work to cross-examine the simulations against various sources of error.

This work opens up several avenues for explorations on theoretical, algorithmic, and hardware implementation fronts. It will be interesting to extend this investigation to models with non-Abelian symmetries and to higher-dimensional gauge theories, in search for unique non-Abelian or 2D/3D features in non-equilibrium properties of matter. For realistic QCD scenarios, finite density and finite temperature must enter the considerations, and the present algorithms for non-equal time correlation functions and topological charges can be combined with efficient thermal-state preparation algorithms towards that goal, see e.g., Refs.~\cite{czajka2022quantum,davoudi2022toward}. Furthermore, a barely explored direction is the entanglement structure of lattice gauge theories, with important applications in the thermalization of Abelian and non-Abelian gauge theories. The reason is that entanglement structure is known to be a powerful diagnostic tool~\cite{mueller2021thermalization}, and may allow characterizing quantum phases~\cite{li2008entanglement}. Finally, to probe entanglement structures of gauge theories on quantum computers, more efficient algorithms for entanglement tomography can be employed
to take advantage of global and local symmetries of the systems~\cite{bringewatt2023randomized}.

\section*{Acknowledgments}
\noindent
We are grateful to Sonika Johri, Matthew Keesan, Denise Ruffner (IonQ), Anthony Tricario (Google Cloud), 
Robert Hugg, Samuel Porter, Sheila M. Zellner-Jenkins (Division of Information Technology, University of Maryland), Ansonia Saunders (Department of Procurement and Strategic Sourcing, University of Maryland), Franz Klein, John Sawyer (Mid-Atlantic Quantum Alliance, University of Maryland), Mark Heil (Burwood), Misha Dhar, Fabrice Frachon, Lorilyn Hall,  George Moussa, Martin Roettler, and Tracy Woods (Microsoft) for their great help and efforts in enabling us, by providing the organizational, technological, and legal groundwork, to use IonQ's quantum hardware through the Google Cloud and Microsoft Azure infrastructures, and for answering our many questions. We thank Torsten Zache for discussions and careful reading of this manuscript.
Z.D. and N.M. acknowledge funding by the U.S. Department of Energy’s Office of Science, Office of Advanced Scientific Computing Research, Accelerated Research in Quantum Computing program award DE-SC0020312 for algorithmic developments for fermionic simulations, and the U.S. Department of Energy’s Office of Science, Office of Nuclear Physics under Award no. DE-SC0021143 via the program on Quantum Horizons: QIS Research and Innovation for Nuclear Science for quantum simulation of gauge-theory dynamics on near-term quantum hardware. N.M. acknowledges funding by the U.S. Department of Energy, Office of Science, Office of Nuclear Physics, InQubator for Quantum Simulation (IQuS) (\url{https://iqus.uw.edu}) under Award Number DOE (NP) Award DE-SC0020970. ZD is further supported by the U.S. Department of Energy, Office of Science, Early Career Award, under award no. DESC0020271.  K.Y.A. and E.D. were supported by the U.S. Department of Energy, Office of Science, National Quantum Information Science Research Centers, Quantum Science Center. K.Y.A. was also supported by MITRE Corporation TechHire and Quantum Horizon programs. The hardware-implementation results were produced using computing credits from IonQ, Google and Microsoft.

\appendix
\section{Analytic Results for
\\
Non-interacting Theory}\label{app:analytics}
\noindent
In this Appendix, analytic results are obtained in the non-interacting theory, i.e., Schwinger model in the $e=0$ limit, for quantities such as the Loschmidt echo, non-equal time correlation functions, R\'enyi entropy, and the entanglement spectrum. This is achieved by analytically diagonalizing the free fermion Hamiltonian in momentum-space computational basis. For $e/|m|>0$, results are numerically determined using exact diagonalization  (\textsc{QuSpin}~\cite{weinberg2017quspin}), as are shown in various plots throughout the main text.

\subsection{Diagonalization of the fermion Hamiltonian}\label{app:diag}
Starting from the Schwinger-model Hamiltonian in position space at $e=0$, \Eq{eq:H0-position}, one arrives at a $2\times 2$ block structure of $H_0$ in momentum space,
 \begin{align}
 \label{eq:defHmom}
      H_0 = \sum_{q=-N/4}^{N/4-1} 
   \Psi^\dagger_q H_q \Psi_q, \qquad H_q\equiv  \begin{pmatrix}
     m & \tilde{p} (q) \\
     \tilde{p}^* (q) & -m
    \end{pmatrix}\,,
 \end{align}
with  $\tilde{p}(q)\equiv a^{-1}e^{i2 \pi q /N} \cos({2 \pi q}/{N} ) $, considering the fermionic Fourier transform defined in \Eq{eq:Fourierdef}, see~\cite{ferris2014fourier,epple2017implementing,babbush2018low,cervera2018exact,kivlichan2020improved} for earlier circuit implementations. Equation~(\ref{eq:defHmom}) obeys the following eigenvalue equation
\begin{align}
     H_q u_q = \omega_q u_q\,, \qquad  H_q v_{-q} = -\omega_q v_{-q}\,,
 \end{align}
leading to \Eq{eq:free}, where $\omega_q \equiv \sqrt{m^2 + |\tilde{p}(q)|^2}$, and the eigenspinors are given as
\begin{align}\label{eq:defspinors}
    u_q \equiv \begin{pmatrix}
        \cos(\beta) \\
        e^{-i\alpha} \sin(\beta)
    \end{pmatrix}  \,,\quad v_{-q} \equiv
    \begin{pmatrix}
        -e^{i\alpha} \sin(\beta)\\\cos(\beta)
    \end{pmatrix}\,,
\end{align}
with $\alpha \equiv \frac{2\pi q}{N}$ and $\beta\equiv \arctan \big( \big[\frac{\omega_q -m}{\omega_q+m} \big]^{{1}/{2}}\big)$.
Equation~(\ref{eq:defHmom}) is diagonalized by a Bogoliubov transformation 
\begin{align}\label{eq:BogApp}
   \Psi_q=U_q \begin{pmatrix}
          a_q\\b_{-q}^\dagger
   \end{pmatrix}\,,
\end{align}
see Eqs.~(\ref{eq:Bog1}) and (\ref{eq:Bog2})), where the $2\times 2$ matrix
$U_q \equiv  (u_q,v_{-q}) $
is defined in terms of the two-component spinor wavefunctions \Eq{eq:defspinors}.

For the time-evolution algorithm presented in the main text, the initial state is prepared in the $(a_q,b_{-q}^\dagger)^T$-eigenbasis of $H_0$ with mass $+m$. This is followed by transforming into the eigenbasis of $\Psi_q=(\psi_q^{\rm even},\psi_q^{\rm odd})^T$ (the inverse transformation relative to \Eq{eq:BogApp}), then transforming back into the eigenbasis of $H_0(-m)$. Both transformations combined lead to the `quench' dynamics studied in this work, see Appendix \ref{app:quenchana} for details. Hereafter, the  transformation from the $(a_q,b_{-q}^\dagger)^T$-eigenbasis of $H_0$ (at either $\pm m$) to the $(\psi_q^{\rm even},\psi_q^{\rm odd})^T$-eigenbasis will be called  `forward' transformation, while its inverse, from the $(\psi_q^{\rm even},\psi_q^{\rm odd})^T$ to the $(a_q,b_{-q}^\dagger)^T$-eigenbasis 
will be called `backward' transformation.

Note that while this work considers only two specific values of the $\theta$ angle, namely $\theta=0$ corresponding to mass $m$ and $\theta=\pi$ corresponding to mass $-m$, the algorithms of this work are applicable to arbitrary values of $\theta$. According to Eq.~(3), the only change compared to the $\theta=0$ case is the introduction of the $\theta$-dependent mass $m\cos\theta$, which leads to different values $\omega_{\bm{q}}$ and $\beta$, without changing the Hamiltonian diagonalization routines and subsequent analyses.

\subsection{Analytics: Loschmidt echo and non-equal time correlation functions}\label{app:quenchana}
The quench studied in the main text is realized by preparing the system in the ground state of $H_0(+m)$ with positive mass $+m$ (\Eq{eq:free}), followed by time evolution using $H_0(-m)$ with negative mass $-m$. Because the dispersion $\omega_q$ is independent of the sign of the mass, $H_0(+m)$ and $H_0(-m)$ are formally identical in their respective momentum-space eigenbasis. The quench is therefore performed by explicit  basis change  from the $H_0(+m)$ to $H_0(-m)$ eigenbases. This transformation is $\prod_q B^\dagger_q(-m) B_q(+m)\equiv Q$, with $B_q(m)$ defined in the previous section.  Time-dependent observables, after this quantum quench, can be computed analytically, which is discussed below~\cite{Zache:2018cqq}.

To simplify notation, the argument $\pm m$ will be dropped, but states and operators of $H_0(+m)$ are labelled with a bar, e.g., the ground state of $\bar{H}_0=H_0(+m)$ is $ | {\rm GS}(m)\rangle \equiv \ket{ \bar{\Omega}} = \prod_q \ket{\bar{0}_q^a}\ket{\bar{0}_q^b}$  (likewise $\bar{a}_q^\dagger$, $\bar{a}_q$, $\bar{b}_{-q}^\dagger$, and $\bar{b}_{-q}$ for the operators). States and operators in the eigenbasis of $H_0\equiv H_0(-m)$ have no bar, e.g.,
$ \ket{ {\Omega} } = \prod_q \ket{ {0}_q^a}\ket{{0}_q^b}$ for the ground state and ${a}_q^\dagger$, ${a}_q$, ${b}_{-q}^\dagger$, and ${b}_{-q}$ for the operators.  The bases are related by forward and backward Bogoliubov transforms (\Eq{eq:Bog1}),
\begin{align}\label{eq:Bogsignmass}
    \begin{pmatrix}
        \bar{a}_q
        \\
        \bar{b}_{-q}^\dagger
    \end{pmatrix} = \bar{U}^\dagger_q {U}_q
    \begin{pmatrix}
        {a}_q
        \\
        {b}_{-q}^\dagger
    \end{pmatrix}
    \,,
\end{align}
where
\begin{align}
 \bar{U}^\dagger_q {U}_q \equiv 
 \begin{pmatrix}
        \bar{u}_q^\dagger \cdot{u}_q &  \bar{u}_q^\dagger\cdot {v}_{-q}\\
        \bar{v}_{-q}^\dagger\cdot {u}_q & \bar{v}_{-q}^\dagger\cdot {v}_{-q}
    \end{pmatrix}\,.
\end{align}
The Loschmidt echo is
\begin{align}
    L(t) \equiv \langle \bar{\Omega} | e^{-iH_0(-m)t} |  \bar{\Omega} \rangle\,,
\end{align}
where  $|\bar{\Omega} \rangle$ obeys $\bar{b}_{-q}|  \bar{\Omega} \rangle = \bar{a}_{q} |\bar{\Omega} \rangle  =0$
and can be written as
\begin{align}
|\bar{\Omega} \rangle = \prod_{q=-N/4}^{N/4-1} \mathcal{N}_q \, \bar{b}_{-q} \bar{a}_q | 0^a_q \rangle| 0^b_q \rangle \,,
\end{align}
 with $\mathcal{N}_q$ a normalization ensuring $L(0)=1$.  
Finally, after abbreviating
\begin{align}
   \mathcal{A}_q &= (v_{-q}^\dagger \cdot \bar{v}_{-q}) (\bar{u}_q^\dagger \cdot v_{-q}) = \frac{({\omega_q^2-m^2})^{\frac{1}{2}}}{\omega_q}\,,\\
    \mathcal{B}_q &= (u_{q}^\dagger \cdot \bar{v}_{-q}) (\bar{u}_q^\dagger \cdot v_{-q})  = 1\,,
\end{align}
explicit computation yields
\begin{align}
    L(t) &= \prod_q |\mathcal{N}_q|^2\left(| \mathcal{A}_q|^2 \, e^{i\omega_q t} +|\mathcal{B}_q|^2  e^{-i\omega_q t} \right)
    \nonumber\\
   &= \prod_q\left[\cos(\omega_q t) + i \frac{\sin(\omega_q t)}{1-2 \omega_q^2/m^2}{} \right]\,,
\end{align}
where $|\mathcal{N}_q|^2 = (| \mathcal{A}_q|^2+| \mathcal{B}_q|^2)^{-1}$. A similar derivation, omitted here, allows to determine the NECFs, $g_q(t)$ in (\Eq{eq:correlator}. Moreover, one finds
\begin{align}
L(t) = \prod_q g_q(t).
\end{align}
in the non-interacting case $e=0$~\cite{Zache:2018cqq}.

\subsection{Analytics: Entanglement entropy}
In the non-interacting theory, the density matrix $\rho(t)\equiv |\Psi(t) \rangle\langle \Psi(t) | $ is Gaussian, and is given by~\cite{chung2001density,cheong2004many}
\begin{align}
\label{eq:gaussianstate}
    \rho(t) = &\det(1-G(t))\, 
    \nonumber\\ &\times e^{  \sum_{n_1,n_2} \log\left[\frac{G(t)}{1-G(t)}\right]_{n_1,n_2} \psi_{n_1}^\dagger \psi_{n_2}}\,.
\end{align}
$ \rho(t)$ is therefore uniquely determined by equal-time (position-space) correlation functions, $G_{n_1,n_2}(t)\equiv\langle \psi_{n_1}^\dagger \psi_{n_2} \rangle$, where $\langle \cdot \rangle \equiv \langle \Psi(t) | \cdot |\Psi(t) \rangle $ and $|\Psi(t)\rangle \equiv e^{-iH_0 t } | \bar{\Omega}\rangle$. At  $t=0$, the correlator is
\begin{align}
\label{eq:Gcorr}
    G_{n_1,n_2}(0) &= \frac{1}{N/2} \sum_{q=-\frac{N}{4}}^{\frac{N}{4}-1}
    e^{i\frac{2\pi q}{N} (n_1 - n_2)}
  \nonumber\\ \times&
    \begin{cases}
    \sin^2(\bar{\beta}) &\text{if }  n_1,n_2\text{ even}\\
    \cos^2(\bar{\beta}) & \text{if }  n_1,n_2 \text{ odd}\qquad, \\
    -\sin(\bar{\beta})\cos(\bar{\beta}) & \text{if }n_1,n_2 \text{ opposite}
    \end{cases}
\end{align}
where $\bar{\beta}$ is evaluated at $+m$. Here, `opposite' means that ${n}_1$ is odd and ${n}_2$ even or vice versa. For $t\ge 0$,
\begin{align}
\label{eq:Gcorr1}
 G_{n_1,n_2}(t) &= \langle \Psi(0)| e^{iH_0t }  \psi_{n_1}^\dagger e^{-iH_0t } e^{iH_0t } \psi_{n_2} e^{-iH_0t } | \Psi(0) \rangle\,,
\end{align}
where
\begin{align}
\label{eq:modestimeevo}
    &e^{iH_0t } \psi_{n} e^{-iH_0t } = 
    \frac{1}{N/2} \sum_{q=-\frac{N}{4}}^{\frac{N}{4}-1}
    e^{i\frac{2\pi}{N/2} q  \mathfrak{n}(n)}
    \nonumber\\
   & \times \begin{cases}
   \cos(\beta) e^{-i \omega_q t} \,a_q - e^{i\alpha} \sin(\beta) e^{i\omega_q t} \,b_{-q}^\dagger & \text{if }n\text{ even}\\
   e^{-i\alpha}\sin(\beta) e^{-i\omega_qt} \, a_q+
   \cos(\beta) e^{i\omega_q t}\, b_{-q}^\dagger & \text{if }n\text{ odd}
   \end{cases}
\end{align}
with supercell index $\mathfrak{n}(n)\equiv n/2$ ($\mathfrak{n}(n)\equiv (n-1)/2$) if $n$ is even (odd), and $\beta$ is evaluated at $-m$, i.e., $\beta\equiv \arctan \big( \big[\frac{\omega_q +m}{\omega_q-m} \big]^{{1}/{2}}\big)$ ($\alpha$ and $\omega_q$ are independent of $\text{sign}(m)$). (A similar expression is found for $ e^{iH_0t } \psi_{n}^\dagger e^{-iH_0t }$.) Next, one uses \Eq{eq:Bogsignmass} to replace
\begin{align}
\label{eq:Uexpl}
   & a_q = c_{\beta,\bar{\beta}}\, \bar{a}_q + e^{i\alpha} d_{\beta,\bar{\beta}}\, \bar{b}_{-q}^\dagger\,,\nonumber
    \\
   & b_{-q}^\dagger = -e^{-i \alpha} d_{\beta,\bar{\beta}}\, \bar{a}_q + c_{\beta,\bar{\beta}} \, \bar{b}_{-q}^\dagger\,,
\end{align}
where $c_{\beta,\bar{\beta}}  \equiv \cos(\beta) \cos(\bar{\beta}) + \sin(\beta) \sin(\bar{\beta})$ and $d_{\beta,\bar{\beta}}  \equiv\cos(\beta) \sin(\bar{\beta}) - \sin(\beta) \cos(\bar{\beta})$, where $\bar{\beta}$ is evaluated at $+m$, i.e., $\bar{\beta}\equiv \arctan \big( \big[\frac{\omega_q -m}{\omega_q+m} \big]^{{1}/{2}}\big)$, and equations for $a_q^\dagger$ and $b_{-q}$ are obtained by Hermitian conjugation. Inserting \Eq{eq:modestimeevo} into \Eq{eq:Gcorr1} and using \Eq{eq:Uexpl} yields
\begin{align}
\label{eqGtcomplete}
    &G_{n_1,n_2}(t) =
    \frac{1}{N/2} \sum_{q=-\frac{N}{4}}^{\frac{N}{4}-1}
    e^{i\frac{2\pi q}{N} (n_1 - n_2)}
  \nonumber\\ \times&
  \begin{cases}
  \big| \cos(\beta) d_{\beta,\bar{\beta}} e^{i\omega_q t} - \sin(\beta) c_{\beta,\bar{\beta}} e^{-i\omega_q t}   \big|^2 &\text{if }  n_1,n_2\text{ e/e}\\
  \big| \sin(\beta) d_{\beta,\bar{\beta}} e^{i\omega_q t} + \cos(\beta) c_{\beta,\bar{\beta}} e^{-i\omega_q t}   \big|^2 &\text{if }  n_1,n_2\text{ o/o}\\
  \big[\cos(\beta) d_{\beta,\bar{\beta}}e^{i\omega_q t} - \sin(\beta) c_{\beta,\bar{\beta}}e^{-i\omega_q t}\big] &  \\
  \times \big[\sin(\beta) d_{\beta,\bar{\beta}}e^{-i\omega_q t} + \cos(\beta) c_{\beta,\bar{\beta}}e^{i\omega_q t}\big] & \text{if }  n_1,n_2\text{ e/o}\\
  \big[\cos(\beta) d_{\beta,\bar{\beta}}e^{-i\omega_q t} - \sin(\beta) c_{\beta,\bar{\beta}}e^{i\omega_q t}\big] & \\
 \times [\sin(\beta) d_{\beta,\bar{\beta}}e^{i\omega_q t} + \cos(\beta) c_{\beta,\bar{\beta}}e^{-i\omega_q t}]&  \text{if } n_1,n_2\text{ o/e }
  \end{cases}
\end{align}
with the abbreviation `e'=`even' and `o'=`odd'. To compute the reduced density matrix $\rho_A \equiv \text{Tr}_B(\rho)$, where $B$ is the complement of $A$, one simply replaces  $G\rightarrow G^A$, where $G^A\equiv G^A_{{n_1},{n}_2}(t) $ is restricted to subsystem $A$ in \Eq{eq:gaussianstate}. 
The entanglement Hamiltonian then follows
\begin{align}\label{eq:entH}
&H_A(t)=-\log[\rho_A(t)] \nonumber\\
&=-\sum_{n_1,n_2=0}^{N_A-1}\log\left[\frac{G^A}{1-G^A}\right]_{n_1,n_2} \psi_{n_1}^\dagger \psi_{n_2}\,.
\end{align}
Here, an immaterial (time-dependent) additive number has been dropped. Because $H_A(t)$ is dimensionless, this constant has no further physical meaning; conventionally the level distribution of $H_A(t)$ is scaled to lie between a standard interval, e.g., [0,1]. 
To obtain the entanglement spectrum, one diagonalizes $H_A$ simply by diagonalizing $G^A$, yielding
\begin{align}\label{eq:EHdiag}
    H_A(t)=-\sum_{n=0}^{N_A-1} \log\left[\frac{\mathcal{G}_n(t)}{1-\mathcal{G}_n(t)} \right] \xi_{n}^\dagger(t) {\xi}_{n} (t)\,,
\end{align}
where $\mathcal{G}_n$ are the $N_A$ eigenvalues of the (single-particle) correlation function $G^A(t)$. and 
${\xi}_{n}^\dagger(t) {\xi}_{n}$(t) is the time-dependent number operator in the eigenbasis of $H_A$ (Schmidt basis). Because it is in the form of a single-particle problem, $G^A(t)$ and thus \Eq{eq:entH} can be numerically diagonalized easily for an arbitrary large system. Computing  the Schmidt spectrum $p_\lambda(t)$, where $\lambda \in[0,2^{N_A-1}]$ labels the time-dependent Schmidt vectors $| \lambda(t)\rangle$, or the corresponding R\'enyi or von Neumann entropies, from \Eq{eq:EHdiag} is a simple counting exercise.

\section{Quantum Algorithms}\label{app:quantumalgo}
\noindent
The algorithmic details of the time-evolution scheme,  Ramsey interferometry circuits, as well as the entanglement tomography scheme are discussed in this Appendix. All circuits are publicly available at \url{https://gitlab.com/Niklas-Mueller1988/dqpt_2210.03089}.

\subsection{Time evolution}
Time evolution is implemented in two bases: in the eigenmode space of $H_0$ in \Eq{eq:free}, and in position space, in which the interaction part, \Eq{eq:positioninteractingH}, can be implemented more straightforwardly. Digitization of the time-evolution operator follows a Trotter scheme, i.e., \Eq{eq:trotterscheme}, which involves Bogoliubov and fermionic Fourier transforms. 

The forward Bogoliubov transformation decomposes into ${N}/{2}$ 2-qubit unitaries, $B\equiv \otimes{}_{q=-N/4}^{N/4-1} B_q$. The matrix representation for the forward transformation, in the two-qubits basis constituting momentum mode $q$, is
\begin{align}
\label{eq:bogstate2}
     B_q = 
     \begin{pmatrix}
     0 & 1 & 0 & 0 \\
    \cos(\beta) & 0 & 0 &  -e^{-i\alpha}\sin(\beta) \\
     e^{i\alpha}\sin(\beta) & 0 & 0 & \cos(\beta)\\
     0 & 0 & 1 & 0 
     \end{pmatrix}.
 \end{align}
where the rows correspond to the Fock basis states $a_q^\dagger b_{-q}^\dagger \ket{0_q^a }\ket{0_q^b}$, $ a_q^\dagger \ket{0_q^a }\ket{0_q^b}$, $b_{-q}^\dagger \ket{0_q^a} \ket{ 0_q^b} $ and $|  0_q^a  \rangle |0_q^b \rangle $ (read \Eq{eq:bogstate2} from top to bottom). The action of $B$ on these states obtains the two-qubit  states ${\psi^{\rm even}_q}^\dagger{\psi^{\rm odd}_q}^\dagger \ket{0^{\rm even}_q}\ket{0^{\rm odd}_q}$, ${\psi^{\rm even}_q}^\dagger \ket{0^{\rm even}_q}\ket{0^{\rm odd}_q}$, ${\psi^{\rm odd}_q}^\dagger \ket{0^{\rm even}_q}\ket{0^{\rm odd}_q}$, and $\ket{0^{\rm even}_q}\ket{0^{\rm odd}_q}$, respectively. These states then correspond to the qubit computational basis $\ket{00}$, $\ket{01}$, $\ket{10}$, $\ket{11}$, respectively. With this convention, one arrives at the following circuit representation for $B_q$
\begin{center}
\begin{adjustbox}{width=0.48\textwidth}
\begin{quantikz}[transparent, row sep=0.3cm]
\qw\qw & \qw  & \qw & \targ{}   & \ctrl{1} & \targ{}  & \qw & \qw  & \qw & \qw
\\
\qw & \gate{  R^z_{ -\alpha} } &  \gate{X} & \ctrl{-1} & \gate{R^y_{-2\beta}} & \ctrl{-1} & \gate{X}&\gate{  R^z_{ \alpha} }& \gate{X} & \qw 
\end{quantikz}
\end{adjustbox}
\end{center}
where $R^{z/y}_\gamma \equiv e^{-i \gamma \sigma^{z/y}/2}$ with $\alpha$ and $\beta$ defined in Appendix \ref{app:diag}.

The (staggered) fermionic Fourier transform  (\Eq{eq:Fourierdef}), from momentum space to position space, $F$, is given by
\begin{center}
\begin{adjustbox}{width=0.4\textwidth}
\begin{quantikz}[transparent, row sep=0.08cm]
 \qw & \gate[8]{\text{fSWAP}}& \gate[4] {\mathcal{F}_{N/2}^\dagger }& \qw & \gate[8]{\text{fSWAP}^\dagger} & \qw  \\
 \qw &                       &                                      & \gate{Z}  &                                & \qw \\
 \qw &                       &                                      & \qw &                                &  \qw\\
  \qw &                       &                                     & \gate{Z}  &                                 &\qw \\
\qw &                         & \gate[4] {\mathcal{F}_{N/2}^\dagger }& \qw &                                &\qw \\
 \qw &                       &                                       &\gate{Z} &                                & \qw \\
  \qw &                       &                                      & \qw &                                &  \qw\\
\qw &                          &                                     &  \gate{Z}  &                               &\qw
\end{quantikz}
\end{adjustbox}
\end{center}
for the $N=8$ example, where $\mathcal{F}_{N/2}^\dagger$ are separate Fourier transforms, from momentum to position space for the $N/2$ even and odd sites.  Pauli-$Z$ gates account for the fact that the Fourier transform is symmetric, i.e., $q \in [-\frac{N}{4},\frac{N}{4}-1]$. In the following, we will discuss the $\mathcal{F}_{N/2}$ components implementing the asymmetric Fourier transform, i.e,. $q \in [0,\frac{N}{2}-1]$, keeping in mind that the aforementioned $Z$ gates must be incorporated subsequently. fSWAP gates permute fermionic modes, accounting for their parity, so that each $\mathcal{F}_{N/2}$ has either even or odd sites as input.
Examples are, in the case of $N=4$ lattice sites,
\begin{center}
\begin{adjustbox}{width=0.3\textwidth}
\begin{quantikz}[transparent, row sep=0.1cm]
    n=0\,\; \gategroup[4,steps=3,style={white},background]{} 
            & \qw                         & \qw \;  n=0\,\rstick[wires=2]{even\\ sites} \\
    n=1\,\;& \gate[swap,style={white}]{} & \qw  \; n=2\, \\
    n=2\,\; &                             & \qw  \; n=1\rstick[wires=2]{odd\\ sites} \\
    n=3\,\; & \qw                         & \qw \;  n=3
\end{quantikz}\,,
\end{adjustbox}
\end{center}
and for $N=8$,
\begin{center}
\begin{adjustbox}{width=0.471\textwidth}
\begin{quantikz}[transparent, row sep=0.1cm]
    n=0\; \gategroup[4,steps=5,style={white},background]{} 
            & \qw                         & \qw                        
            & \qw                         & \qw   \; n=0 \rstick[wires=4]{even\\ sites} \\
        n=1\; & \gate[swap,style={white}]{} & \qw                        
            & \qw                         & \qw   \; n=2 \\
   n=2\; &                             & \gate[swap,style={white}]{}
            & \qw                         & \qw  \; n=4 \\
    n=3\; & \gate[swap,style={white}]{} &                            
            & \gate[swap,style={white}]{} & \qw   \; n=6 \\
    n=4\; &                             & \gate[swap,style={white}]{}
            &                             & \qw   \; n=1 \rstick[wires=4]{odd\\ sites}\\
    n=5\; & \gate[swap,style={white}]{} &                            
            & \qw                         & \qw   \; n=3\\
    n=6\; &                             & \qw                        
            & \qw                         & \qw   \; n=5 \\
    n=7\; & \qw                         & \qw                        
            & \qw                         & \qw   \; n=7
\end{quantikz}\,.
\end{adjustbox}
\end{center}
Crossed lines denote the 2-qubit fermionic swap,
\begin{align}\label{eq:fsawp}
   \text{fSWAP}_2\equiv  \begin{pmatrix}
           -1 & 0 & 0 & 0 \\
            0 & 0 & 1 & 0 \\
             0 & 1 & 0 & 0 \\
             0 & 0 & 0 & 1
    \end{pmatrix}\,,
\end{align} 
where rows and columns denote basis states in the following order: $ \psi_0^\dagger \psi_1^\dagger  |  0_0  \rangle |0_1 \rangle$, $ \psi_0^\dagger |  0_0  \rangle |0_1 \rangle$, $ \psi_1^\dagger  |  0_0  \rangle |0_1 \rangle$, and $ |  0_0  \rangle |0_1 \rangle$. 
Here, $\psi_0$ and $\psi_1$ stand for any two fermionic modes the sub-circuit acts on, the notation indicating that qubit $0$ comes before qubit $1$ when circuits are read from top to bottom. fSWAP${}_2$, up to an overall constant phase, has the
circuit representation
\begin{center}
\begin{adjustbox}{width=0.48\textwidth}
\begin{quantikz}[transparent, row sep=0.3cm]
\qw & \gate{X} & \gate{R_{\frac{\pi}{2}}^x} & \ctrl{1} & \gate{R^x_{\frac{\pi}{2}}} & \ctrl{1} &  \gate{R^x_{-\frac{\pi}{2}}} &  \gate{R^z_{\frac{\pi}{2}}} & \gate{X} &\qw\\
\qw & \gate{X} & \gate{R_{\frac{\pi}{2}}^x} & \targ{} & \gate{R^z_{\frac{\pi}{2}}} & \targ{} & \gate{R^x_{-\frac{\pi}{2}}} &\gate{R^z_{\frac{\pi}{2}}} & \gate{X} & \qw
\end{quantikz}\,,
\end{adjustbox}
\end{center}
where  $R^x_\gamma \equiv e^{-i \gamma \sigma^x/2}$. 

The fermionic Fourier transform $\mathcal{F}_{N/2}$, from position to momentum space (the inverse is used in the main text) for a staggered component, even or odd, is
\begin{align}
    {\psi}_q^\dagger = \frac{1}{\sqrt{N/2}} \sum_{\mathfrak{n}=0}^{N/2-1} \psi_{\mathfrak{n}}^\dagger \, e^{-i \frac{2\pi q \mathfrak{n}}{N/2}}\,,
\end{align}
where $\mathfrak{n}\in [0,{N}/2-1]$ labels $N/2$ even or odd sites and $q\in[-N/4,N/4-1]$.   To simplify the discussion, we will from now on
assume $\mathfrak{n} \in [0,N/2-1]$ as well as $q\in [0,N/2-1]$. The symmetric case, i.e., $q\in [-N/4,N/4-1]$,  is realized through additional $Z$-gates as discussed above.

The transformation is defined recursively and consists of 2-qubit unitaries. For the simplest case $N/2=2$, the transformation is
\begin{equation}
\label{eq:FTop}
{\psi}_{q=0}^\dagger = \frac{1}{\sqrt{2}} \big( \psi_{\mathfrak{n}=0}^\dagger + \psi_{\mathfrak{n}=1}^\dagger  \big)\,,~ {\psi}_{q=1}^\dagger = \frac{1}{\sqrt{2}} \big( \psi_{\mathfrak{n}=0}^\dagger - \psi_{\mathfrak{n}=1}^\dagger \big)\,.
\end{equation}
Using explicit notation, the transformation acts as follows on  (two-mode) states,  from (staggered) position, $\mathfrak{n}$ to momentum, $q$ space,
\begin{align}
\label{eq:FTstates}
&{\psi}_{ q=0}^\dagger {\psi}_{q=1}^\dagger  |  0_{q=0}  \rangle |0_{q=1} \rangle =  - 
  {\psi}_{\mathfrak{n}=0}^\dagger {\psi}_{\mathfrak{n}=1}^\dagger  |  0_{\mathfrak{n}=0}  \rangle |0_{\mathfrak{n}=1} \rangle,\nonumber\\
  &{\psi}_{q=0}^\dagger  |  0_{q=0}  \rangle |0_{q=1} \rangle =  \frac{1}{\sqrt{2}} \big( \psi_{\mathfrak{n}=0}^\dagger + \psi_{\mathfrak{n}=1}^\dagger  \big)|  0_{\mathfrak{n}=0}  \rangle |0_{\mathfrak{n}=1} \rangle\,,
  \nonumber\\
   &{\psi}_{q=1}^\dagger  |  0_{q=0}  \rangle |0_{q=1} \rangle =  \frac{1}{\sqrt{2}} \big( \psi_{\mathfrak{n}=0}^\dagger - \psi_{\mathfrak{n}=1}^\dagger  \big)|  0_{\mathfrak{n}=0}  \rangle |0_{\mathfrak{n}=1} \rangle\,,\nonumber\\
    & | 0_{q=0}  \rangle |0_{q=1} \rangle  = |  0_{\mathfrak{n}=0}  \rangle |0_{\mathfrak{n}=1}  \rangle\,, 
\end{align}
where $|  0_{q=0}  \rangle |0_{q=1} \rangle$ and $|0_{\mathfrak{n}=0}  \rangle |0_{\mathfrak{n}=1} \rangle$ are zero-fermion states in momentum and position space, respectively. Equations~(\ref{eq:FTop}-\ref{eq:FTstates}) are realized by
\begin{align}
\label{eqF20}
    F_2^{0} \equiv \begin{pmatrix}
           -1 & 0 & 0 & 0 \\
           0 & \frac{1}{\sqrt{2}} & \frac{1}{\sqrt{2}} & 0\\
           0 & \frac{1}{\sqrt{2}} & -\frac{1}{\sqrt{2}} & 0\\
           0 & 0 & 0 & 1
    \end{pmatrix}\,,
\end{align}
where the row ordering is as in \Eq{eq:FTstates} (from top to bottom).
For $N=4$, ($N/2=2$), \Eq{eqF20} is the full (even- or odd-site)  position-to-momentum Fourier transform. Note that in our staggered formulation with $N=4$ sites, this matrix is surrounded by swap gates ensuring that even (odd) sites are separately transformed, and $Z$ gates ensuring $q\in [-N/4,N/4-1]$.

For $N=8$  ($N/2=4$), the Fourier transform of four (even or odd) modes is
\begin{align}
\label{eq:N8fFT}
    \psi_{q=0}^\dagger &= \frac{1}{\sqrt{2}} \left[
    \frac{1}{\sqrt{2}} (\psi_{\mathfrak{n}=0}^\dagger +\psi_{\mathfrak{n}=2}^\dagger ) + \frac{1}{\sqrt{2}}(\psi_{\mathfrak{n}=1}^\dagger +\psi_{\mathfrak{n}=3}^\dagger )\right] \nonumber\,,
    \\
    \psi_{q=1}^\dagger &= \frac{1}{\sqrt{2}} \left[
    \frac{1}{\sqrt{2}} (\psi_{\mathfrak{n}=0}^\dagger -\psi_{\mathfrak{n}=2}^\dagger ) +  \frac{W^{1}_{4}}{\sqrt{2}}(\psi_{\mathfrak{n}=1}^\dagger -\psi_{\mathfrak{n}=3}^\dagger )\right]\nonumber\,,
    \\
    \psi_{q=2}^\dagger &=
    \frac{1}{\sqrt{2}} \left[
    \frac{1}{\sqrt{2}} (\psi_{\mathfrak{n}=0}^\dagger +\psi_{\mathfrak{n}=2}^\dagger ) +  \frac{W^{2}_{4}}{\sqrt{2}}(\psi_{\mathfrak{n}=1}^\dagger +\psi_{\mathfrak{n}=3}^\dagger )\right]\nonumber\,,
    \\
    \psi_{q=3}^\dagger &=
    \frac{1}{\sqrt{2}} \left[
    \frac{1}{\sqrt{2}} (\psi_{\mathfrak{n}=0}^\dagger -\psi_{\mathfrak{n}=2}^\dagger ) +  \frac{W^{3}_{4}}{\sqrt{2}}(\psi_{\mathfrak{n}=1}^\dagger -\psi_{\mathfrak{n}=3}^\dagger )\right]\,.
\end{align}
where $W_{{N}/2}^k\equiv e^{ -i \frac{2\pi k}{ {N}/2} }$, $k\in [0,N/2-1]$.

A recursive pattern is evident, which can be generalized to arbitrary even $N>8$.
Concretely, for $N=8$ and considering that $W^2_4=-1$ and $W^3_4=-W^1_4$, the position-to-momentum Fourier transform $\mathcal{F}_{4}$ is,
\begin{center}
\begin{adjustbox}{width=0.4\textwidth}
\begin{quantikz}[transparent, row sep=0.3cm]
 & \qw                           & \gate[2]{F_4^{0}} &  \qw                        &  \gate[2]{F_4^{0}} &  \qw & \qw \\
   & \gate[swap,style={white}]{} &                 & \gate[swap,style={white}]{} &                  &   \gate[swap,style={white}]{}& \qw\\
   &                             & \gate[2]{F_4^{0}} &                             &  \gate[2]{F_4^{1}} &                              &\qw\\
   & \qw                         &                 & \qw                         &                  & \qw & \qw
\end{quantikz}
\end{adjustbox}
\end{center}
where
\begin{align}
    F_{{N}/2}^k \equiv \begin{pmatrix}
           -W_{{N}/2}^k & 0 & 0 & 0 \\
           0 & \frac{1}{\sqrt{2}} & \frac{1}{\sqrt{2}}W_{{N}/2}^k & 0\\
           0 & \frac{1}{\sqrt{2}} & -\frac{1}{\sqrt{2}}W_{{N}/2}^k & 0\\
           0 & 0 & 0 & 1
    \end{pmatrix}
\end{align}
with circuit representation
\begin{center}
\begin{adjustbox}{width=0.48\textwidth}
\begin{quantikz}[transparent, row sep=0.3cm]
\qw & \gate{X} & \qw & \targ{} & \ctrl{1} & \targ{} & \gate{Z} &  \gate{X}& \qw \\
\qw & \gate{X} & \gate{P\big({ - \frac{2\pi k}{{N}/2}}\big)}  & \ctrl{-1} & \gate{H} & \ctrl{-1} & \ctrl{-1} & \gate{X} &\qw
\end{quantikz}\,,
\end{adjustbox}
\end{center}
where $P(\gamma) \equiv \text{diag}(1,e^{i\gamma})$ is a phase gate. This concludes the construction of the fermion Fourier transform, from position to momentum space.

As part of the time-evolution scheme, first
the ground state of $H_0(+m)$, $\ket{ \bar{\Omega}} = \prod_q \ket{\bar{0}_q^a}\ket{\bar{0}_q^b}$, is prepared in its eigenbasis. Evolving with $H(-m)$, a change of basis between the eigenbasis of $H_0$ with $\pm m$ is performed, $Q = \bigotimes_q Q_q$, where
\begin{equation}
Q_q \equiv {B}_q^\dagger  \bar{B}_q= 
     \begin{pmatrix}
     \sin(2\beta) & 0 & 0 & e^{-i\alpha}\cos(2\beta) \\
    0 & 1 & 0 &  0 \\
     0 & 0 & 1 & 0\\
     -e^{i\alpha}\cos(2\beta) & 0 & 0 & \sin(2\beta) 
     \end{pmatrix}\,,
\end{equation}
with
 $B_q$ and $\bar{B}_q$ given by \Eq{eq:bogstate2}. The transformation $Q_q$ has the following circuit representation,
\begin{center}
\begin{adjustbox}{width=0.52\textwidth}
\begin{quantikz}[transparent, row sep=0.6cm]
\qw&\gate{X}                            & \ctrl{1} & \gate{R^x_{\gamma_1} } & \gate{H} &  \ctrl{1} & \gate{S}                          & \gate{H} & \ctrl{1} & \gate{R^x_{ -\frac{\pi}{2}}} & \gate{X}  \\
\qw&\gate{R^z_{\gamma_2} } & \targ{}  & \qw                                & \qw      & \targ{} & \qw & \gate{R^z_{-\gamma_1}}   .    &  \targ{}                         & \gate{R^x_{ \frac{\pi}{2}}} & \gate{R^z_{-\gamma_2}}           
\end{quantikz}\,,
\end{adjustbox}
\end{center}
where $\gamma_1\equiv 2\beta -\frac{\pi}{2}$, $\gamma_2\equiv -\alpha -\frac{\pi}{2}$, and $\alpha$ and $\beta$ are evaluated at $|m|$.

For $e=0$, the time-evolution circuit is (c.f. \Fig{fig:overview}(a)) 
\begin{center}
\begin{adjustbox}{width=0.42\textwidth}
\begin{quantikz}[transparent, row sep=0.1cm]
\lstick[wires=4]{ $\bigotimes_q | \bar{0}_q^a \rangle| \bar{0}_q^b \rangle $} & \gate[4]{Q} & \gate[4]{e^{-iH_0(-m])t}} &\qw \rstick[wires=4]{$|\psi(t) \rangle$} \\
 &  &\qw & \qw\\ 
 &  &\qw &\qw\\
 & &  &\qw
\end{quantikz}
\end{adjustbox}
\end{center}
with $ H_0(-m) =  \sum_{q=-\frac{N}{4}}^{\frac{N}{4}-1} \omega_q( a_q^\dagger  a_q - b_{-q} b_{-q}^\dagger )$. This is realized, apart from a global phase, as follows
\begin{align}
    e^{-iH_0(-m)t} = \bigotimes_q [ R^z(2\omega_q t)] _a  \otimes [ R^z(2\omega_q t)  ]_b,
\end{align}
where $R^z$ are single-qubit $Z$ rotations, and  $[\cdot]_{a/b}$ denotes that two modes (2 qubits), $a$ and $b$,  are involved for every $q$. 

In the interacting case, $e/|m|>0$, following the basis change $Q$, the following Trotter scheme, corresponding to \Eq{eq:trotterscheme}, is employed
\begin{center}
\begin{adjustbox}{width=0.471\textwidth}
\begin{quantikz}[transparent, row sep=0.1cm]
& 
\gate[4]{e^{-iH_0\delta t} } &  \gate[4]{B}  &   \gate[4]{F} & \gate[4]{e^{-iH_I \delta t}} & \gate[4]{F^\dagger} & \gate[4]{B^\dagger} &
\\
 & & & & & & & \qw\\
 & & & & & & &\qw \\
 & & & &  & & &\qw 
\end{quantikz}
\end{adjustbox}
\end{center}
$N_T$ times with $\delta t = t/N_{\rm T}$. The time-evolution operator for $e^{-iH_I \delta t}$ includes $
H_I\equiv ae^2\sum_{n,m=0}^{N-1} \nu(d_{nm}) \mathcal{Q}_{n} \mathcal{Q}_m   \,, 
$, where $\nu(d_{nm})$ are defined in \Eq{eq:defnu} and  $\mathcal{Q}_n\equiv \psi_n^\dagger \psi_n - [(-1)^n-1]/2$. In position space the interaction part is diagonal and  consists
 of 1-qubit $R^z$ and 2-qubit $R^{zz}$ rotations. Its implementation requires careful bookkeeping  but is otherwise straightforward.

\subsection{Ramsey interferometry}\label{sec:RS}
Figure \ref{fig:overview}(b) summarizes the Ramsey interferometry scheme utilized in Sec.~\ref{sec:nonequaltime} to compute Loschmidt echos and non-equal time correlation functions.
To quantum compute Loschmidt echos, one can use
\begin{center}
\begin{adjustbox}{width=0.48\textwidth}
\begin{quantikz}
\lstick{$| 0 \rangle$ }                                                         & \gate{H}                        & \qw          &\ctrl{1}  &  \meter{x/y} & \rstick[wires=1]{ $\begin{cases}
\text{Re}[{L(t)}]\\
\text{Im}[{L(t)}]
\end{cases}$
}\\
&\lstick{ $ | \bar{0}_q^a \rangle| \bar{0}_q^b \rangle^{\otimes \frac{N}{2}}  $} & \gate{Q} \qwbundle[alternate]{} &  \gate{e^{-iHt}}  \qwbundle[alternate]{} &   \meter{}  \qwbundle[alternate]{}\vcw{-1} & \rstick[wires=1]{symm-based\\postselection} 
\end{quantikz}\,,
\end{adjustbox}
\end{center}
where measuring $\langle \sigma^x \rangle $ ($\langle \sigma^y \rangle $) returns the real (imaginary) part of $L(t)$. The controlled time-evolution operator $e^{-iHt}$ is either the non-interacting one, $e=0$, which involves no Trotterization (`fast-forwarding'), or uses the Trotter scheme at  $e/|m|>0$. We emphasize that, $Q$, as well as (for $e/|m|>0$) Bogoliubov and Fourier transforms are not controlled by the ancilla. The circuit involves controlled $R^z$ rotations for $e=0$ with simple circuit implementation, and controlled  $R^z$ and $R^{zz}$ rotations
at $e/|m|>0$, also with simple implementation but significantly larger circuit depth.
Again, `symm-based postselection' indicates measurements as part of an error-mitigation scheme, whereby `unphysical', i.e., particle-number-violating measurements, are discarded.

Quantum computation of the NECFs, $g_q(t)$, makes use of the decomposition
\begin{align}
\label{eq:ccorrapp}
    g_q(t)\equiv & \langle \bar{{\Omega}} | \psi_q^\dagger(t) \psi_q(0) |  \bar{{\Omega}}\rangle
   = \langle \bar{{\Omega}} | e^{iH t } a_q^\dagger  e^{-iH t } a_q  |  \bar{{\Omega}}\rangle+ \nonumber\\
    & 
    \langle \bar{{\Omega}} | e^{iH t } b_{-q}  e^{-iH t } b_{-q}^\dagger  |  \bar{{\Omega}}\rangle
    \rangle\,,
\end{align}
where $|  \bar{{\Omega}}  \rangle \equiv \prod_q | \bar{0}_q^a \rangle| \bar{0}_q^b \rangle$. A circuit to compute both terms separately, acting on  the same input states as before, is
\begin{center}
\begin{adjustbox}{width=0.48\textwidth}
\begin{quantikz}
 & \gate{H}   &    \ctrl{1}         & \qw       &   \ctrl{1}       & \qw &   \meter{x/y}  & \\
  & \gate{Q}\qwbundle[alternate]{} &    \gate{a_q / b_{-q}^\dagger} \qwbundle[alternate]{} & \gate{e^{-iHt}} \qwbundle[alternate]{}  &   \gate{a_q^\dagger / b_{-q}}\qwbundle[alternate]{} &     \gate{e^{iHt}}  \qwbundle[alternate]{}   &     \meter{}     \qwbundle[alternate]{} \vcw{-1}      &
\end{quantikz}.
\end{adjustbox}
\end{center}
Because $a_q$, $a_q^\dagger$, $b_{-q}$, $b_{-q}^\dagger$ are non-unitary, they cannot be implemented without the use of additional ancillas. This is beyond the capabilities of the device and a different strategy is employed. First, via a Jordan-Wigner transformation, 
\begin{align}
    a_q^{(\dagger)} = \prod_{q'=-\frac{N}{4};a/b}^{q-1} \big(-\sigma^z_{q'}\big) \sigma^{-(+)}_{a,q} \,,
   \nonumber \\
     b_{-q}^{(\dagger)} = \prod_{q'=-\frac{N}{4};a/b}^{q-1} \big(-\sigma^z_{q'}\big) \sigma^{-(+)}_{b,q} \,,
\end{align}
consistent with the fermionic mode ordering, fermion Fock operators are mapped onto spins. Note that ${q';a/b}$ indicates that the Jordan-Wigner string includes $\sigma^z_q$ Pauli operators corresponding to both $a$ and $b$ modes, see the discussion of mode ordering in Sec.~\ref{sec:DQPTbasics} of the main text. The following decomposition
\begin{align}\label{eq:unitaryopdec}
    \sigma^\pm_q = \frac{1}{2}(\sigma^x_q \pm i \sigma^y_q)
\end{align}
allows to write every term of \Eq{eq:ccorrapp} as a sum of four unitary operators, which are separately computed. For example,
\begin{eqnarray}
    &&\langle \bar{{\Omega}} | e^{iH_0 t } a_q^\dagger  e^{-iH_0 t } a_q  |  \bar{{\Omega}}\rangle \equiv \frac{1}{4} \times
    \nonumber\\
    &&\langle \bar{{\Omega}} |\bigg[ e^{iH_0 t } \prod_{q';a/b} \big(-\sigma^z_{q'}\big) \, \sigma^x_{a,q} \,  e^{-iH_0 t } \,  \prod_{q^{''};a/b} \big(-\sigma^z_{q^{''}}\big) \, \sigma^x_{a,q}
    \nonumber\\
    && \hspace{3 mm} -ie^{iH_0 t } \prod_{q';a/b} \big(-\sigma^z_{q'}\big) \, \sigma^x_{a,q} \,  e^{-iH_0 t } \,  \prod_{q^{''};a/b} \big(-\sigma^z_{q^{''}}\big) \, \sigma^y_{a,q}
    \nonumber\\
    && \hspace{3 mm}-i e^{iH_0 t } \prod_{q';a/b} \big(-\sigma^z_{q'}\big) \, \sigma^y_{a,q} \,  e^{-iH_0 t } \,  \prod_{q^{''};a/b} \big(-\sigma^z_{q^{''}}\big) \, \sigma^x_{a,q} +
    \nonumber\\
    && \hspace{3 mm}e^{iH_0 t } \prod_{q';a/b} \big(-\sigma^z_{q'}\big) \, \sigma^y_{a,q} \,  e^{-iH_0 t } \,  \prod_{q^{''};a/b} \big(-\sigma^z_{q^{''}}\big) \, \sigma^y_{a,q}\bigg] |  \bar{{\Omega}}\rangle\,,
    \nonumber\\
\end{eqnarray}
The second term in \Eq{eq:ccorrapp} is obtained in the same way. Note that the Pauli operations (and not time-evolution operators) will need to be controlled in the full Ramsey interferometry circuit shown above. Putting everything together, obtaining $g_q(t)$ involves 16 circuits, including real and imaginary parts. Because of the significant gate complexity in the interacting case, only the results for  $e=0$ are obtained in the main text. In this case,
\begin{align}
    e^{\pm iH t } = \prod_{q=-\frac{N}{4}}^{\frac{N}{4}-1} e^{\pm iH_q t} \equiv  \prod_{q=-\frac{N}{4}}^{\frac{N}{4}-1} e^{\pm i \omega_q t \big( a_q^\dagger a_q - b_{-q} b^\dagger_{-q}  \big)} \,,
\end{align}
so that time evolution is performed mode-by-mode, each requiring two qubits and one ancilla, with 
$a^\dagger_q = \sigma^+_{a,q}$, $ b^\dagger_{-q} = -\sigma^z_{a,q}\sigma^+_{b,-q} $, etc. A symmetry-based error-mitigation scheme is employed, identical to that for $L(t)$, see Appendix \ref{app:symmerr} for details.
\begin{table*}[ht]
\centering
\begin{tabular}
{|l|c|c|c|}\hline
 \textbf{Component} & Per mode $q$ & {$N=4$}& {$N=8$}\\
 \hline\hline
 Quench gate & 11 / 3  &22 / 6 & 44 / 12 \\\hline
 Free time evolution & 2 / 0 & 4 / 0 & 8 / 0 \\\hline
  Free time evolution (controlled) & 2 / 4 & 4 / 8 & 8 / 16\\\hline
  Interacting time evolution & -- &10 / 12 & 36 / 56 \\\hline
 Interacting time evolution  (controlled) & -- & 128 / 92 & 576 / 408 \\\hline
  Bogoliubov transformation & 7 / 4 & 14 / 8 & 28 / 16 \\\hline
  Fermionic Fourier transform & -- & 44 / 12 & 236 / 68 \\\hline
\end{tabular}
\caption{1-/2-qubit gate count of (not transpiled) sub-circuits used in this manuscript.}
\label{table:complexitycomponents}
\end{table*}

\subsection{Entanglement tomography}
The entanglement tomography scheme is adapted from Refs.~\cite{brydges2019probing,kokail2021entanglement} and is summarized in \Fig{fig:overview}(c). It involves random measurements via basis transformations consisting of  single-qubit rotations $\mathscr{U} = \otimes_i {u}_i  $,
where ${u}_i $ is drawn from a circular unitary ensemble (CUE). In practice ${u}_i$
is given by
\begin{center}
\begin{adjustbox}{width=0.4\textwidth}
\begin{quantikz}
\gate{u_i} & 
\end{quantikz}
=
\begin{quantikz}
\qw& \gate{R^z_{\gamma_1}} & \gate{R^y_{\gamma_2}}  & \gate{R^z_{\gamma_3}}  & \qw
\end{quantikz}\,,
\end{adjustbox}
\end{center}
and the matrices ${u}_i$ are drawn randomly with the parametrization
\begin{align}
    u_i\equiv \begin{pmatrix}
           x & y\\
           -y^* & x^*
    \end{pmatrix}.
\end{align}
The matrices $u_i$ are generated with {\verb unitary_group.rvs(2) } 
of the Python module \textsc{unitary\_group} from the package \textsc{scipy.stats}, based on the algorithm in Ref.~\cite{mezzadri2006generate}.
The angles $\gamma_1$, $\gamma_2$ and $\gamma_3$ in the circuit are determined by~\cite{brydges2019probing}
\begin{align}
    \gamma_1& = \text{Re}\Big\{ \tan^{-1} \left[\frac{-i(x^* -x)}{x^* + x} \right] - \tan^{-1} \left[ \frac{-i(y^* - y)}{y^* + y}\right] \Big\}\,,\nonumber\\
\gamma_2&=\text{Re} \Big\{ 2\tan^{-1} \left[ \frac{|y| }{|x|} \right] \Big\}\,,\nonumber\\
\gamma_3&=\text{Re} \Big\{  \tan^{-1} \left[\frac{-i(x^* -x)}{x^* + x} \right] + \tan^{-1} \left[ \frac{-i(y^* - y)}{y^* + y}\right]\Big\}\,.
\end{align}
The probabilities of measuring the bit string $s$, $P_{\mathscr{U}}(s)$, in the random basis $\mathscr{U}$ are input for the (classical) postprocessing procedure to obtain R\'enyi entropies, fidelities, and the entanglement Hamiltonian via the BW ansatz, see \Eq{eq:EHTprotocol} and \eqref{eq:variation} and the discussions in the main text.
\begin{table*}[t]
\centering
\begin{tabular}{|c|c|c|c|c|c|}
\hline
\textbf{Quantity} & \textbf{Figures} & \textbf{Date} & \textbf{Qubits} & \multicolumn{2}{|c|}{\textbf{Gates}} \\
& & mm/dd/yy & & raw &transpiled \\
\hline\hline
$\Gamma(t)$ ($N=4$) & \Fig{fig:LoschmidtN4free} & 06/06/22 & 5 & 36 / 14  & 34 / 14  \\\hline
$\Gamma(t)$ ($N=8$) & \Fig{fig:LoschmidtN8free} & 06/07/22 & 9 & 
70 / 28  & 66 / 28 \\\hline
$\Gamma(t)$ ($N=4$, $e/|m|=1$) &  \Fig{fig:LoschmidtInteracting}&  06/16/22 & 5 & 206 / 126  & 213 / 126 
\\\hline
 $g_q(t)$, $\nu(t)$ $(N=4)$ & \Figs{fig:corr}{fig:TO} & 08/02/22 & 2 & 19 / 5 & 16 / 5 
 \\\hline
 $g_q(t)$, $\nu(t)$ $(N=8)$ & \Figs{fig:corr}{fig:TO} & 08/02/22 & 2 & 19 / 5 & 16 / 5  
  \\\hline
  $S^{(2)}(t)$ ($N=4$)& \Fig{fig:RenyiFidelityN4} & 06/13/22 & 4 & 84 / 26 & 74 / 26
    \\\hline
 $\rho_A(t)$ $(N=4)$ & \Fig{fig:EHTN4}& 06/13/22 & 4 & 84 / 26 & 74 / 26 \\\hline
 $S^{(2)}(t)$ ($N=8$)& \Fig{fig:RenyiFidelityN8} & 05/23/22 & 8 & 284 / 96 & 254 / 96\\ \hline
\end{tabular}
\caption{Qubit and 1-/2-qubit gate count for all computations performed in this manuscript. Also shown are the dates on which data were obtained on the hardware.
}
\label{table:complexity}
\end{table*}

\subsection{Circuit construction and data generation}\label{app:data_circuitcounts}
All circuits are implemented in Qiskit~\cite{QISKIT} using standard gates and were submitted through use of the Qiskit IonQ Provider, \textsc{Qiskit\_ionq}~\cite{Qiskitionq}.

Table \ref{table:complexitycomponents} shows the 1- and 2-qubit gate count of the subcircuit components employed in this manuscript. The number of gates refers to fundamental gates in Qiskit,  obtained by iteratively decomposing the circuits using {\verb decompose() }. 2-qubit gates count CNOT gates and 1-qubit gates are single-qubit rotations.

In contrast, Table \ref{table:complexity} displays the total 1- and 2-qubit count of all quantum computations. `Raw' refers to the circuits implemented as in Qiskit with gate count as in Table \ref{table:complexitycomponents}, and `transpiled' (`transp.') stands for the gate count of circuits that were transpiled specifically for the utilized hardware, using {\verb transpile() }, and sent to the device through the Cloud. The gate count reflects this transpilations, with 2-qubit gates standing for CNOTs and 1-qubit gates standing for single-qubit rotations.
We note that  native-gate implementation, qubit-ion assignment, etc. is performed after one sends jobs to the Cloud, thus is beyond the user's control.

\section{Experimental Hardware}\label{app:hardware}
\noindent
Experiments were conducted on the 11-qubit IonQ trapped-ion quantum computer, accessed through Google Cloud~\cite{GCP} and Microsoft Azure~\cite{Azure} services. More information about the device can be found in Refs.~\cite{wright2019benchmarking, kawashima2021optimizing}; see also Refs.~\cite{cirac1995quantum,monroe1995demonstration,james1997quantum,steane1998quantum} for ground-laying works and pedagogical introduction of trapped-ion  digital quantum computing.

The trapped-ion quantum computer consists of 13 $^{171}$Yb$^+$ ions which are aligned to form a linear crystal, suspended in a chip trap with a radial pseudopotential frequency of $\approx 3.1$MHz, and spacing between the ions of about 4$\mu$m. In this setup, the 11 ions in the middle, which are more uniformly spaced, were utilized as qubits. In the experiments of this work for $N=4$ ($N=8$) sites, 5 (9) of these qubits were used.  The gates are implemented by using counter-propagating Raman laser beams capable of illuminating individual ions and utilizing the ion-ion coupling mediated by the collective motional modes. The native gate set that is physically executed on IonQ hardware is as follows~\cite{IonQGates}. The single-qubit gates are
\begin{equation}
    {\rm GPI}(\phi) = \begin{bmatrix}
        0 & e^{-i\phi} \\
        e^{i\phi} & 0
    \end{bmatrix}\,, \ {\rm GPI2}(\phi)=\frac{1}{\sqrt{2}}\begin{bmatrix}
        1 & -ie^{-i\phi} \\
        -ie^{i\phi} & 1
    \end{bmatrix}\,,
\end{equation}
where ${\rm GPI}(\phi)$ can be considered as a bit-flip rotation with an embedded phase and ${\rm GPI2}(\phi)$ can be considered as an $R^x(\pi/2)$ gate with an embedded phase. The rotations around the $Z$ axis of the Bloch sphere can also be performed, but these require no quantum operation and only involve a classically advancing or retarding the phase of the following operation in the circuit. The 2-qubit native gate is the M\o{}lmer-S\o{}renson (MS) gate which can be expressed as
\begin{align}
    &{\rm MS}(\phi_0,\phi_1)=\frac{1}{\sqrt{2}} \times 
    \nonumber\\
    &\begin{bmatrix}
    1 & 0 & 0 & -ie^{-i(\phi_0+\phi_1)} \\
   0 & 1 & -ie^{-i(\phi_0-\phi_1)} & 0 \\
    0 & -ie^{i(\phi_0-\phi_1)} & 1 & 0 \\
    -ie^{i(\phi_0+\phi_1)} & 0 & 0 & 1
\end{bmatrix}\,.
\end{align}

Table~\ref{table:IonQ_info} provides reported fidelity and timing for the IonQ system used in the experiments.
\begin{table}[h!]
\begin{tabular}{l|l|}
\hline
\multicolumn{1}{|c|}{\textbf{Fidelity}} & \multicolumn{1}{c|}{\textbf{Timing}} \\ \hline
\multicolumn{1}{|l|}{1Q: 0.99717}       & T1: 10,000 s                         \\ \hline
\multicolumn{1}{|l|}{2Q: 0.9696}        & T2: 0.2 s                            \\ \hline
\multicolumn{1}{|l|}{SPAM: 0.9961}      & 1Q: 0.000011 s                       \\ \hline
                                        & 2Q: 0.00021 s                        \\ \cline{2-2} 
                                        & Readout: 0.000175 s                  \\ \cline{2-2} 
                                        & Reset: 0.000035 s                    \\ \cline{2-2} 
\end{tabular}
\caption{Reported average fidelity values for 1-qubit (1Q), 2-qubit (2Q) gates, and state preparation and measurement (SPAM) errors for the IonQ device used in the experiments (left column), as well as reported qubit relaxation (T1) and coherence (T2) times and 1Q, 2Q, readout, and `reset operations' timing values (right column).}
\label{table:IonQ_info}
\end{table}

\section{Symmetry-based Error-Mitigation Protocol}\label{app:symmerr}
\noindent
A symmetry-based error-mitigation scheme is used in all Ramsey interferometry circuits in Secs. \ref{sec:nonequaltime} and \ref{sec:TO}. For example, computing $L(t)\equiv \langle\Psi(0)   | \Psi(t) \rangle = \langle {\rm GS}(m) |e^{-i H(-m) \, t} | {\rm GS}(m)\rangle$ (\Eq{Eq:loschmidt}) for $e=0$, the Ramsey scheme produces the following superposition of time-evolved and non-time-evolved state
\begin{align}\label{eq:freeerrormit}
  |\Psi\rangle = | 0 \rangle \, Q | {\rm GS}(m) \rangle + 
  | 1 \rangle\, e^{-i H_0(-m) \, t}  Q | {\rm GS}(m)\rangle,
\end{align}
where $ | {\rm GS}(m)\rangle\equiv | \bar{\Omega} \rangle\equiv \prod_{q=-N/4}^{N/4-1} | \bar{0}^a_q \rangle | \bar{0}^b_q \rangle $ is the ground state of the non-interacting Hamiltonian $\bar{H}_0$ with mass $m$ in momentum space, $Q\equiv B^\dagger(-m) B(m)$ with $B(m)$ being the Bogoliubov transformation discussed in \Eq{eq:Bog1} and Appendices \ref{app:analytics} and \ref{app:quantumalgo}.
\begin{figure}[t]
\begin{center}
\includegraphics[scale=0.525]{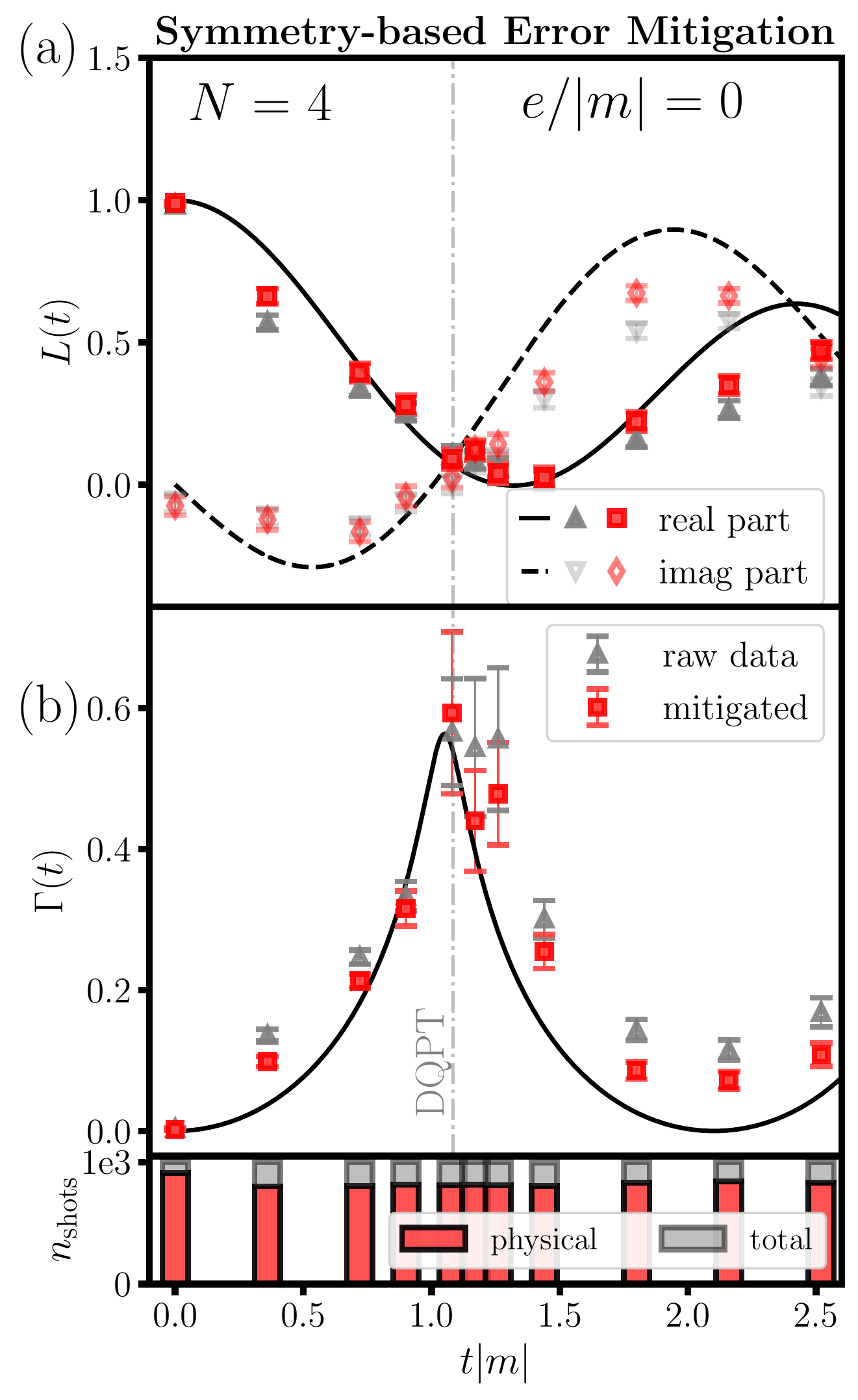}
\caption{(a) Real and imaginary parts of the Loschmidt echo $L(t)$, quantum-computed with IonQ, for $N=4$, $|m|  a=0.9$ and $e=0$. Unmitigated raw data (gray triangles) are compared with symmetry-based error-mitigated results (red squares). (b) Resulting rate function $\Gamma(t)$. To obtain any data point, $n_{\rm shots}=1000$ shots are performed.}
\label{fig:LoschmidtN4freeSymErrMit}
\end{center}
\end{figure}
\begin{figure}[t]
\begin{center}
\includegraphics[scale=0.525]{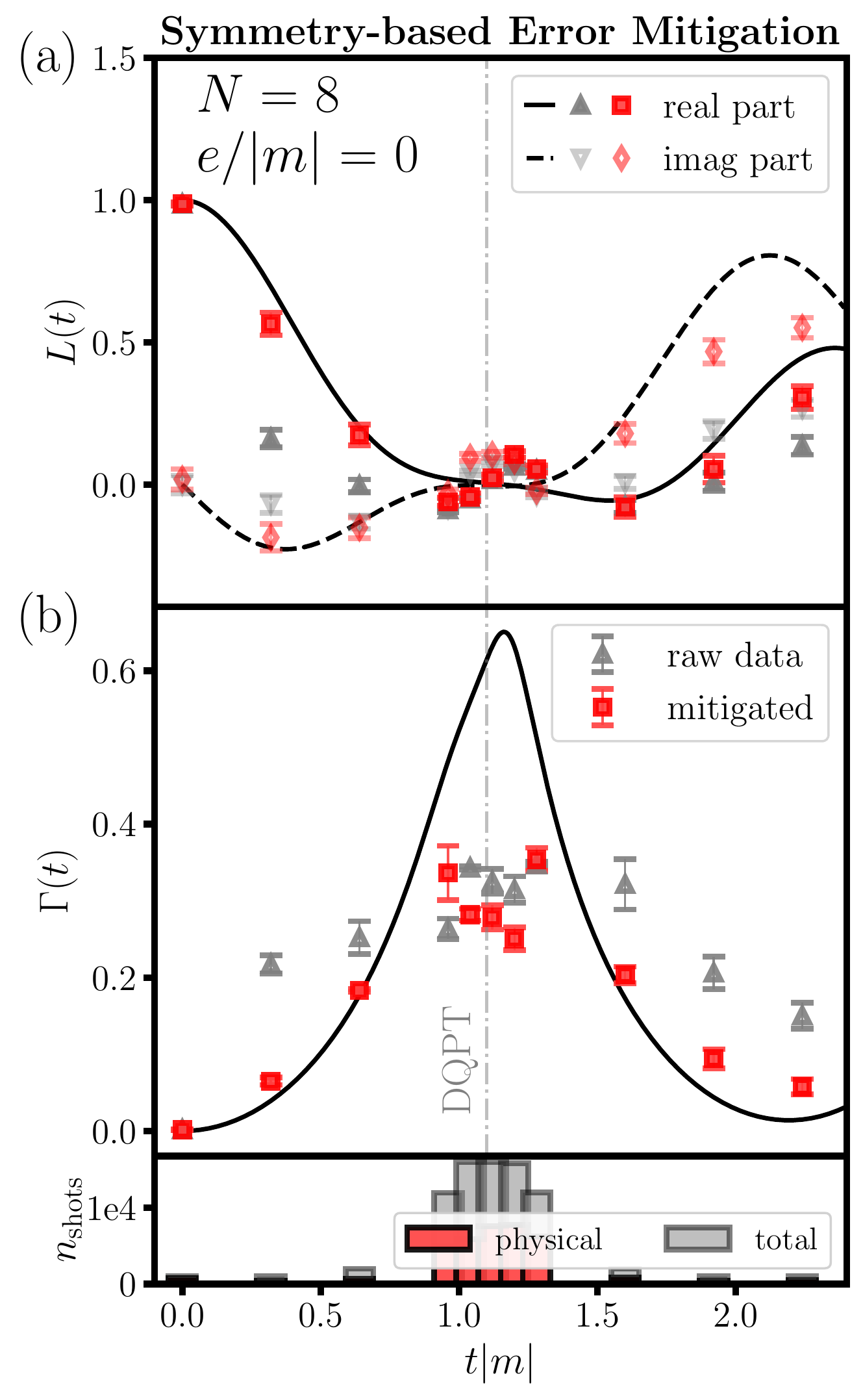}
\caption{(a) Real and imaginary parts of the Loschmidt echo $L(t)$, quantum-computed with IonQ, for $N=8$, $|m|  a=0.8$ and $e=0$. Unmitigated raw data (gray triangles) are compared with symmetry-based error-mitigated results (red squares). (b) Resulting rate function $\Gamma(t)$. The number of shots yielding physical and unphysical results is shown in the bottom panel.}
\label{fig:LoschmidtN8freeSymErrMit}
\end{center}
\end{figure}
Measuring $ \sigma^x$ and $\sigma^y$ on the ancilla qubit yields real and imaginary part of $L(t)$, respectively. Additionally measuring the $N$ qubits representing the system, which are in a superposition of time-evolved and non-time-evolved states, allows us to perform a simple error-mitigation scheme: For the quench gate $Q = \prod_{q=-N/4}^{N/4-1} Q_q$, every $Q_q$ only mixes 0- and 2-particle states, and furthermore, $H_0$ preserves particle number. Therefore, the system's $N$ qubits encoding $| \Psi\rangle$ are in a product state, with the 2-qubits per $q$ being in either in a 0- or 2-particle state. Measurements that yield 1-particle states indicate a bit-flip error of the device. These can simply be discarded and the remaining measurements re-weighted. The results of this procedure for $L(t)$ at $e=0$ are summarized in \Figs{fig:LoschmidtN4freeSymErrMit}{fig:LoschmidtN8freeSymErrMit}, demonstrating significant improvement. We also show the statistical error assuming a binomial distribution, which slightly increases with postselection as the overall statistics is reduced. 

\begin{figure}[t]
\begin{center}
\includegraphics[scale=0.525]{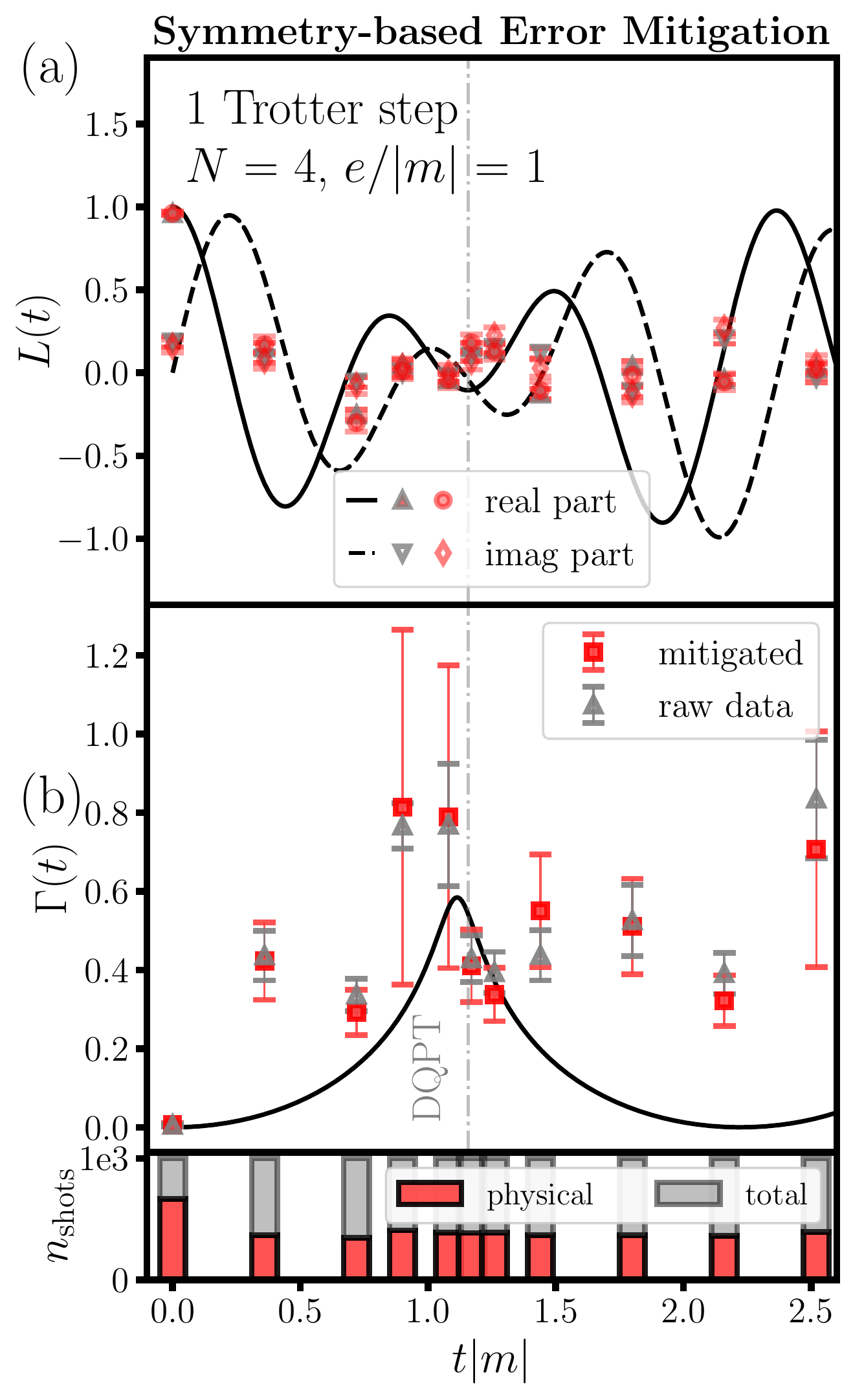}
\caption{(a) Real and imaginary parts of the Loschmidt echo $L(t)$, quantum-computed with IonQ, for $N=4$, $|m|  a=0.9$ and the non-interacting case $e/|m|=1$. Unmitigated raw data (gray triangles) are compared with symmetry-based error-mitigated results (red squares). (b) Resulting rate function $\Gamma(t)$. To obtain any data point, $n_{\rm shots}=1000$ shots are performed.}
\label{fig:LoschmidtInteractingSymmErrMit}
\end{center}
\end{figure}

For $e/|m|>0$ and one Trotter step,
\begin{align}\label{eq:interrormit}
  |\Psi\rangle =& | 0 \rangle \, F B(-m) Q \, | {\rm GS}(m) \rangle
  \nonumber  + 
 | 1 \rangle\, e^{-i H_I \, t}  F B(-m) \\
 & \times e^{-i H_0(-m) \, t}  Q \, | {\rm}GS(m)\rangle\,,
\end{align}
where $F$ is the fermionic Fourier transform and $H_I$ is the interaction part of the Hamiltonian in positions space. The system state, $|\Psi \rangle$, is in a superposition in position space, where both the time-evolved and non-evolved components have total particle number $N/2$ (`half-filling'). Again, particle-number measurement allows
identifying bit-flip errors which can be discarded. Results are shown in \Fig{fig:LoschmidtInteractingSymmErrMit}, yielding no improvement.  

The same error-mitigation scheme is applied in Sec.~\ref{sec:TO}, when computing the NECFs $g_q(t)$ (\Eq{eq:correlator}) and  the topological charge $\nu(t)$ (\Eq{eq:deftopologicalorderparam}).  In the non-interacting limit $e=0$, where  $g_q(t)$ is computed mode-by-mode with shallow (3-qubit) circuits, the effect of this symmetry-based error-mitigation scheme is minimal (only $\approx 5-15\%$ of events contain number-violating errors, see \Fig{fig:corr}).

We note that this error-mitigation procedure was previously explored in trapped-ion computation
of the Schwinger model with up to six qubits~\cite{nguyen2022digital}. A technique based on Ref.~\cite{tran2021faster} was also tried in the same work, which inserts random rotations between Trotter steps of evolution to average away the error. Nonetheless, no significant improvement was observed, pointing to the time de-correlated incoherent nature of noise in the current trapped-ion hardware.
A number of different (hardware-suitable) error-mitigation schemes are available to potentially improve our computation, see e.g., Refs.~\cite{temme2017error,li2017efficient} for a recent overview of various techniques. These include randomized compiling and dynamical decoupling~\cite{viola1998dynamical,duan1999suppressing,viola1999dynamical}. Randomized compiling can potentially improve our results, albeit it significantly increases simulation cost, which we therefore have not explored. Dynamical decoupling may be less profitable given the architecture and the dominant error source that we identify below to be inaccurate gate operations. The effect of gate errors could, in fact, potentially be amplified by this error-mitigation scheme. As a result, the effective and computationally economic symmetry-based post-selection is the only error-mitigation scheme tried in this work.

\section{Statistical Uncertainties}\label{app:staterror}
\noindent
This section presents details of our statistical measurement-error estimation. Only one data set per observable has been obtained in this work, hence the statistical uncertainty cannot be inferred directly by comparing many independent measurements. In part, this is because  calibration drifts make it  difficult to compare  data sets at different times. We balanced getting data in quick succession with relatively stable systematic uncertainties, versus getting a large amount of data for statistical analysis.

\begin{figure}[t]
\begin{center}
\includegraphics[scale=0.45]{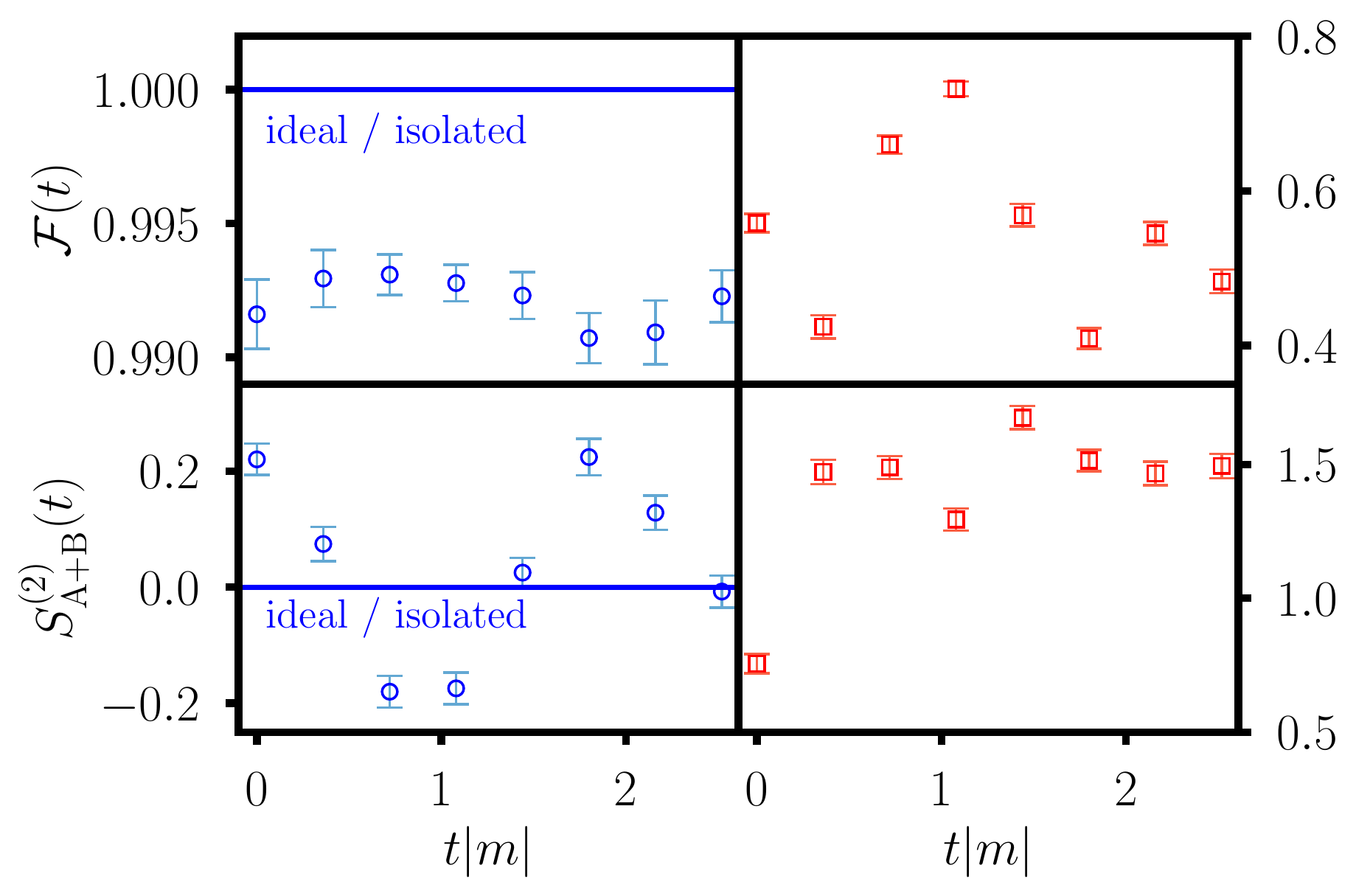}
\caption{Close-up of the lower two panels of \Fig{fig:RenyiFidelityN4}, showing the nearly negligible statistical shot-noise uncertainty. For the simulator data (left panel), the only other source of deviation from the expected $\mathcal{F}=1$ and $S^{(2)}_{A+B}=0$ is due to the finite number of random basis changes $n_{\rm CUE}$. While our allocation with IonQ did not allow us to take more data, this figure shows that the combined statistical error is very small.}
\label{fig:RenyEntrN4Err}
\end{center}
\end{figure}

Statistical uncertainties are estimated as follows. For observables based on single-qubit measurements, such as the Loschmidt echo and topological index in Secs.~\ref{sec:nonequaltime} and \ref{sec:TO}, a binominal distribution is assumed, i.e. the uncertainty is $\Delta p =\sqrt{p_0 p_1 n_{\rm shots}}$, where $p_0$ ($p_1=1-p_0$)  is  the probability of measuring $0$ ($1$). Random-measurement-based results, such as Renyi entropies and fidelities, and the EHT analysis are investigated as follows: Given the probabilities $P_{U}(s)$, $n_b=100$ bootstrap copies are created, i.e., one can sample from $P_{U}(s)$ to create additional `measurements'. Statistical uncertainties are obtained for $S^{(2)}$,  $\mathcal{F}$, and $S^{(2)}_{A+B}$ by computing \Eq{eq:EHTprotocol} and \Eqs{eq:defrenyientropy}{eq:fidelity} for every bootstrap sample individually. The uncertainty is the standard deviation in the limit $n_b\rightarrow \infty$, e.g., $\Delta \mathcal{F}\equiv  \sqrt{\sum_{i_b=0}^{n_b-1} ( \bar{\mathcal{F}} - \mathcal{F}_{i_b}   )^2/n_b}$
where $\bar{\mathcal{F}}=\frac{1}{n_b}\sum_{i_b=0} ^{n_b-1}\mathcal{F}_{i_b}$ where the sum is over bootstrap samples. Shot-noise uncertainties are very small and difficult to discern in the lower two panels  of \Fig{fig:RenyiFidelityN4}. A close-up of the data is shown \Fig{fig:RenyEntrN4Err}.
The  procedure is repeated for the EHT analysis. $n_b=100$ samples are used, except for $N=8$ where due to the immense classical post-processing cost, only $n_b=10$ samples are tried. 

\section{Simple Error Models and Entanglement Tomography}\label{app:error}
\noindent
Because of the indirect Cloud access to the hardware, it is difficult to identify the origin of deviation of experimental results from the theoretical expectations, which may be due to drifting machine parameters as well as ion movement and heating. In this appendix, via additional entanglement-tomography studies for $N=4$ lattice sites, we address, in a simple error model, local decoherence. As will be shown, this simple model is not sufficient to describe the observed reduced fidelities.
\begin{figure}[t]
\begin{centering}
\includegraphics[scale=0.525]{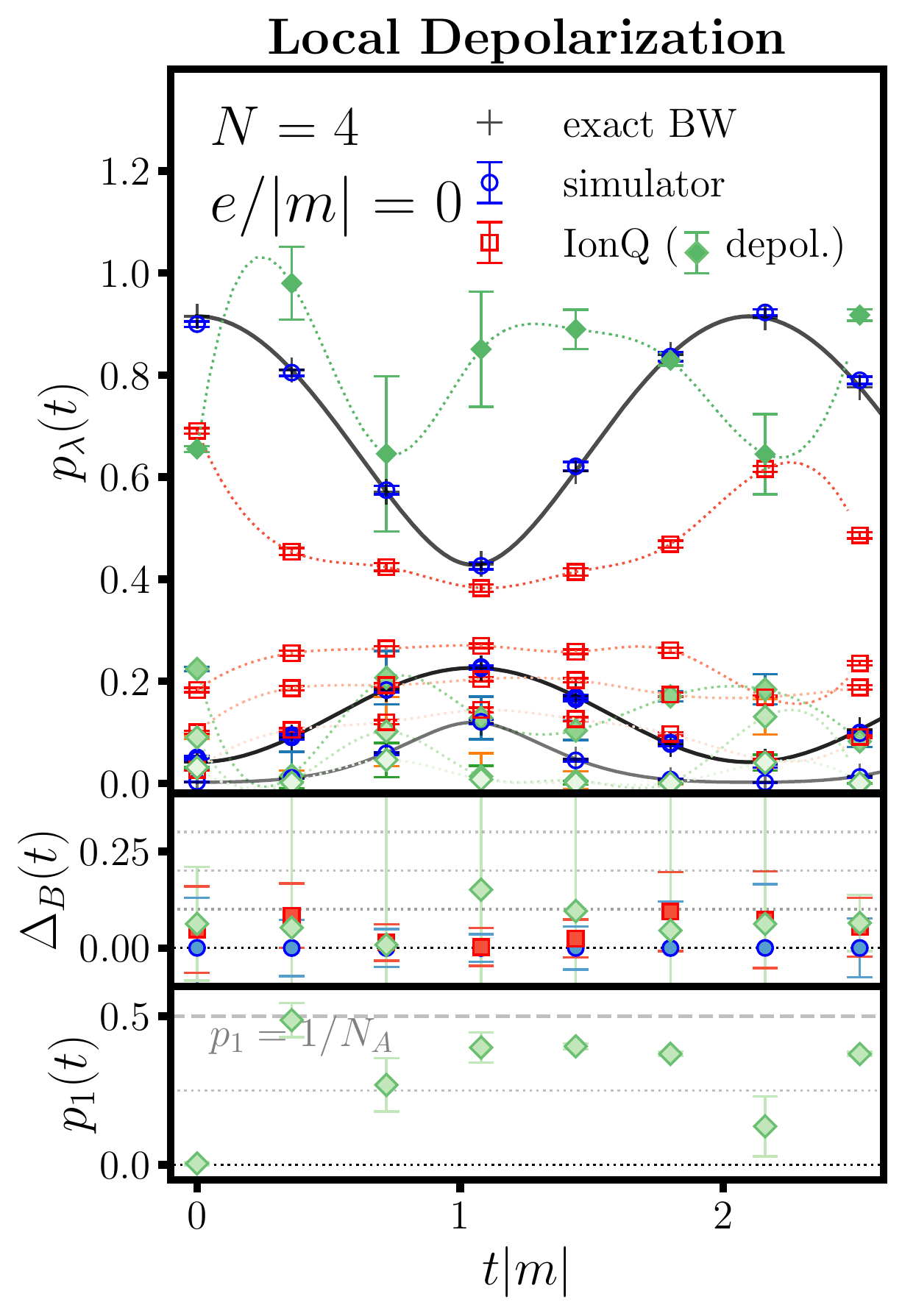}
\caption{
The top panel shows the Schmidt spectrum $p_\lambda(t)$ of $\rho_A(t)$ of a bi-partition of $N_A=N/2$ lattice sites. We compare IonQ data assuming no depolarization $p_1=0$ (red symbols) with the same data but allowing for depolarization $p_1 >0$ in the BW ansatz in \Eq{eq:locdep} (green symbols) as well as exact BW ansatz (black plus symbols), for $n_{\rm CUE}=25$ and $n_{\rm shots }=1000$. The middle and bottom panels show the Bhattacharyya distance $\Delta_B(t)$ relative to the exact BW results, and the depolarization infidelity $p_1(t)$, respectively. The dashed line in the bottom panel represent the maximum value of the depolarization probability for a system of size $N_A$.}
\label{fig:EHTN4_depol}
\end{centering}
\end{figure}

The three simple local decoherence error models employed replace the density matrix with modified ones as~\cite{kokail2021entanglement}
\begin{align}
\label{eq:locdep}
   \mathcal{M}^{\rm depol.}
   [\rho] = (1-p_1 N) \rho + p_1 \sum_{i=1}^N \text{Tr}_i (\rho) \frac{\mathbb{I}_i}{2}\,,
   \\\label{eq:locbitflip}
   \mathcal{M}^{\rm bit~flip}
   [\rho] = (1-p_2 N) \rho + p_2 \sum_{i=1}^N \sigma^x_i \, \rho \, \sigma^x_i\,, 
   \\\label{eq:locphase}
   \mathcal{M}^{\rm phase}
   [\rho] = (1-p_3 N) \rho + p_3 \sum_{i=1}^N \sigma^z_i \, \rho \, \sigma^z_i \,,
\end{align}
where $p_{1,2,3} \in[0,1/N]$, $\text{Tr}_i$ is the partial trace over qubit $i$, and $N$ is the number of qubits. A more complete approach would be solving a Lindblad equation of the quantum circuit, see e.g., Ref.~\cite{jensen2022dynamical} for a study of the presented model, but this route will not be explored here. 
\begin{figure}[t]
\begin{centering}
\includegraphics[scale=0.525]{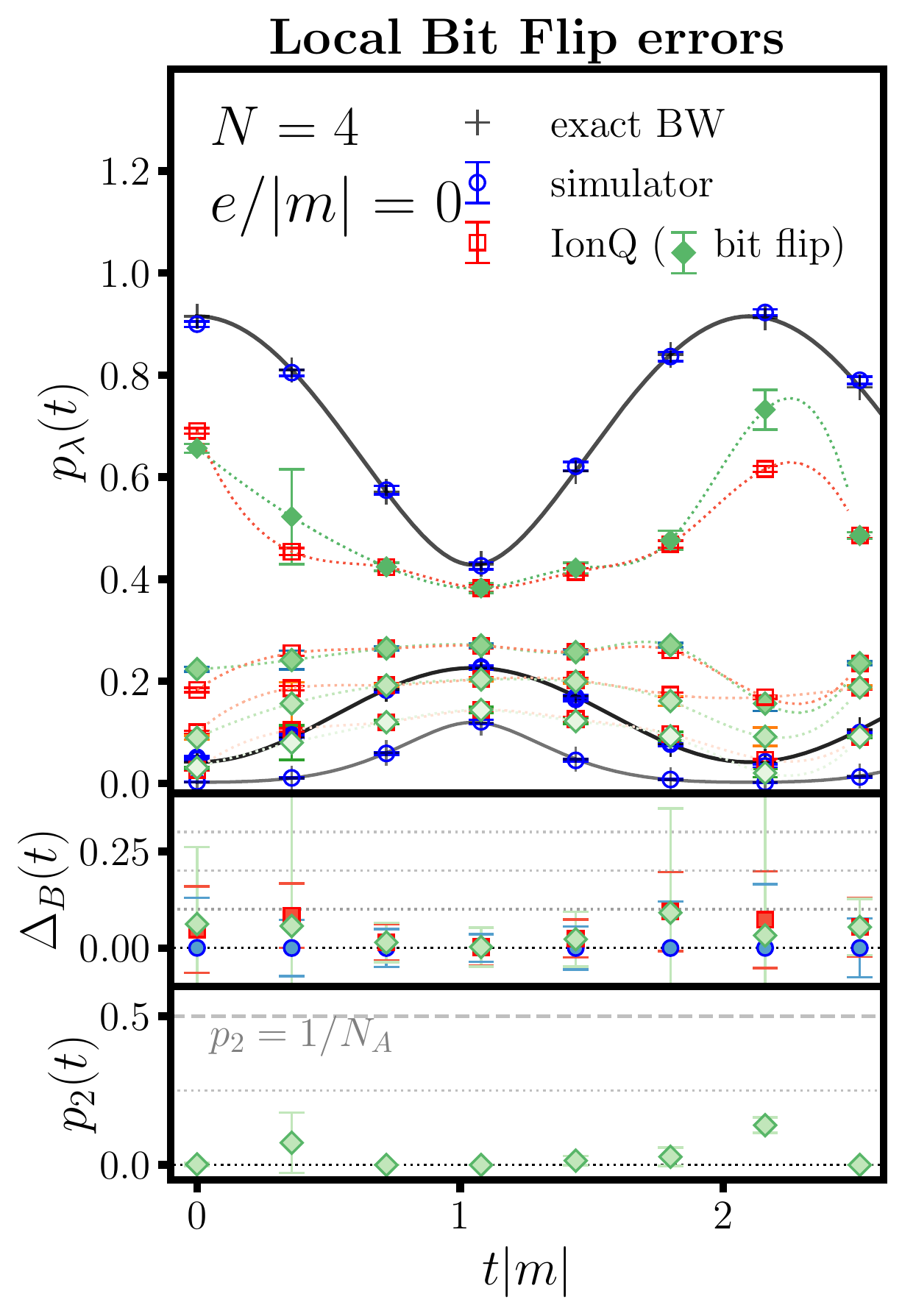}
\caption{
The top panel shows the Schmidt spectrum $p_\lambda(t)$ of $\rho_A(t)$ of a bi-partition of $N_A=N/2$ lattice sites. We compare IonQ data assuming no bit-flip error $p_2=0$ (red symbols) with the same data but allowing for bit-flip error $p_2 >0$ in the BW ansatz in \Eq{eq:locdep} (green symbols) as well as exact BW ansatz (black plus symbols), for $n_{\rm CUE}=25$ and $n_{\rm shots }=1000$. The middle and bottom panels show the Bhattacharyya distance $\Delta_B(t)$ relative to the exact BW results, and the bit-flip-error infidelity $p_2(t)$, respectively. The dashed line in the bottom panel represent the maximum value of the bit-flip-error probability for a system of size $N_A$.}
\label{fig:EHTN4_X}
\end{centering}
\end{figure}
\begin{figure}[t]
\begin{centering}
\includegraphics[scale=0.525]{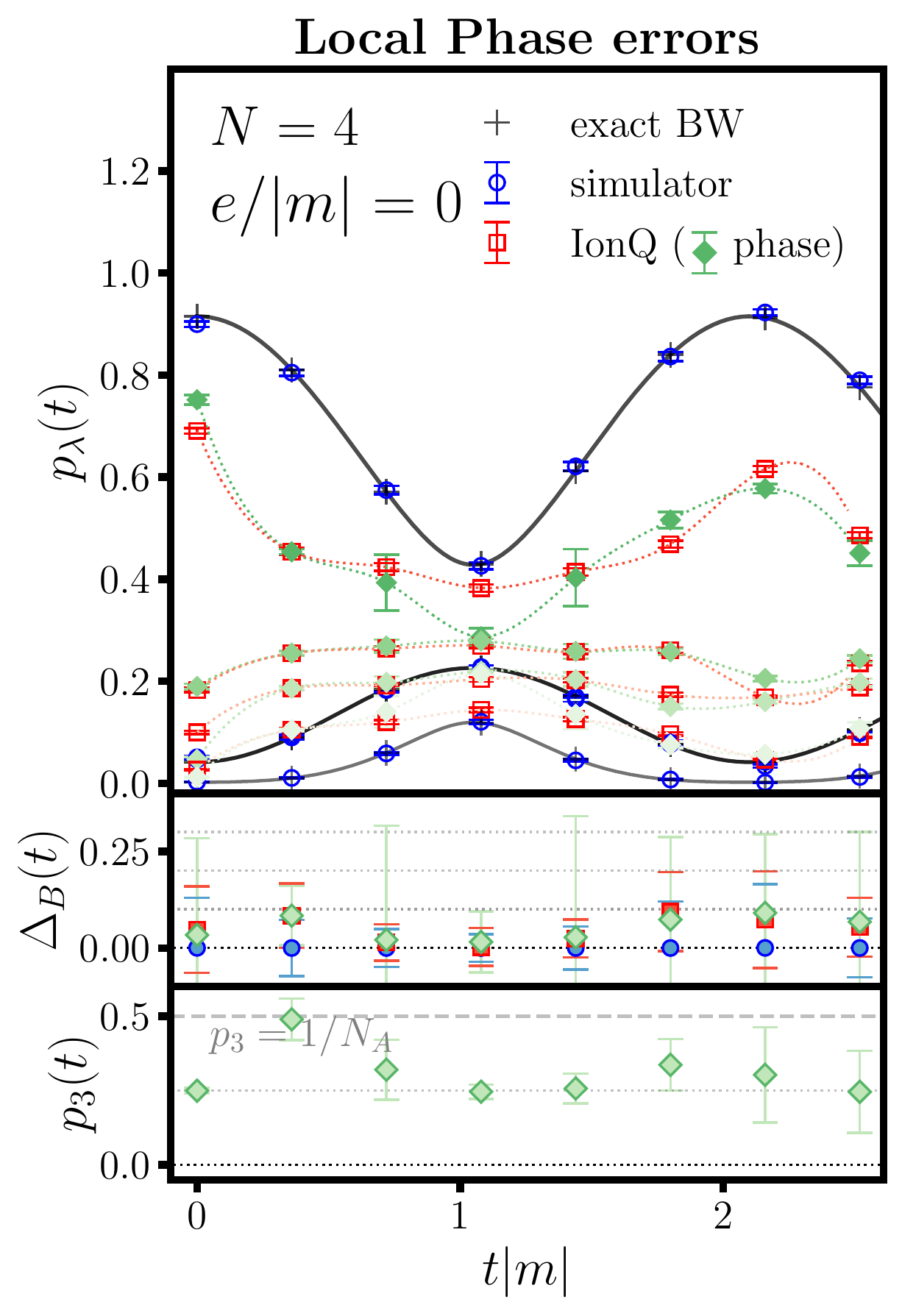}
\caption{
The top panel shows the Schmidt spectrum $p_\lambda(t)$ of $\rho_A(t)$ of a bi-partition of $N_A=N/2$ lattice sites. We compare IonQ data assuming no phase error $p_3=0$ (red symbols) with the same data but allowing for phase error $p_3 >0$ in the BW ansatz in \Eq{eq:locdep} (green symbols) as well as exact BW ansatz (black plus symbols), for $n_{\rm CUE}=25$ and $n_{\rm shots }=1000$. The middle and bottom panels show the Bhattacharyya distance $\Delta_B(t)$ relative to the exact BW results, and the phase-error infidelity $p_3(t)$, respectively. The dashed line in the bottom panel represent the maximum value of the phase-error probability for a system of size $N_A$.}
\label{fig:EHTN4_Z}
\end{centering}
\end{figure}
\begin{figure}[t]
\begin{centering}
\includegraphics[scale=0.525]{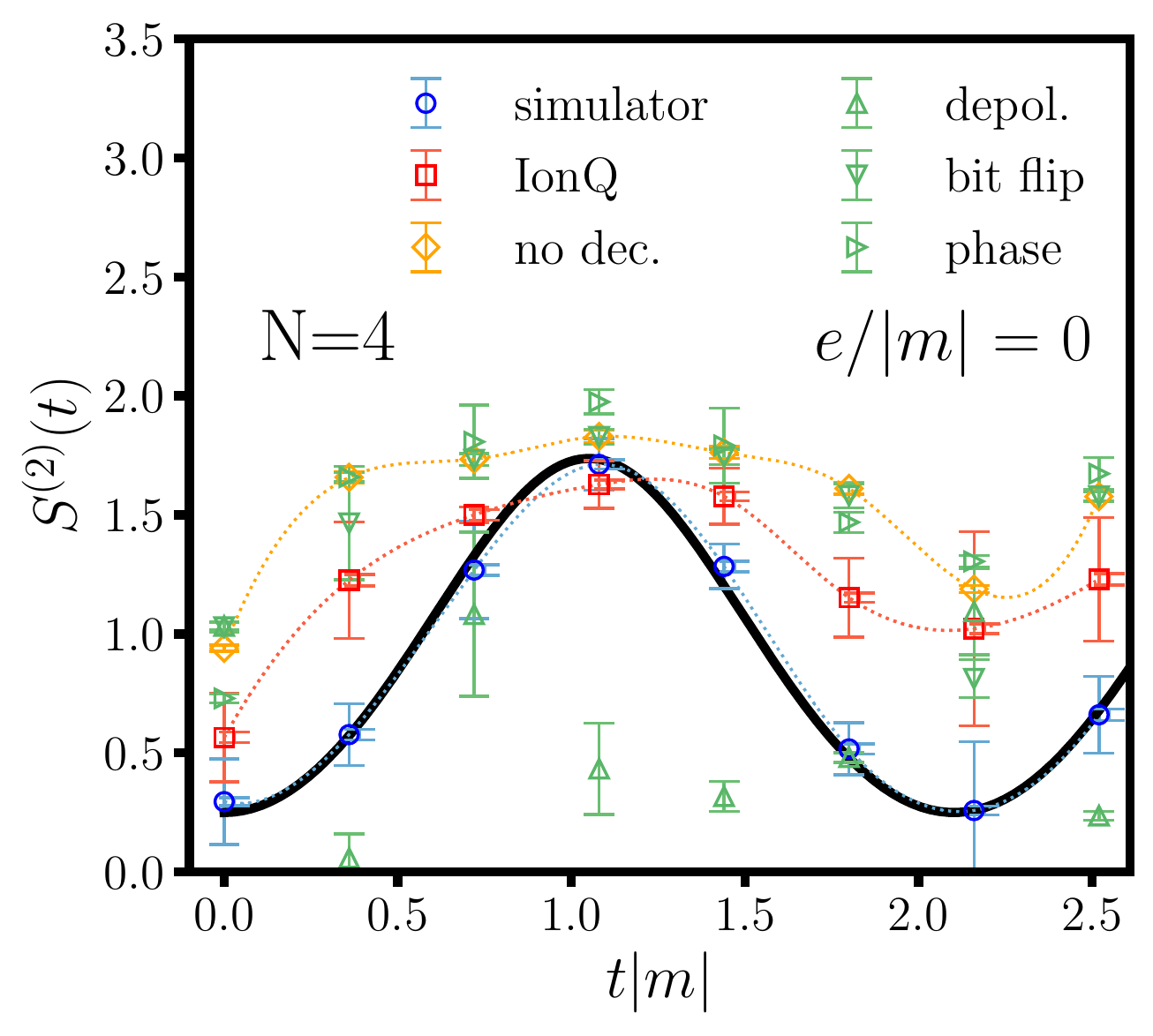}
\caption{
The second-order R\'enyi entropy extracted from data via \Eq{eq:EHTprotocol} (red symbols) versus that reconstructed from the BW ansatz via Eqs.~(\ref{eq:HA-def}-\ref{eq:variation}) with no decoherence effect included (orange diamonds), and including local depolarization, bit-flip, and phase error channels \Eqs{eq:locdep}{eq:locphase} (green symbols). Blue symbols show simulator data, while black curves represent the exact results.}
\label{fig:RenyiFidelityN4_dec}
\end{centering}
\end{figure}
\begin{figure}[t!]
\begin{centering}
\includegraphics[scale=0.525]{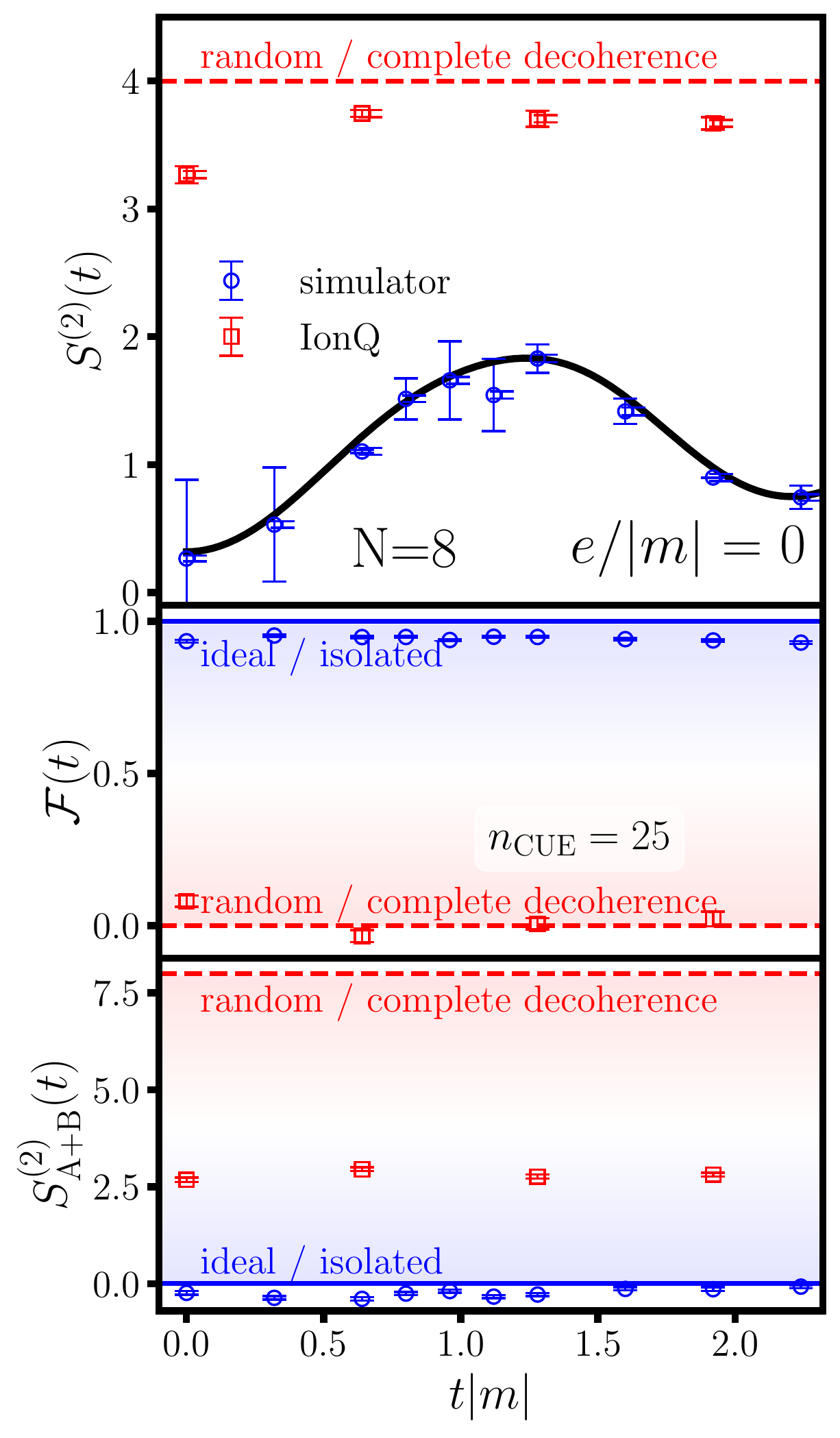}
\caption{The top panel shows R\'enyi entropy averaged over subsystems $A$ and $B$, $S^{(2)}(t)\equiv\frac{1}{2}(S_A^{(2)}(t)+S_B^{(2)}(t))$ (with $S_A^{(2)}$ and similarly $S_B^{(2)}$ defined in \Eq{eq:defrenyientropy}), for $N=8$, $|m|  a =0.8$, $e=0$, $n_{\rm CUE}=25$, and $n_{\rm shots}=1000$, including simulator (blue symbols) and  IonQ (red symbols) results. The middle panel depicts  fidelity $\mathcal{F}(t)$ (\Eq{eq:fidelity}). The bottom panel shows the  R\'enyi entropy of the full system, $S^{(2)}_{\rm A+B}$ (relative to the environment). Blue horizontal lines in the middle and bottom panels indicate ideal results, a horizontal red line indicates zero fidelity or maximal entropy ($\rho(t) = \mathbb{I}/2^N$), respectively. See \Fig{fig:RenyEntrN8Err} for a close-up of the lower two panels.}
\label{fig:RenyiFidelityN8}
\end{centering}
\end{figure}
\begin{figure}[t]
\begin{center}
\includegraphics[scale=0.45]{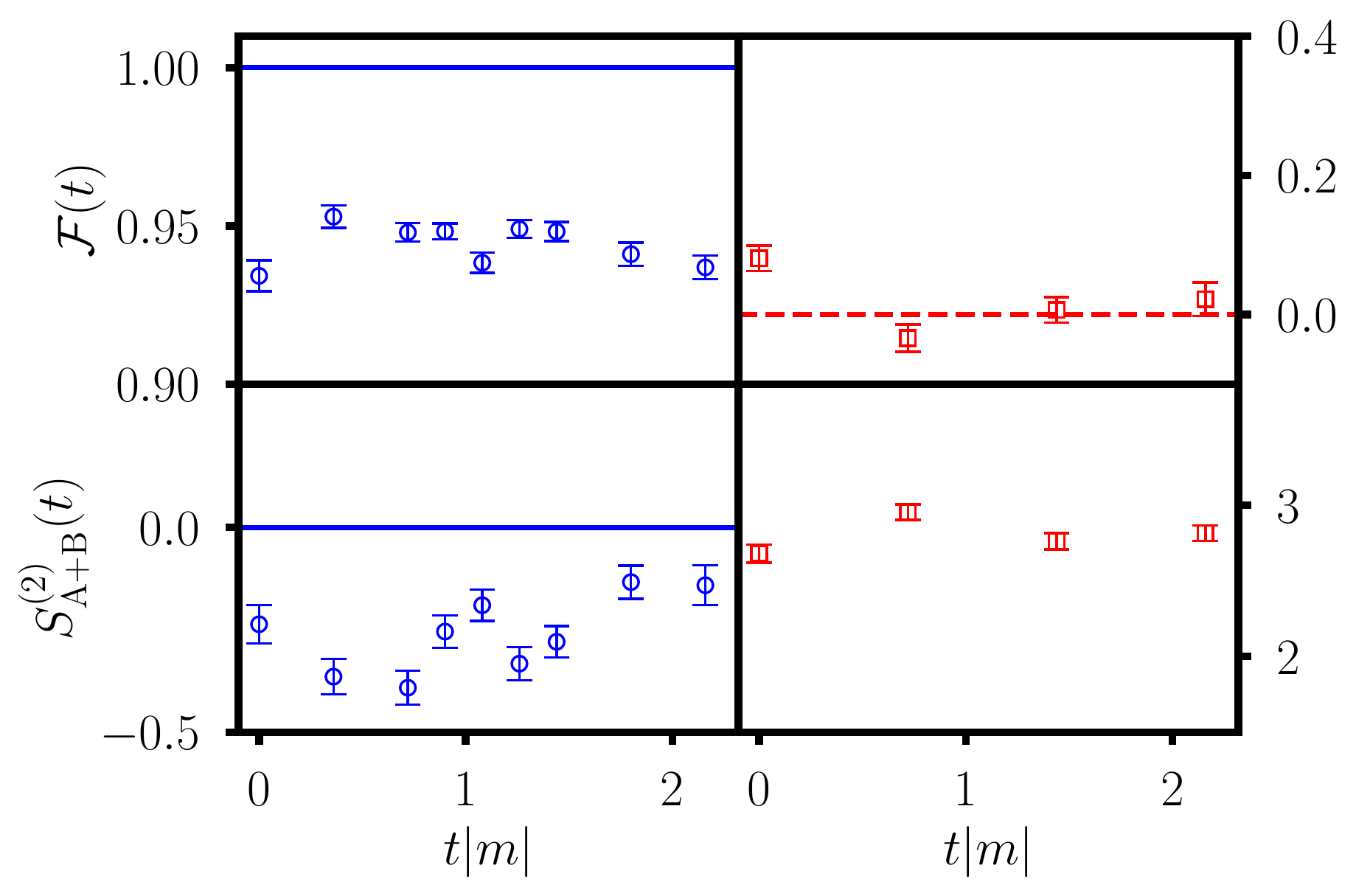}
\caption{Close-up of the lower two panels of \Fig{fig:RenyiFidelityN8}.}
\label{fig:RenyEntrN8Err}
\end{center}
\end{figure}

\begin{figure}[t!]
\begin{centering}
\includegraphics[scale=0.525]{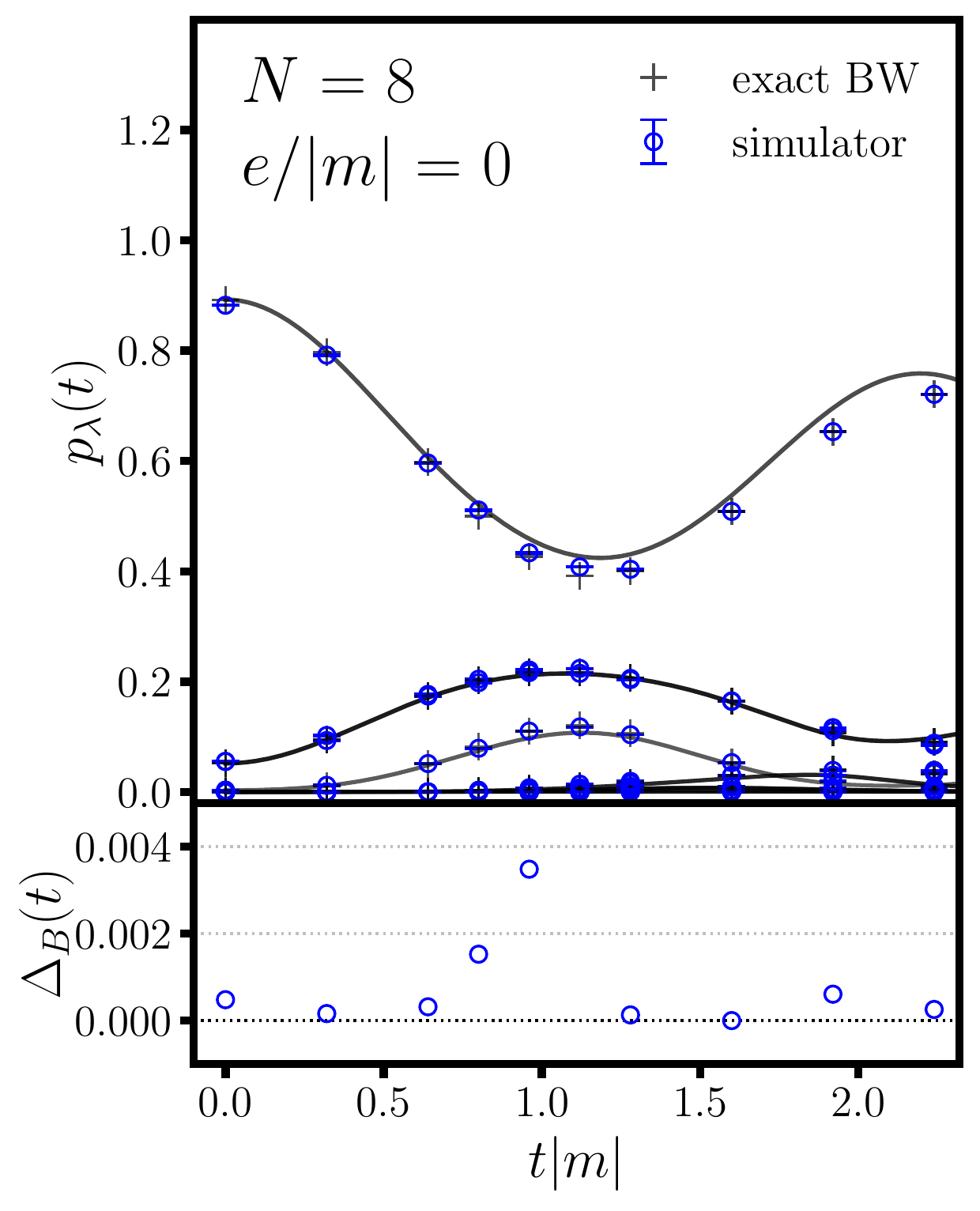}
\caption{The top panel shows the eigenvalue spectrum $P_\lambda(t)$ of the reduced density matrix $\rho_{A}(t)$ of a bi-partition, with $N=8$ and $e=0$, as a function of time, compared with the exact results (black lines), exact results reconstructed from the BW ansatz (black plus symbols), and simulator (blue  circles), with $n_{\rm shots}=2000$ and $n_U=25$. The bottom panel shows the Bhattacharyya distance $\Delta_B(t)$ between exact BW results (blue circles). We do not plot the IonQ data in this figure because the quality of the results was too low to allow reconstructing the EH using the BW ansatz.}
\label{fig:EHTN8}
\end{centering}
\end{figure}
\begin{figure}[t!]
\begin{center}
\includegraphics[scale=0.525]{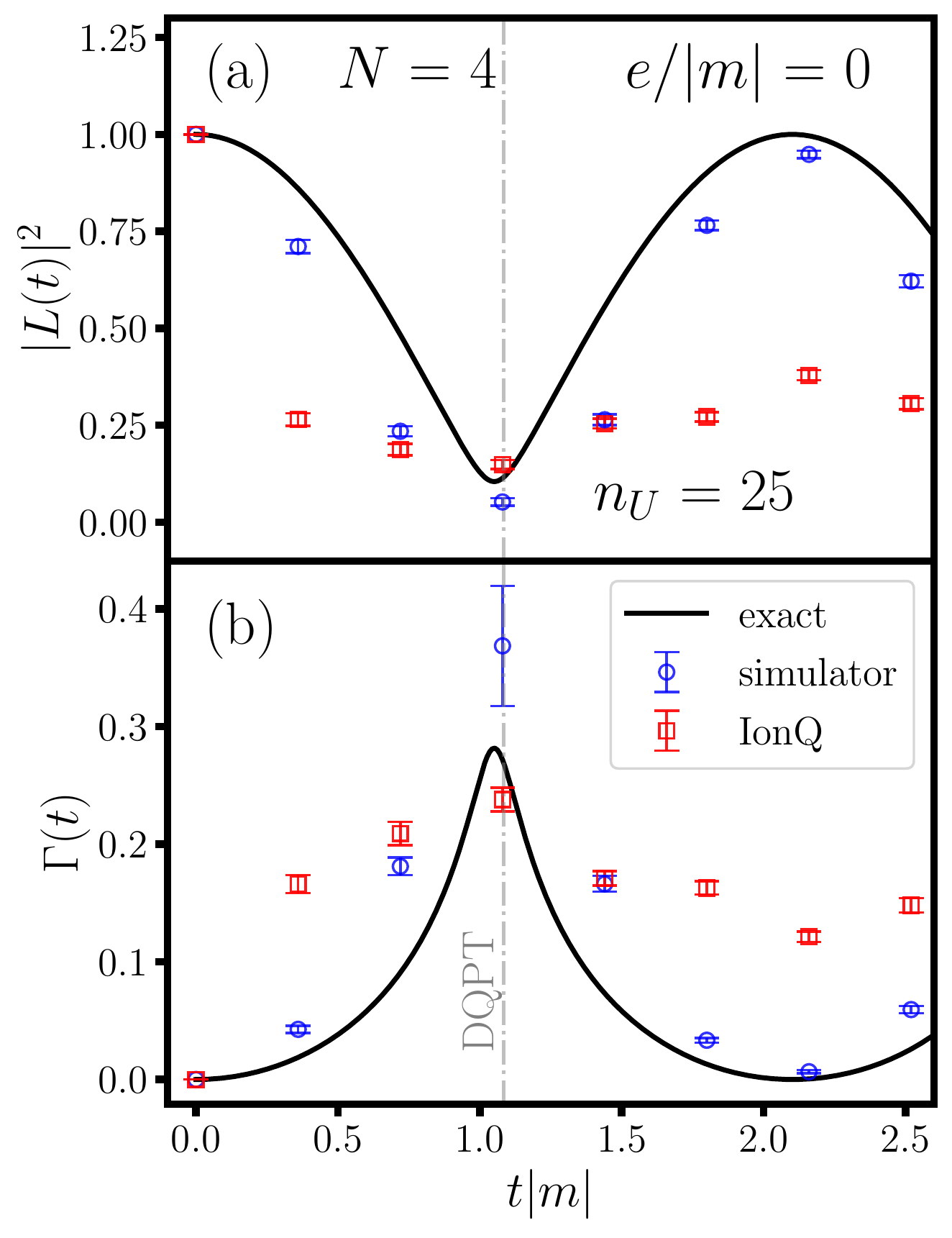}
\caption{(a) The absolute-value squared of the Loschmidt echo, $|L(t)|^2=\text{Tr}[\rho(0)\rho(t)]$, reconstructed using the ET protocol, \Eq{eq:EHTprotocol}, thus avoiding the Ramsey scheme employed and requiring no ancilla qubits. (b) The rate function for the same data. To compute the two-time observable, the same data presented before in Figs.~\ref{fig:RenyiFidelityN4} and \ref{fig:EHTN4} are used.}
\label{fig:LoschmidtEHT}
\end{center}
\end{figure}

The EHT analysis using the BW ansatz presented in section \ref{sec:EHT} is repeated using $\mathcal{M}[\rho_A]$ instead of $\rho_A$ in the fit, separately for three local decoherence error models \Eqs{eq:locdep}{eq:locphase}. Here, $\rho_A$ is given by the BW ansatz in \Eq{eq:ansatzT} as before, and $p_{1,2,3}$ are additional free fit parameters. Figure~\ref{fig:EHTN4_depol} shows the results this study, using the same data as in the main text for $|m|  a =0.9$,  $e=0$, $n_{\rm CUE}=25$, and $n_{\rm shots }=1000$, generalized to include local depolarization via \Eq{eq:locdep}. Other error sources are not considered. IonQ data without allowing for depolarization noise in the fit (red symbols) are compared to results using \Eq{eq:locdep} (green symbols). Blue symbols are simulator data and black curves denote exact results. The middle and bottom panels show Bhattacharyya distance $\Delta_B(t)$ (relative to the exact BW results) and the depolarization probability $p_1(t)$. The quality of the data is not improved by the ansatz and $p_1(t)$ varies significantly from data point to data point. Figures~\ref{fig:EHTN4_X} and \ref{fig:EHTN4_Z} show the same analysis for the bit-flip (\Fig{fig:EHTN4_X}) and phase (\Fig{fig:EHTN4_Z}) error channels. Likewise, including the bit-flip and phase error channels does not substantially improve the reconstruction of $\rho_A(t)$. Finally, Fig.~\ref{fig:RenyiFidelityN4_dec} shows R\'enyi entropy, extracted from \Eq{eq:EHTprotocol}, versus that computed from  $\rho_A(t)$ with the BW ansatz. Shown are fits without  error channels (orange diamonds), local depolarization (green filled diamonds), local bit-flip errors (green empty diamonds), and phase errors (green empty circles). Blue symbols show the simulator data while black curves represent exact results.

Overall, these plots demonstrate that simple ans\"atze in \Eqs{eq:locdep}{eq:locphase} fail to explain the deviations of the IonQ results. Instead, a major source of uncertainty appear to be imperfections of 1- and 2-qubit gates. Especially, 2-qubit entangling operations involve long pulse durations, and the ions may heat and change their positions as well machine parameters may drift during long circuit operations. Such effects require a more detailed modeling of the noise, which nonetheless requires more insight into the machine characteristics and the underlying operational framework that are generally not available to Cloud users. Finally, it is worth noting that the precision of 1-qubit rotations is important for the accuracy of the EHT scheme. Following the strategy of Ref.~\cite{brydges2019probing}, improving robustness against miscalibration and drift by concatenating random unitaries did not lead to a significant improvement 
but will be investigated in greater detail in future work.
Finally, state preparation and readout errors may need to be studied in this context too.

\section{Further Results on Entanglement Tomography}\label{app:EHTdetails}
\noindent
The results of the entanglement tomography analysis for the largest system with $N=8$ lattice sites are presented in this appendix. The significant gate count, as evident from Table~\ref{table:complexity}, is dominated by the fermion Fourier transform.
Results are summarized in \Fig{fig:RenyiFidelityN8}, where the top panel shows R\'enyi entropy of a bi-partition of $N_A=N/2=4$ sites comparing exact results (black curves), simulator data (blue symbols), and IonQ data (red symbols). The middle and bottom panel show fidelity and total entanglement entropy, respectively. The horizontal red line indicates the expectation from a maximally mixed state, $\rho_A(t) = \mathbb{I}/2^{N_A}$. 
Near complete failure to quantum compute the desired target state is apparent.

Regardless of hardware implementation, the BW ansatz for the reconstruction of the reduced density matrix still works reasonably well for the case of $N=8$. As shown in \Fig{fig:EHTN8}, using a (perfect) hardware simulator with $n_{\rm CUE}=25$, $n_{\rm shots}=2000$, minimal deviations between the exact (black curves), and the BW ansatz (ideal: black plus symbols, simulator: blue circles) are observed for the reconstruction of the time-evolved state $\rho_A(t)$. While the agreement is excellent overall, these deviation can systematically be improved by including slightly less local terms in the BW ansatz in \Eq{eq:ansatzT}, as discussed in Ref.~\cite{kokail2021entanglement}.

Finally, performing random measurements provides another avenue for computing non-equal time observables. Specifically, noting that
\begin{align}
|L(t)|^2 = |\langle \psi(0) | \psi(t) \rangle|^2 = \text{Tr}[\rho(0) \rho(t)]\,,
\end{align}
Loschmidt echo and rate function are computed from \Eq{eq:EHTprotocol} using $P_{\mathscr{U}}(s)$ obtained at different times, instead of an ancilla-based interferometry scheme. The result is shown in \Fig{fig:LoschmidtEHT}, reconstructed from the same data as in Figs.~\ref{fig:RenyiFidelityN4} and \ref{fig:EHTN4}. The IonQ results (red symbols) are significantly less accurate than with the Ramsey scheme, because of a larger gate count compared to the direct computation in section \ref{sec:nonequaltime}.

\pagebreak
\newpage
\bibliography{references}

\end{document}